\documentclass[a4paper,12pt]{article}
\usepackage{jcappub}
\usepackage{amsthm,graphicx}
\usepackage{epsfig}
\usepackage{latexsym, amssymb} 
\usepackage{amsmath}
\def\beq{\begin{equation}}
\def\eeq{\end{equation}}
\def\br{\begin{eqnarray}}
\def\er{\end{eqnarray}}
\def\benu{\begin{enumerate}}
\def\efnu{\end{enumerate}}
\def\nn{\nonumber}

\def\l{\left}
\def\r{\right}

\begin{document}
\title{Consistency of the Planck CMB data and $\Lambda$CDM cosmology}
 \author[a,b]{Arman Shafieloo}
 \author[c]{Dhiraj Kumar Hazra} 

\affiliation[a]{Korea Astronomy and Space Science Institute, Daejeon 34055, Korea}
\affiliation[b]{University of Science and Technology, Daejeon 34113, Korea}
\affiliation[c]{AstroParticule et Cosmologie (APC)/Paris Centre for Cosmological Physics, Universit\'e
Paris Diderot, CNRS/IN2P3, CEA/lrfu, Observatoire de Paris, Sorbonne Paris Cit\'e, 10, rue Alice Domon et Leonie Duquet, 75205 Paris Cedex 13, France}

\emailAdd{shafieloo@kasi.re.kr, dhiraj.kumar.hazra@apc.univ-paris7.fr}

\abstract 
{We test the consistency between Planck temperature and polarization power spectra and the concordance model of $\Lambda$ Cold Dark Matter cosmology ($\Lambda$CDM) within the 
framework of Crossing statistics. We find that Planck TT best fit $\Lambda$CDM power spectrum is completely consistent with EE power spectrum data while EE best fit $\Lambda$CDM 
power spectrum is not consistent with TT data. However, this does not point to any systematic or model-data discrepancy since in the Planck EE data, uncertainties are much larger 
compared to the TT data. We also investigate the possibility of any deviation from $\Lambda$CDM model analyzing the Planck 2015 data. Results from TT,  TE and EE data analysis 
indicate that no deviation is required beyond the flexibility of the concordance $\Lambda$CDM model. Our analysis thus rules out any strong evidence for beyond the concordance 
model in the Planck spectra data. We also report a mild amplitude difference comparing temperature and polarization data, where temperature data seems to have slightly lower 
amplitude than expected (consistently at all multiples), as we assume both temperature and polarization data are realizations of the same underlying cosmology.}

%\color{red} Setting a prior for the Hubble parameter $H_0=73.24 \pm -1.74$ as suggested by some local observations results to about $2 \sim 3\sigma$ inconsistency between the concordance model and Planck data.\color{black}

\maketitle

\section{Introduction}
%~\cite{isotropy}~\cite{flatness}~\cite{lambda}~\cite{powerlaw}
Looking for evidences for deviation from the predictions of the concordance model is one of the most important tasks of current cosmology. Hence rigorous tests of different
aspects of the concordance model is of vital importance. This can include testing isotropy and homogeneity of the Universe, its metric and flatness, whether dark energy is cosmological constant and also if primordial fluctuations can be described by a power-law spectrum. While some cosmological data are sensitive to a special characteristic of a cosmological model, some other data can be in fact function of combination of model properties. Cosmic Microwave Background data is sensitive to various model properties including metric and curvature, model of dark energy, primordial spectrum and model of re-ionization. While Cosmic Microwave Background (CMB) data can be used trivially to estimate the parameters of a given cosmological model (such as $\Lambda$CDM model), it is still important to check if a model considering all its flexibility and degrees of freedom is consistent to a data or not. In this paper we confront the concordance model of cosmology with Planck 2015 temperature and polarization data~\cite{Planck15data} to test if data suggest any modification to our assumed model. We use Crossing statistics in our analysis as it has been used earlier in similar contexts. In this approach we test consistency of the concordance model to Planck temperature and polarization data by comparing it with its own variations. Finding some variations preferred by the data (with statistical significance) can point towards the necessity of a venture beyond the predictions of the concordance model. This novel approach has been shown to be efficient to deal with the data with complicated correlations and in different contexts~\cite{Shafieloo:2010xm,Shafieloo:2012jb,Shafieloo:2012yh,Shafieloo:2012pm,Hazra:2013oqa,Hazra:2014hma}. In this analysis we use the likelihood code provided by the Planck team and we show that there is no signature of any strong deviation from expectations of the concordance model in Planck temperature and polarization data. This is different from the results of our first analysis in this context using Planck 2013 temperature data~\cite{Planck:2013kta,Ade:2013zuv}
(first release of the data and before correcting the systematics) where we found 2-3$\sigma$ deviations from the concordance model~\cite{Hazra:2014hma}. 
%{Shafieloo:2011zv} with eric 2011
The paper is organized as the following. In the next section we will briefly describe the formalism. Next we present the results of our analysis and we conclude at the end. 

%%%%%%%%%%%%%%%%%%%%%%%%%%%%%%%%%%%%%%%%%%%%%%%%%%%%%%%%%%%%%%%%%%%%%%%%%%%%%%%
\section{Formalism}~\label{sec:formalism}
 
The aim of our paper is two folds. First, we would like to test within the framework of the concordance model of cosmology, whether the temperature and polarization data from Planck are in agreement with each other. Secondly, we look for signatures of deviation from the standard model of cosmology using the Planck angular power spectrum data. To address the first goal, we use the baseline $\Lambda$CDM best fit angular power spectrum (we refer as mean function) from temperature data  modified with a fifth order crossing function as follows :
\beq
 {\cal C}_{\ell}^{\rm TT}\mid_{\rm modified}^{N} ={\cal C}_{\ell}^{\rm TT}\mid_{\rm \Omega_{\rm b}, \Omega_{\rm CDM}, H_0, \tau, A_{\rm s}, n_{\rm s}}~\times ~ T_{N}(C_0,C_1,C_2,...,C_N,\ell).
\label{eq:main} 
\eeq
$T_{N}(C_0,C_1,C_2,...,C_N,\ell)$ is the crossing function and is expanded in the basis of orthogonal Chebyshev polynomials as :
\begin{eqnarray}
T_{\rm 0}(C_0,x)&=&C_0 \nn\\
T_{\rm I}(C_0,C_1,x)&=&T_{\rm 0}(C_0,x)+C_1~x\nn\\
T_{\rm II}(C_0,C_1,C_2,x)&=&T_{\rm I}(C_0,C_1,x)+C_2(2x^2-1)\nn\\
T_{\rm III}(C_0,C_1,C_2,C_3,x)&=&T_{\rm II}(C_0,C_1,C_2,x)+C_3(4x^3-3x)\nn\\
T_{\rm IV}(C_0,C_1,C_2,C_3,C_4,x)&=&T_{\rm III}(C_0,C_1,C_2,C_3,x)+C_4(8x^4-8x^2+1)\nn\\
T_{\rm V}(C_0,C_1,C_2,C_3,C_4,C_5,x)&=&T_{\rm IV}(C_0,C_1,C_2,C_3,C_4,x)+C_5(16x^5-20x^3+5x),~~~~~\label{eq:Crossing-function}  
\end{eqnarray}
where $C_{0-5}$ are the crossing hyperparameters.  $x=\ell/\ell_{\rm max}$ denotes the multipole range for the Chebyshev expansion. 
We set $\ell_{\rm max}=2508$ while comparing with the TT data and to 1996 when comparing with the EE data. Note that for $C_0=1$ and $C_{1-5}=0$,
the Crossing function is 1 and reflects no modification to the mean function. We confront the modified spectrum with the EE polarization data. If
the polarization data is in agreement with the temperature data, the baseline model (which fits the temperature data very well) will be well within 
the confidence ball of crossing hyperparameters. The distance of the mean function from the center of the confidence ball of $C_{0-5}$ determines 
the agreement between the two datasets. We should mention that this approach can work very well to compare two datasets when the assumed baseline 
model has a relatively good fit to one of the datasets which is the case in our study as we know $\Lambda$CDM model has a reasonable fit to CMB data. 
Similarly, we also use the best fit base line angular power spectrum from Planck E-mode polarization data and confront with the temperature data from 
Planck. The reason for this twofold comparison is due to the very different signal to noise ratio of the detection of the temperature and polarization 
anisotropies. Twofold consistency check allows us to understand whether a possible mismatch can be due to different amplitude of uncertainties in 
the two data. We had performed similar analysis previously comparing Planck 2013 temperature and WMAP-9 year temperature data~\cite{Hazra:2013oqa}.  In this analysis we also examine to what extent TT + lowT best fit $\Lambda$CDM model is supported by the TE cross spectrum data.

Secondly, using the Crossing statistics, we examine whether the concordance model of cosmology, that we essentially use as the baseline cosmology for parameter 
estimation, is consistent with the data or the data indicates towards some non-standard cosmology. We have explored similar possibilities with the Planck 2013 temperature
anisotropy data in~\cite{Hazra:2014hma}. We found that Planck 2013 data required a large scale and small scale suppression, significant at 2-3$\sigma$ 
level \footnote {Since then there have been various developments. We followed our analysis with direct reconstruction of the primordial power spectrum from the
Planck 2013 datasets~\cite{Hazra:2014jwa}. While the large scale suppression seemed to be present at low-$\ell$ with low significance, at small scales two relatively 
significant features in the primordial power spectrum were indicated, one around $\ell\sim750-850$ and another one around $\ell\sim 1800$. We performed a reanalysis of our 
consistency test removing the 217 GHz power spectra from the Planck data and found that the concordance model becomes completely consistent with the Planck 2013 
data~\cite{Hazra:2014hma,Aghamousa:2015fja}. Hence, we could explain that the small scale power suppression was entirely coming from the 217 GHz in Planck. 
Since this channel is the best channel in Planck that probes the intermediate to small scale physics, some more careful analysis were required. The strong 
feature around $\ell\sim 1800$ was later on explained by Planck team as the 4K cooler line systematic in Planck data~\cite{Ade:2015dga}. The feature around 
$\ell\sim750-850$ remained as a possibility of a real feature though it does not have a high statistical significance.}. 
%results obtained upon removing this channel does not necessarily mean that the suppression was entirely caused by systematics.
%The strong feature around $\ell\sim 1800$ was explained as the 4K cooler line systematic in Planck data~\cite{} while the feature around $\ell\sim750-850$ remained as a possibility of a real feature.

%We mentioned that along with the possible hint of a new physics, the suppression could also be originated from some systematics in the data. 
%While the large scale suppression was definitely caused by the lack of power at low-$\ell$. In our reconstruction of primordial power spectrum from the Planck 2013 datasets~\cite{}, at small scales two significant features in the primordial power spectrum were indicated, one around $\ell\sim750-850$ and another around $\ell\sim 1800$, while the later one was explained as the 4K cooler line systematic in Planck data~\cite{}. We performed a reanalysis of our consistency test removing the 217 GHz power spectra from the Planck data and found that the concordance model becomes completely consistent with the datasets. Hence, it is explained that the small 
%scale power suppression entirely coming from the 217 GHz in Planck. Since this channel is the best channel in Planck that probes the intermediate to small scale physics,
%results obtained upon removing this channel does not necessarily mean that the suppression was entirely caused by systematics. 
With Planck 2015 data, it is important to revisit the consistency test of the concordance model because, the $\ell\sim1800$ systematics is removed and at the same time we
have data from E mode polarization and its correlation with temperature. Here we compare TT best fit with TT data with crossing modification (also we compare EE best 
fit with EE data and TE best fit with TE data). Similar to the data consistency check, the position of the mean function {\it  w.r.t} the centre of the confidence ball 
of crossing hyperparameters will determine the significance of a required modification to the concordance model. Another plausible situation considering the crossing
modification is that a different region in the parameter space of the concordance model becomes favored by the data. To include above situation, we allow the baseline
model parameters to vary along with the crossing hyperparameters. This analysis enables us to have a consistency of the framework of concordance model with the Planck 
CMB datasets that is not just limited to the best fit model and its parameter values. 

In all the analysis, we have allowed the corresponding nuisance parameters to the particular datasets to vary as fast parameters. We use the publicly available
{\tt CAMB}~\cite{Lewis:1999bs,cambsite} and {\tt CosmoMC}~\cite{Lewis:2002ah,cosmomcsite} to generate the angular power spectra and compare them with the Planck 
data using the Planck likelihood code~\cite{PLC}. We use the binned Plik likelihood for the high-$\ell$ TT and EE data. For low-$\ell$ TT data, we use the commander 
likelihood and for the polarization low-$\ell$ likelihood we use the lowTEB likelihood. We denote the temperature likelihood as TT + lowT and the polarization as EE + lowTEB
or EE + lowEB hereafter. Note that since lowEB likelihood is not available, to compare with EE data we will use EE + lowTEB likelihood, but when we compare TT + low T data 
with best fit polarization spectrum, we use the best fit values from EE + lowEB data which are publicly available. We use the best fit values of the cosmological parameters 
from the Planck chains provided~\cite{Planckchains}. 

%%%%%%%%%%%%%%%%%%%%%%%%%%%%%%%%%%%%%%%%%%%%%%%%%%%%%%%%%%%%%%%%%%%%%%%%%%%%%%%
\section{Results and discussions}\label{sec:results}

%%%%%%%%%%%%%%%%%%%%%%%%%%%%%%%%%%%%%%%%%%%%%%%%%%%%%%%%%%%%%%%%%%%%%%%%%%%%%%%
\subsection{Consistency of the data}\label{subsec:condata}

\begin{figure*}[!htb]
\begin{center} 
\resizebox{142pt}{142pt}{\includegraphics{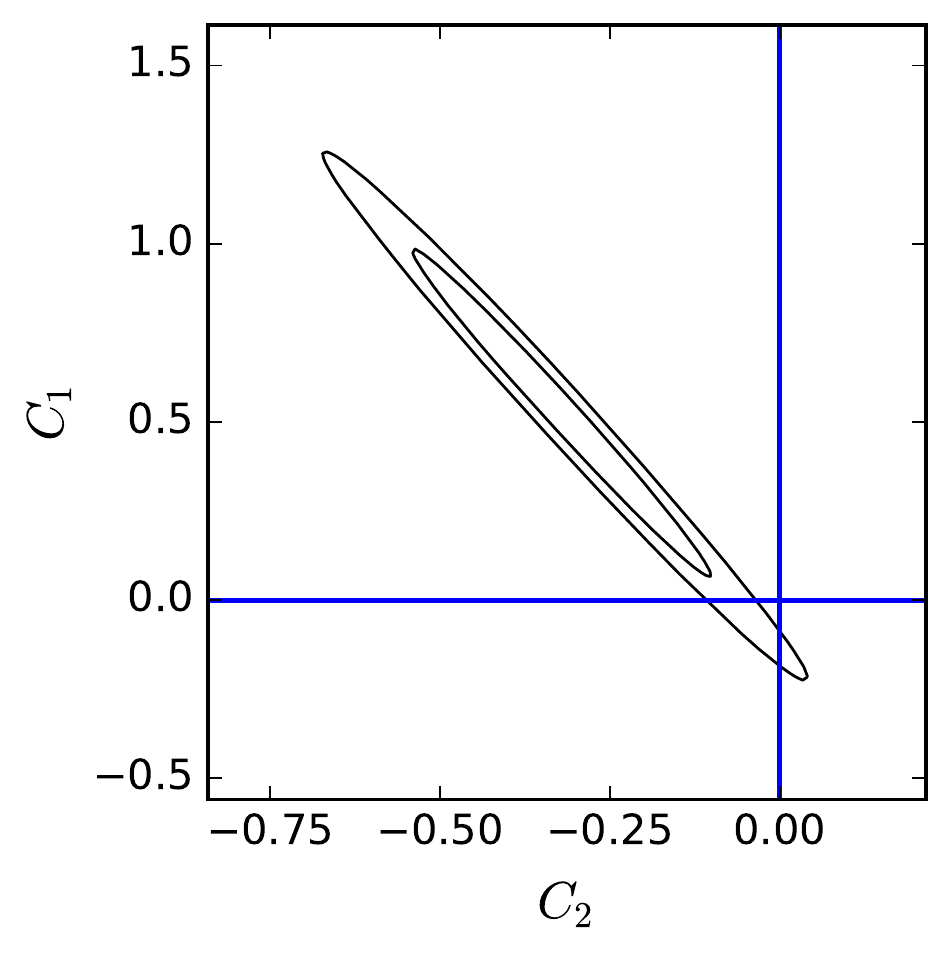}} 
\resizebox{142pt}{142pt}{\includegraphics{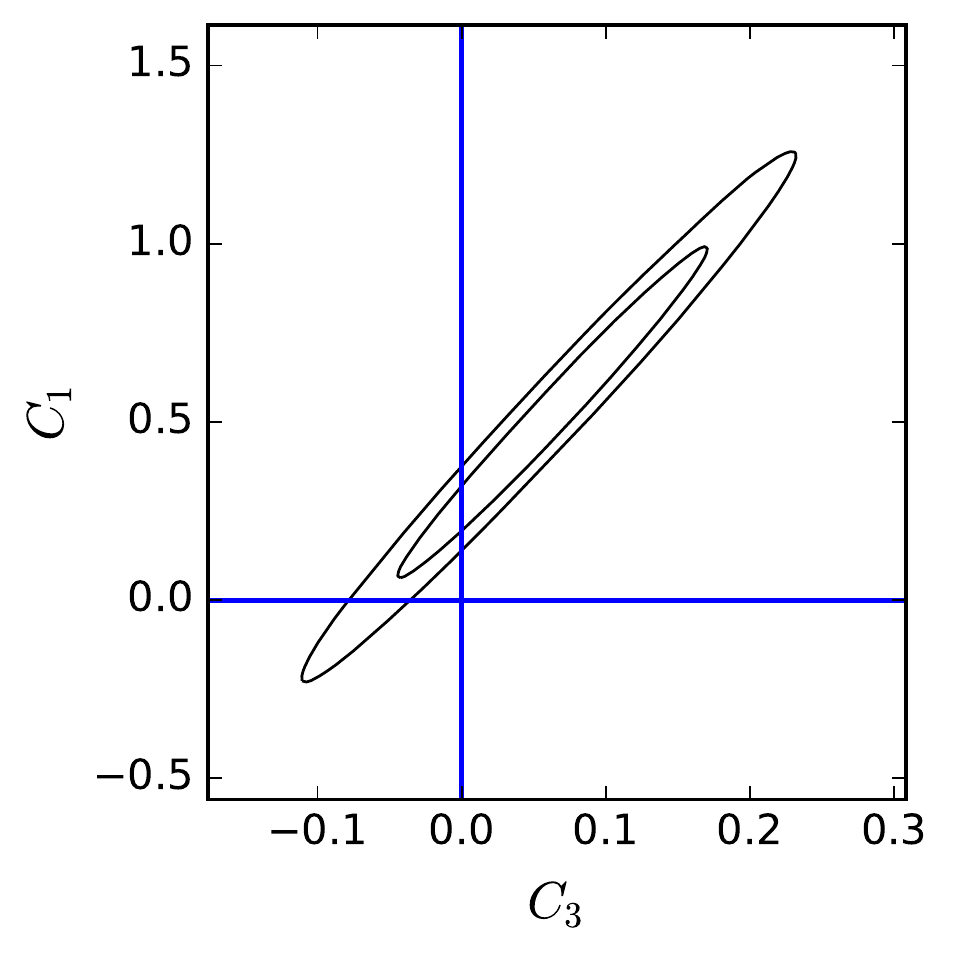}} 
\resizebox{142pt}{142pt}{\includegraphics{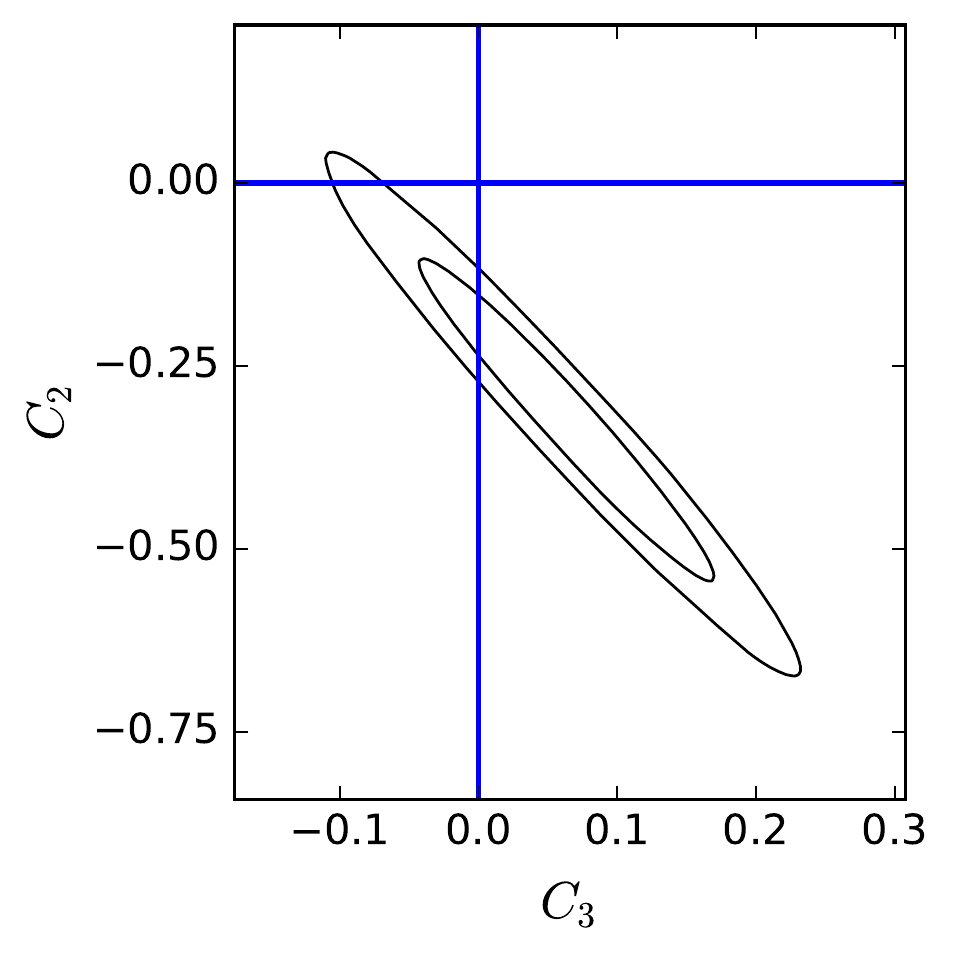}} 
\end{center}
\caption{\footnotesize\label{fig:EEtoTTC3} Confronting EE + lowEB best fit $\Lambda$CDM model (as a mean function) to TT + lowT data considering third order Crossing function. The marginalized contours of the crossing hyperparameters are plotted. The location of the mean function, {\it i.e.} the ${\cal C}_{\ell}^{TT}$ of the best fit $\Lambda$CDM model from EE+ lowEB data is located significantly far from the center of the contours that indicates the best fit model from EE + lowEB data is inconsistent with the TT + lowT data.}
\end{figure*}

To begin with, we present the results of the consistency test between Planck TT and EE datasets. 
In Fig.~\ref{fig:EEtoTTC3} we plot the contours of crossing hyperparameters when EE + lowEB best fit mean
function is compared with the TT + lowT datasets. In this case we have modified the mean function with upto 
third order of Crossing function. Each plot is marginalized over other Crossing hyperparameters. The mean function,
{\it i.e.} the best fit from EE + lowEB data is outside the confidence contours of the hyperparameters. This result 
signifies that the best fit $\Lambda$CDM cosmology from EE + lowEB data is inconsistent with the 
TT + lowT data. Even with third order modification, the best fit $\chi^2$ is 83 worse than the $\Lambda$CDM best fit 
$\chi^2$ from TT + lowT. We define the difference in $\chi^2$ from the modified mean spectra considering Crossing functions
from one dataset (from EE + lowEB) and the $\Lambda$CDM best fit (to TT + lowT ) as $\Delta\chi^2_{CF-BF}$. 

\begin{figure*}[t]
\begin{center} 
\resizebox{120pt}{120pt}{\includegraphics{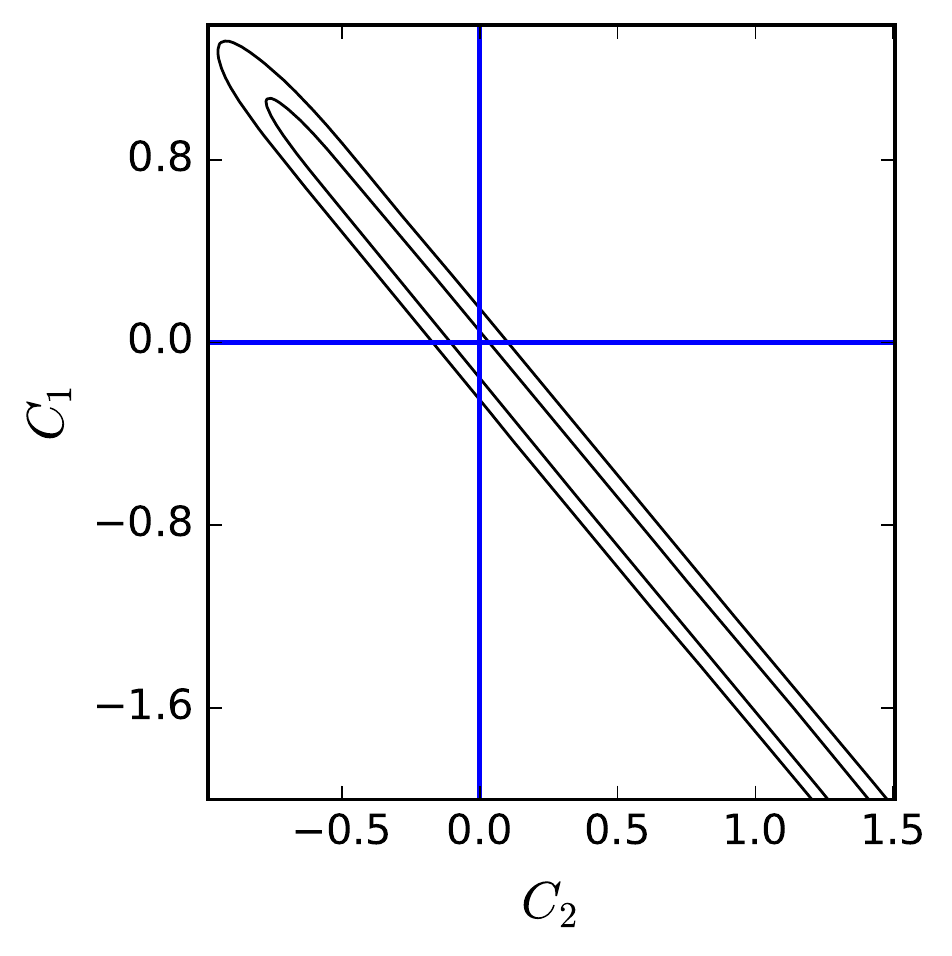}} 
\resizebox{120pt}{120pt}{\includegraphics{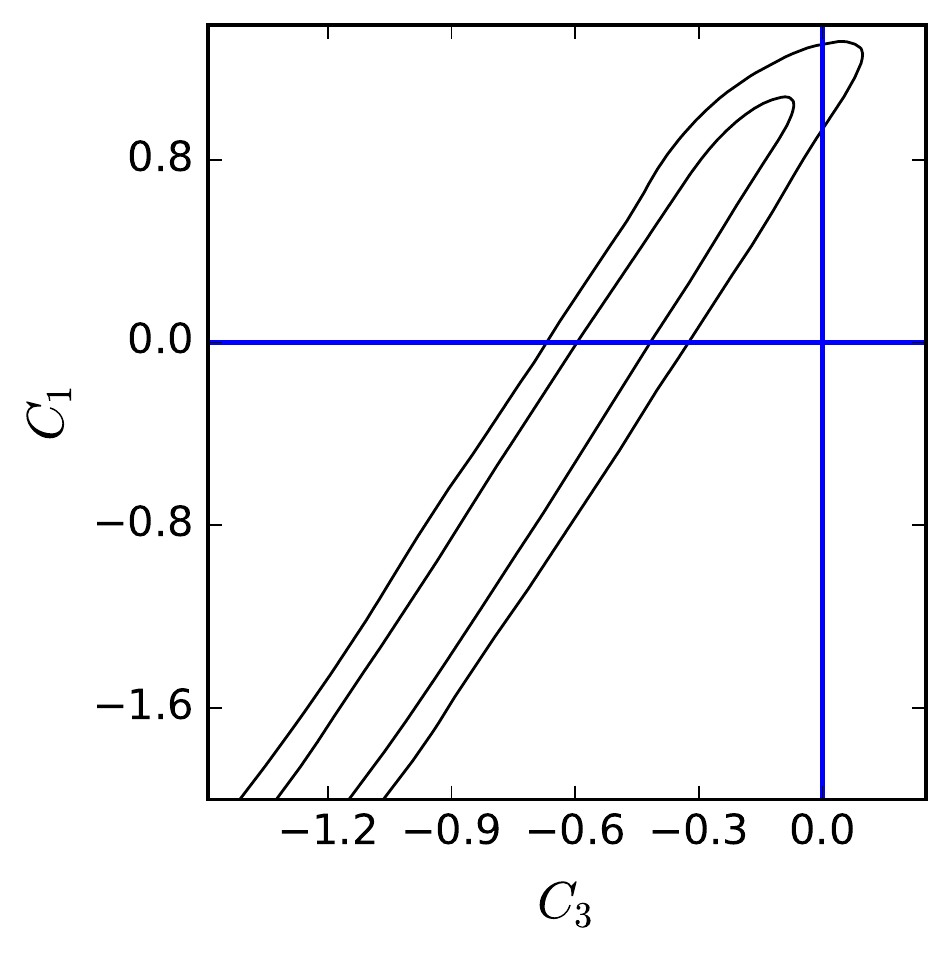}} 
\resizebox{120pt}{120pt}{\includegraphics{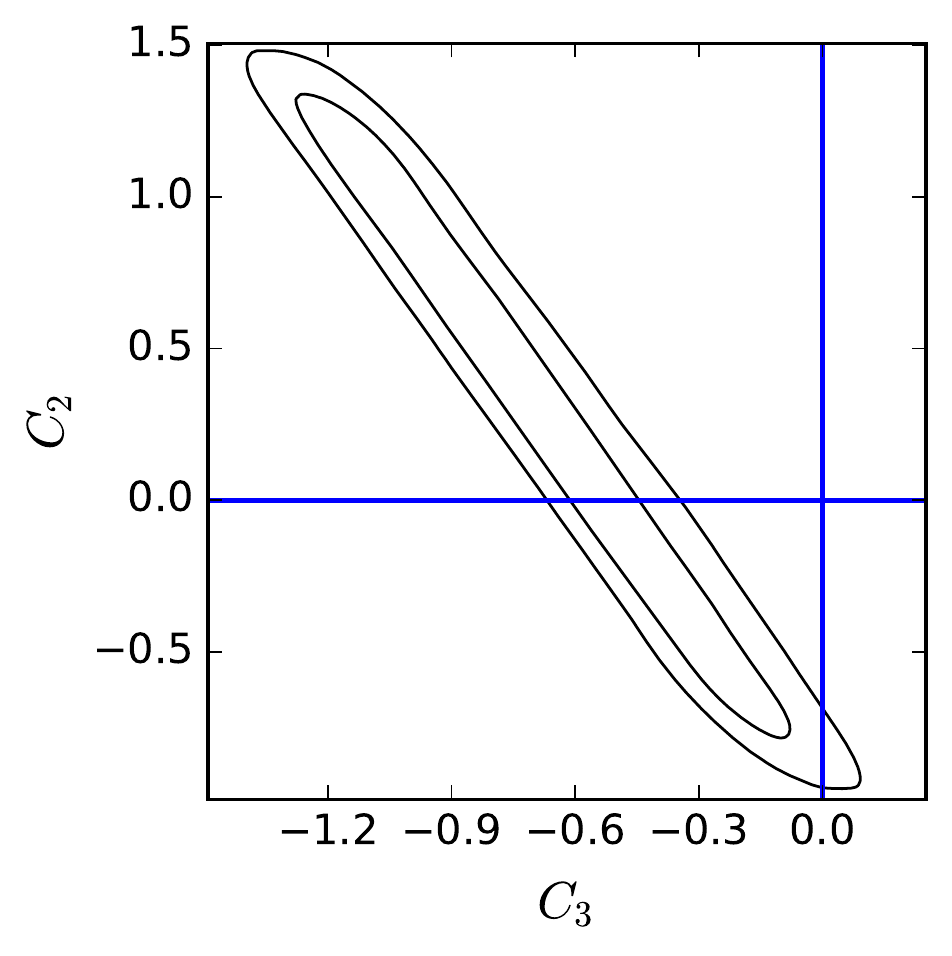}} 

\resizebox{120pt}{120pt}{\includegraphics{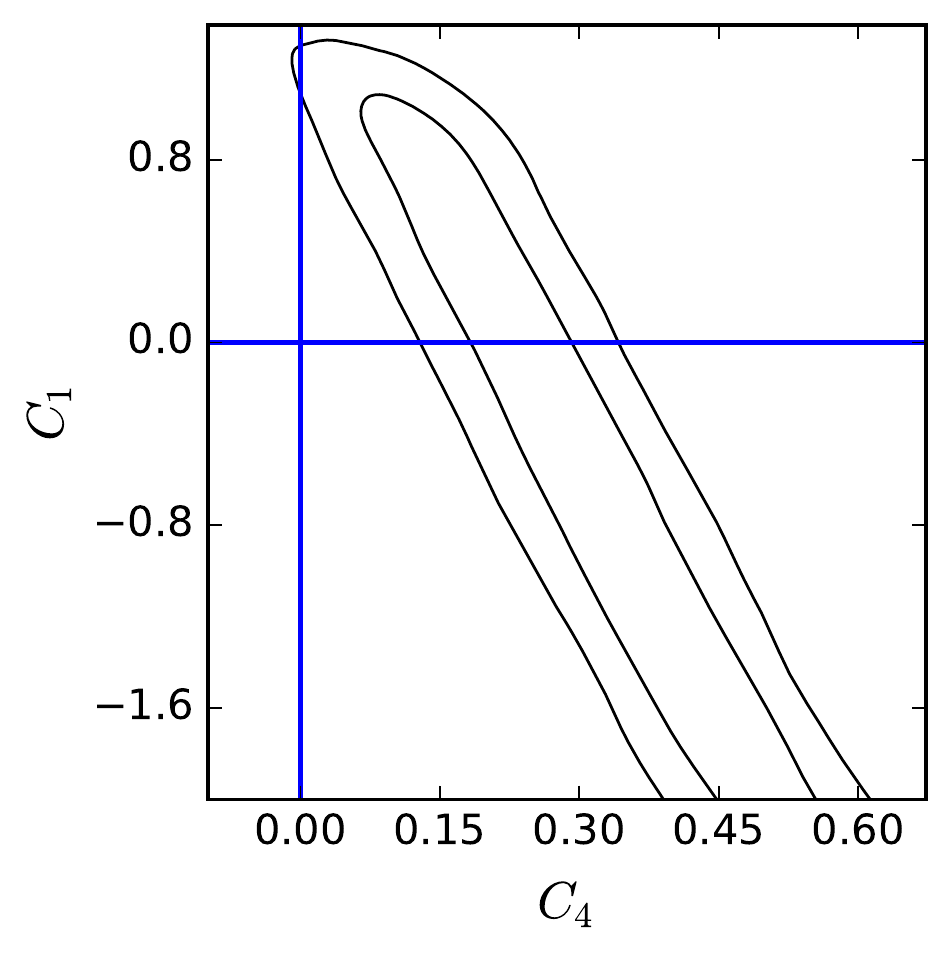}} 
\resizebox{120pt}{120pt}{\includegraphics{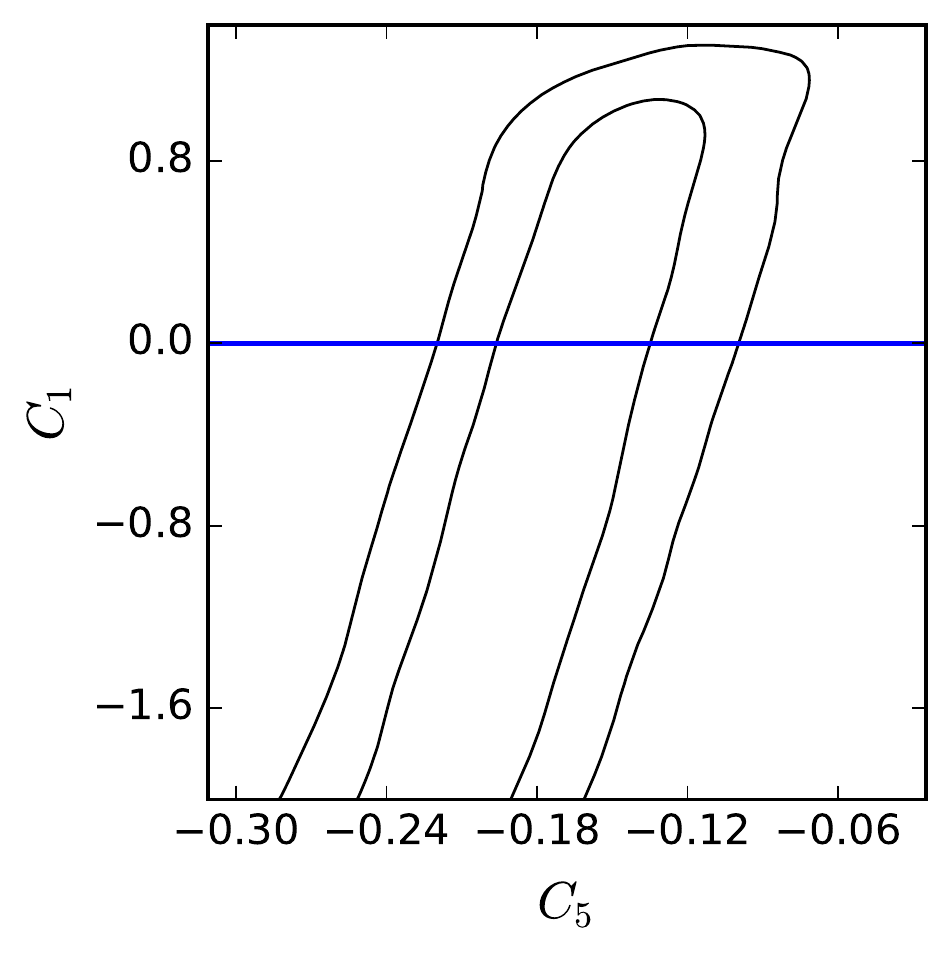}} 
\resizebox{120pt}{120pt}{\includegraphics{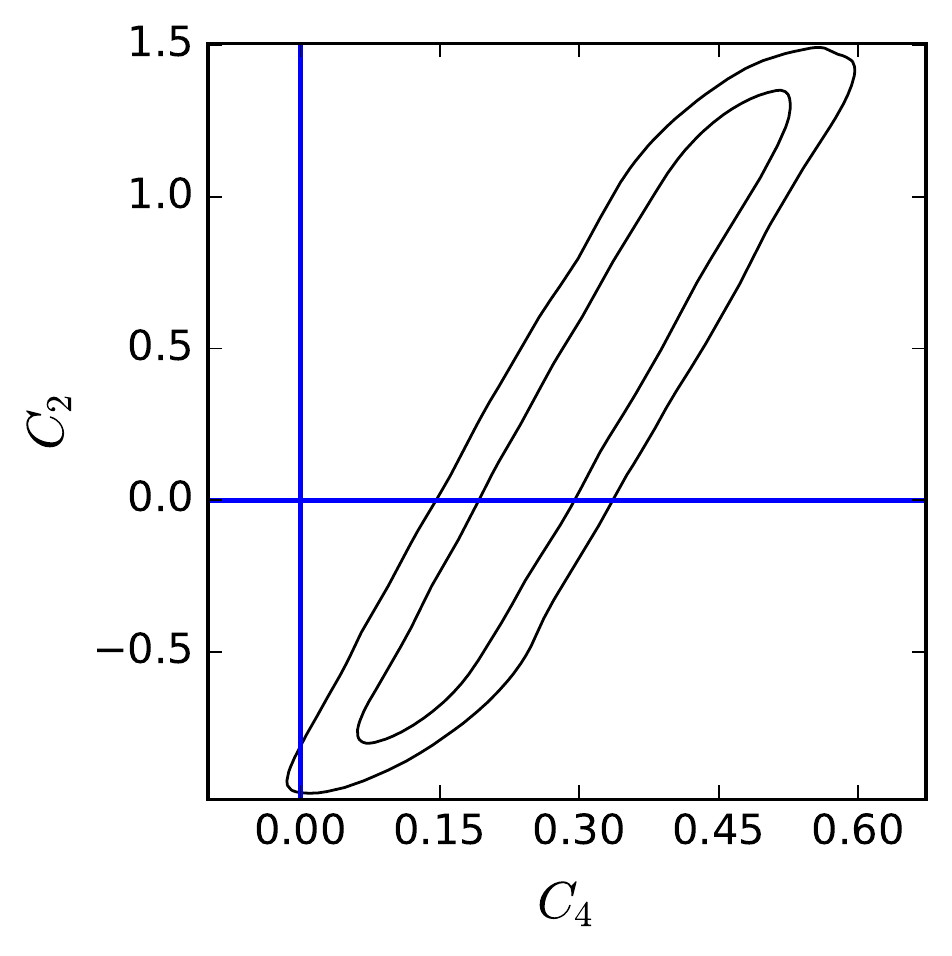}} 

\resizebox{120pt}{120pt}{\includegraphics{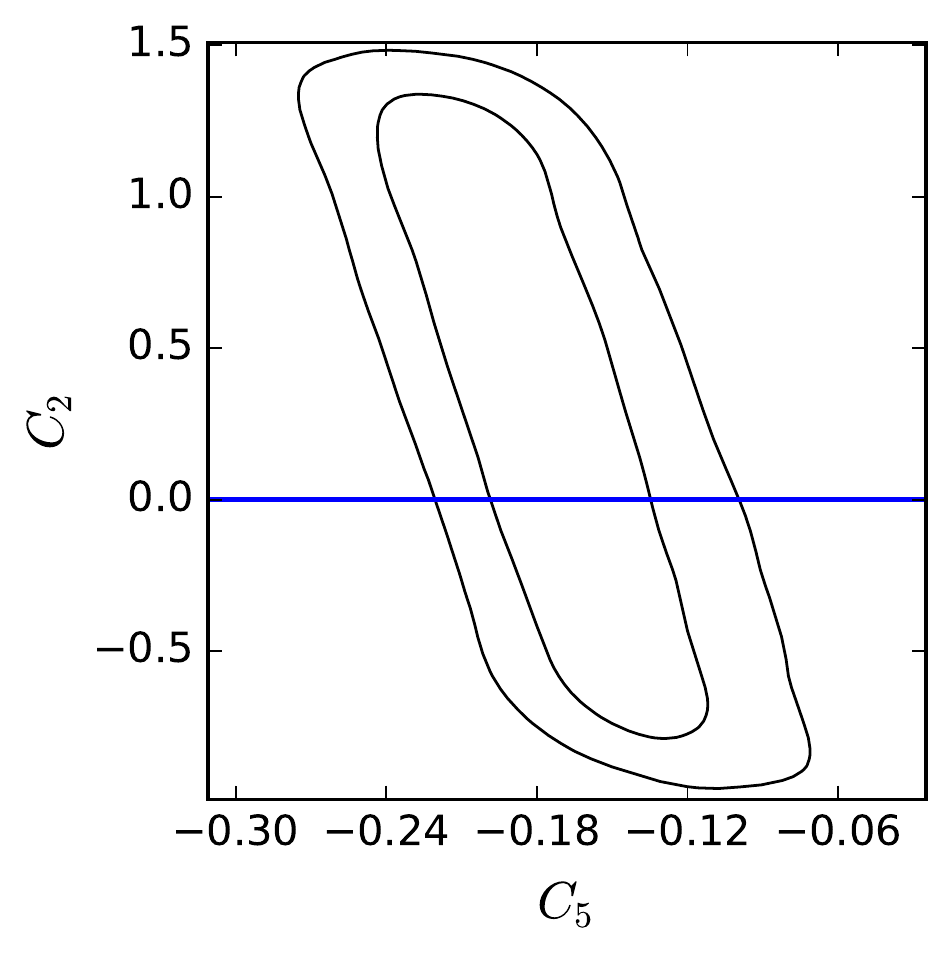}} 
\resizebox{120pt}{120pt}{\includegraphics{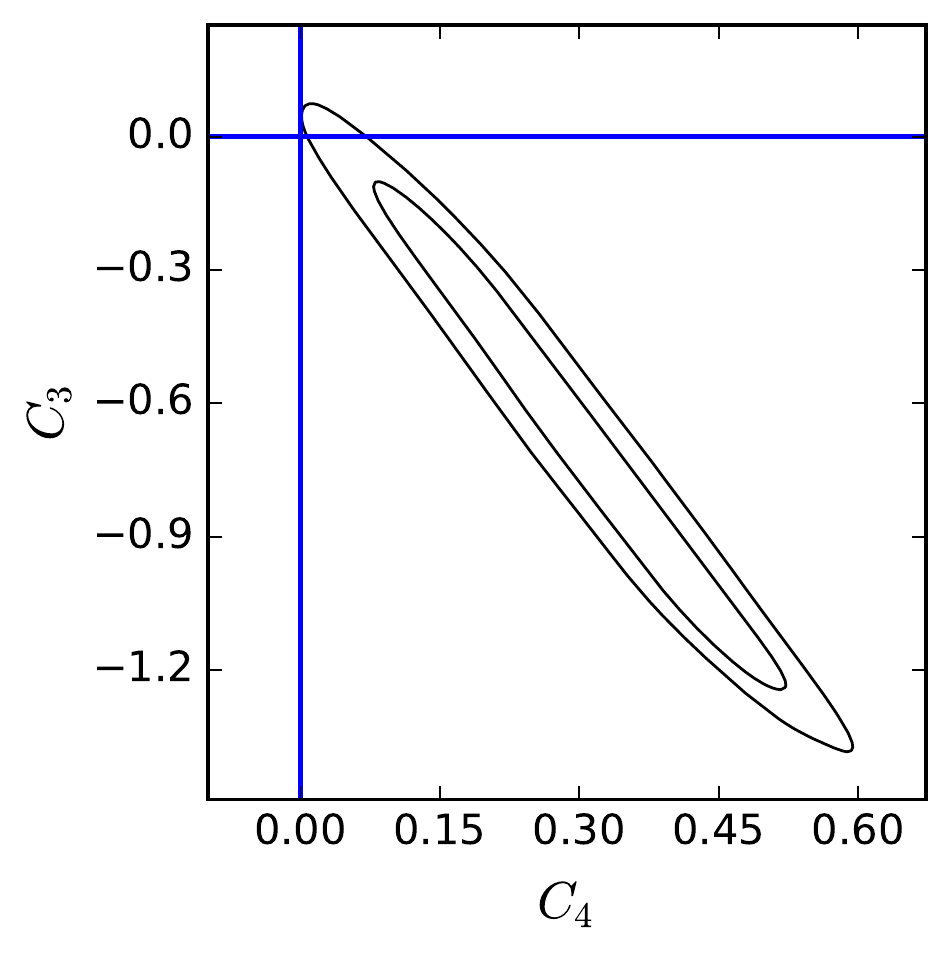}} 
\resizebox{120pt}{120pt}{\includegraphics{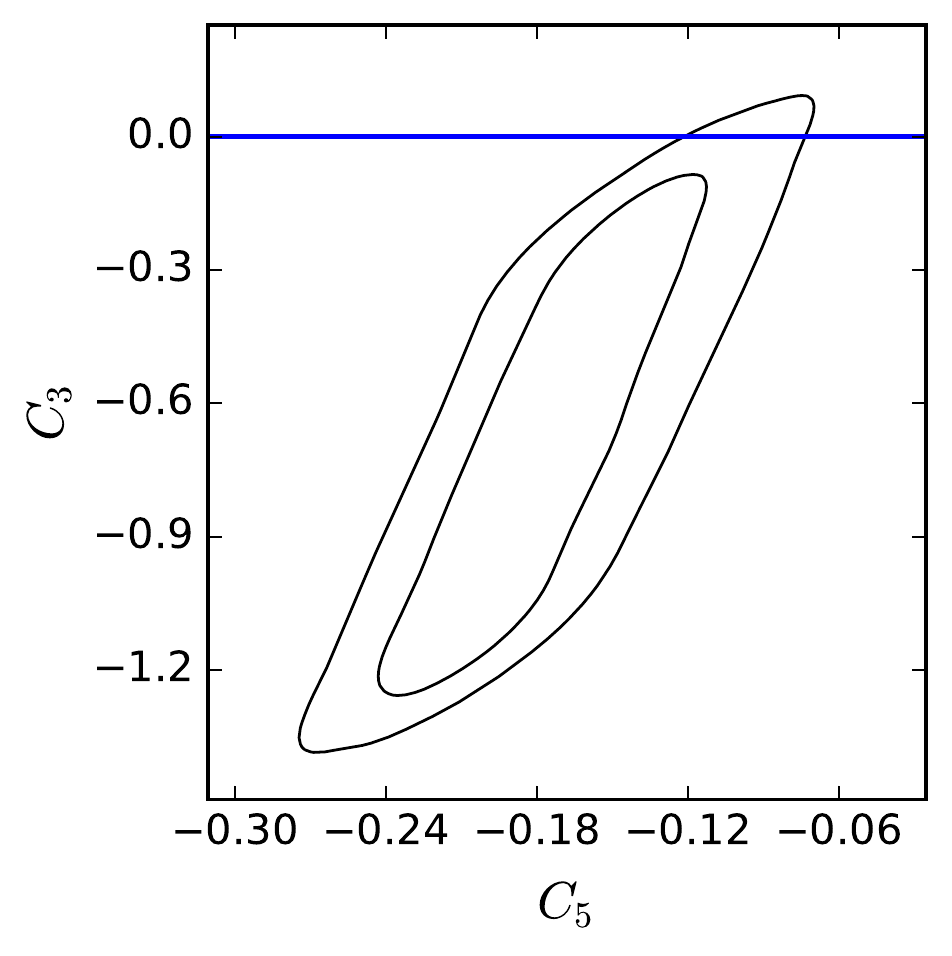}} 

\resizebox{142pt}{142pt}{\includegraphics{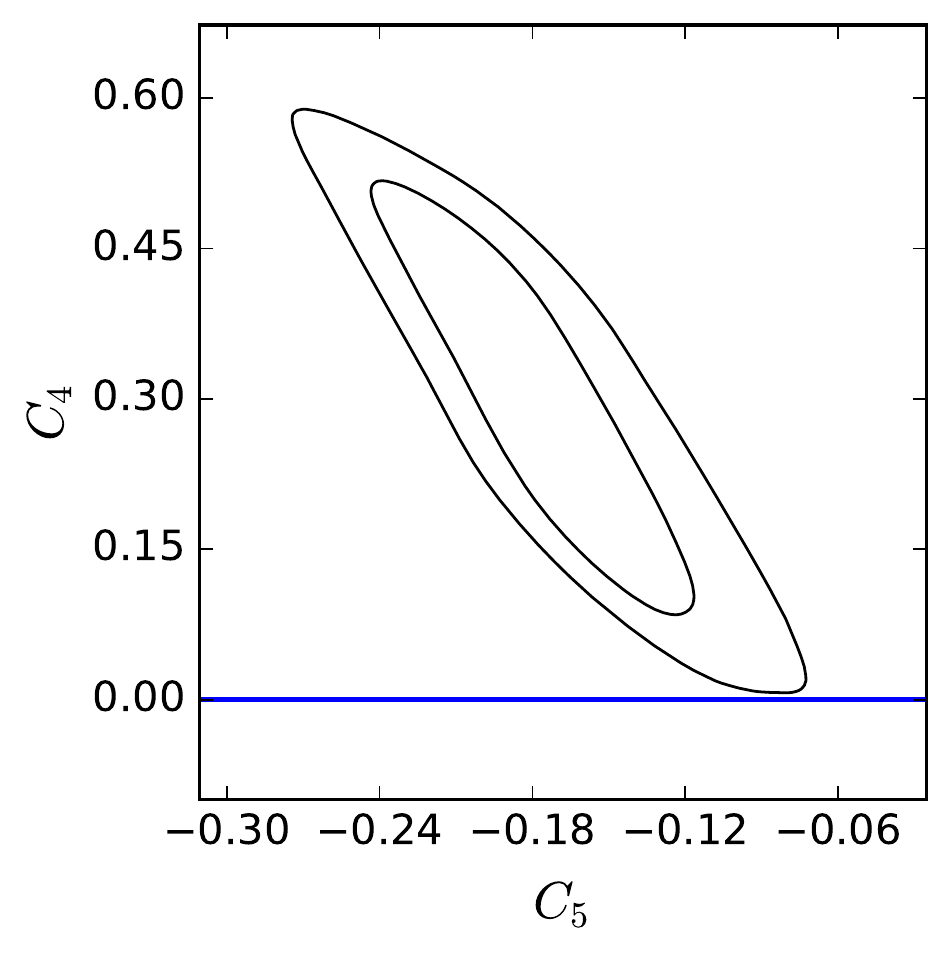}} 
\end{center}
\caption{\footnotesize\label{fig:EEtoTTC5} Confronting EE + lowEB best fit $\Lambda$CDM model as a mean function to TT + lowT data considering fifth order Crossing function. The marginalized contours of the crossing hyperparameters are plotted. Similar to the third order results here too, we find significant departures
from consistency. The mean function is distantly located from the center of the confidence ball of the hyperparameters.}
\end{figure*}

We repeat the analysis modifying the mean function with fifth order Crossing hyperfunction. It is important
to extend this analysis to higher order to understand the differences between the two datasets beyond the third order 
modifications. In Fig.~\ref{fig:EEtoTTC5} we plot the marginalized contours of Crossing hyperparameters. Here too, we notice
the mean function is significantly away from the center of all the contours. With fifth order modifications, the difference   
in the $\Delta\chi^2_{CF-BF}$ comes down to $\sim60$. It is evident that even with fifth order Crossing modifications, it is not yet possible 
to fit the temperature data with polarization best fit model up to a good extent.
%However, we notice that with increasing order, there is  a significant improvement in likelihood. 

%%%%%%%%%%%%%%%%%%%%%%%%%%%%%%%%%%%%%%%%%%%%%%%%%%%%%%%%%%%%%%%%%%%%%%%%%%%%%%%
\begin{figure*}[!htb]
\begin{center} 
\resizebox{142pt}{142pt}{\includegraphics{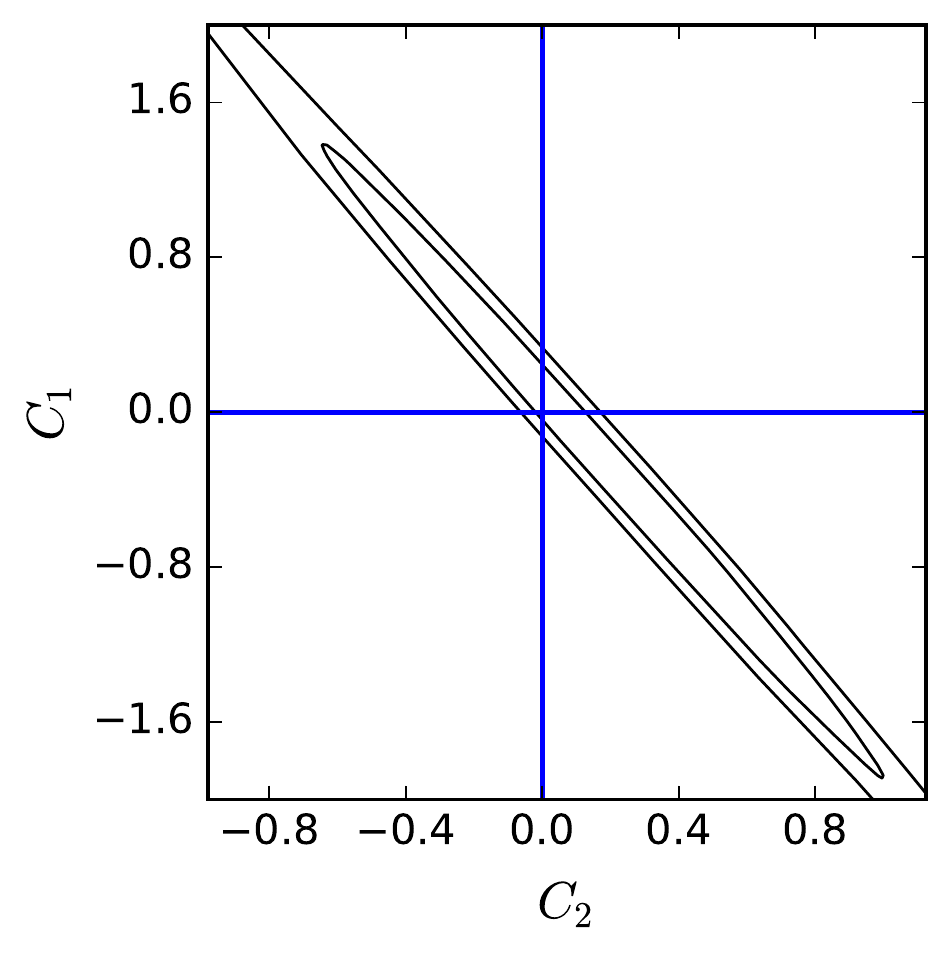}} 
\resizebox{142pt}{142pt}{\includegraphics{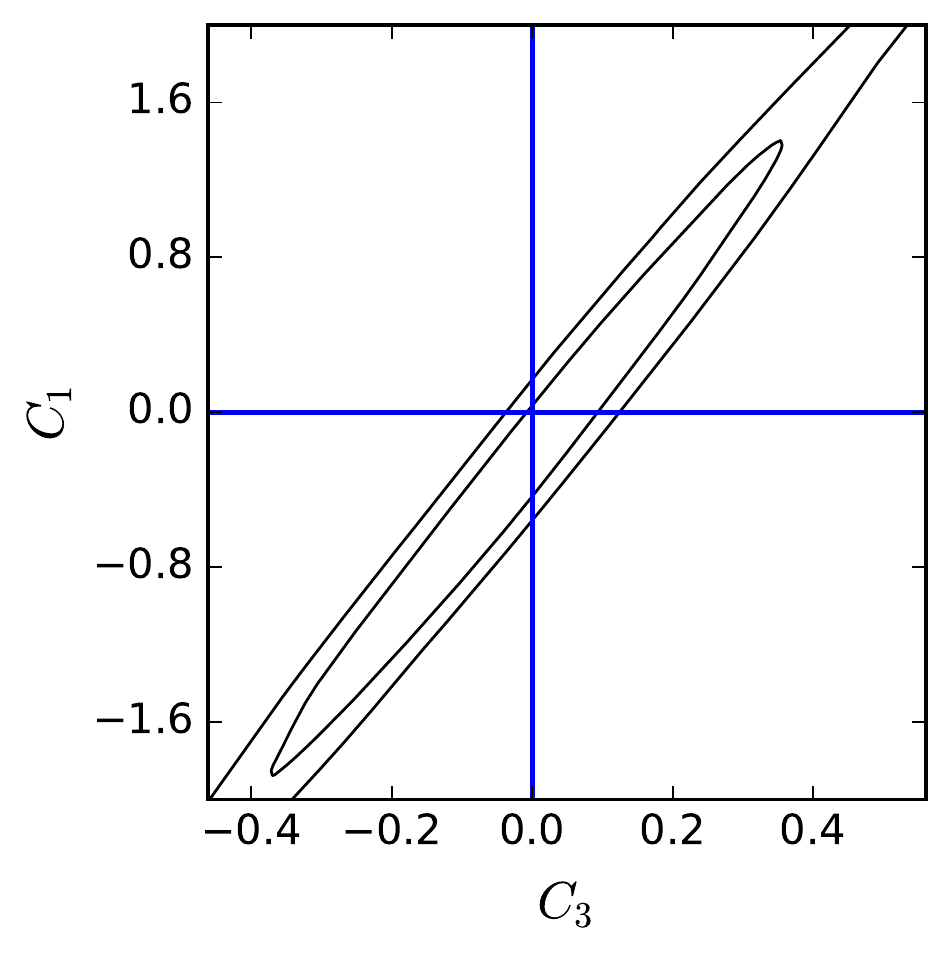}} 
\resizebox{142pt}{142pt}{\includegraphics{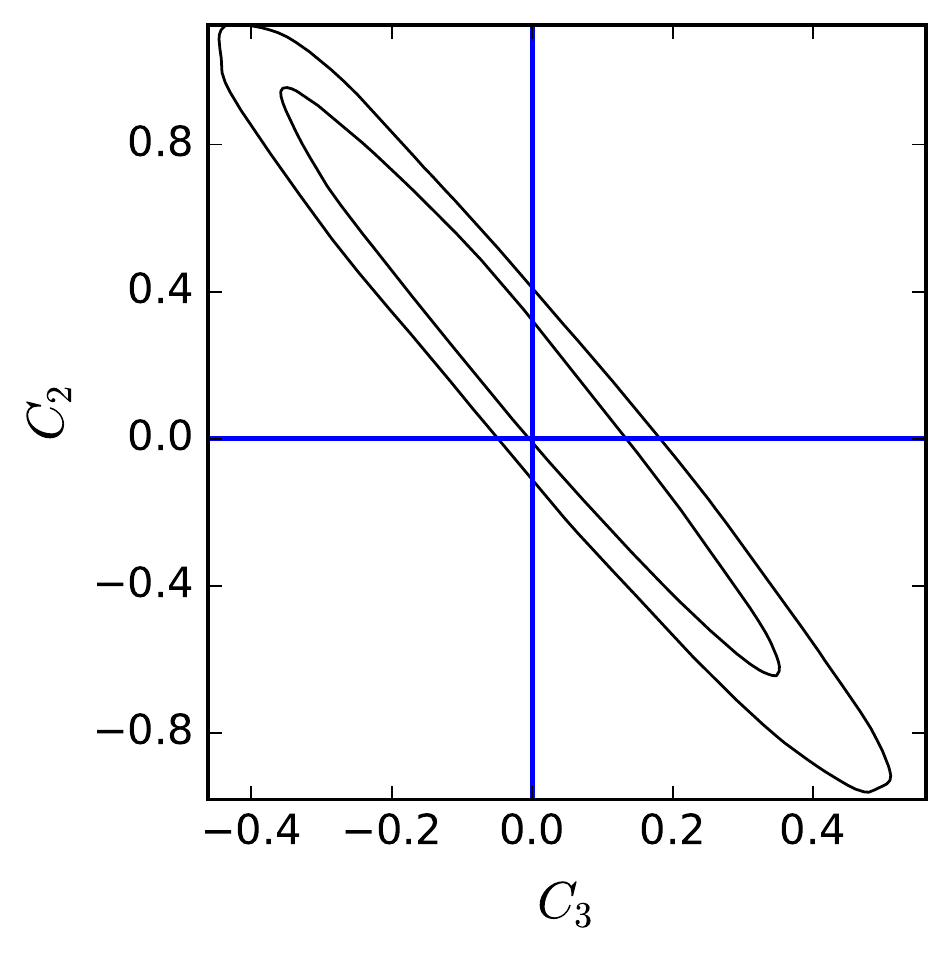}} 
\end{center}
\caption{\footnotesize\label{fig:TTtoEEC3} Confronting TT + lowT best fit $\Lambda$CDM model (as a mean function) to EE + lowTEB data considering third order Crossing function. The marginalized contours of the crossing hyperparameters are plotted. The location of the mean function, {\it i.e.} the best fit ${\cal C}_{\ell}^{EE}$ from TT+ lowT data is located within 1$\sigma$ confidence contours that indicates the best fit model from TT + lowT data is completely consistent with the EE + lowTEB data.}
\end{figure*}

\begin{figure*}[t]
\begin{center} 
\resizebox{120pt}{120pt}{\includegraphics{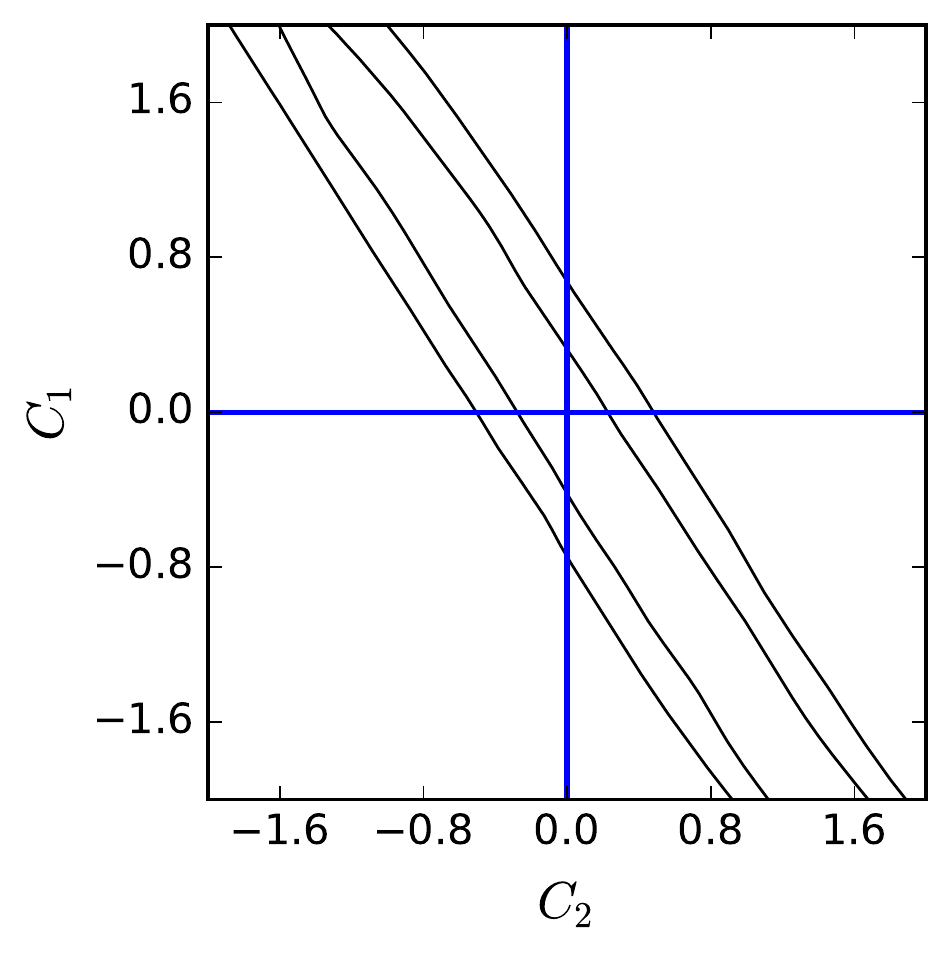}} 
\resizebox{120pt}{120pt}{\includegraphics{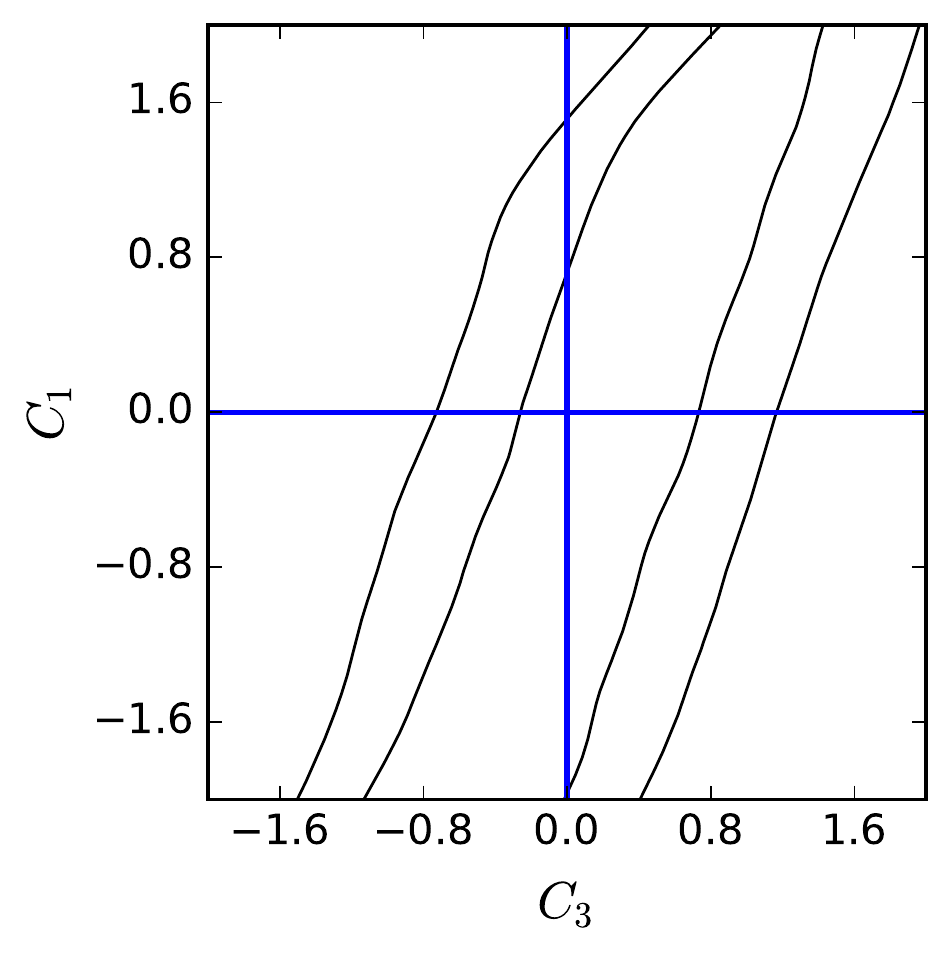}} 
\resizebox{120pt}{120pt}{\includegraphics{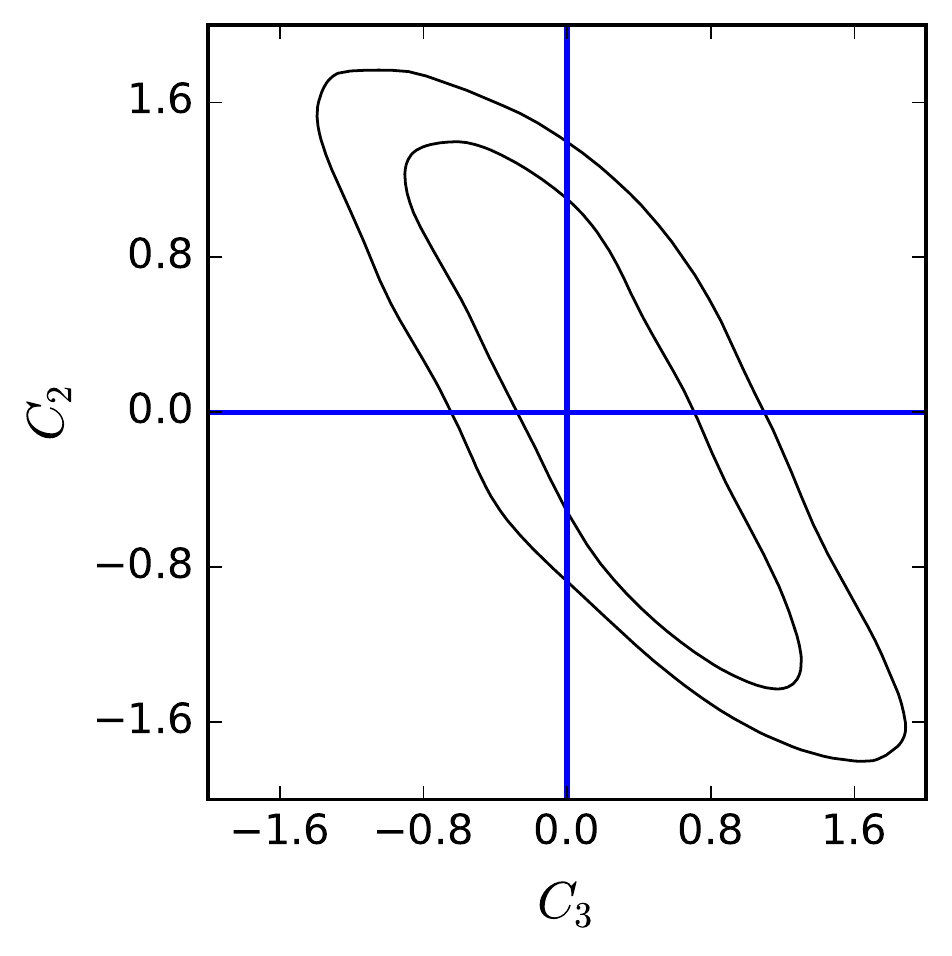}}

\resizebox{120pt}{120pt}{\includegraphics{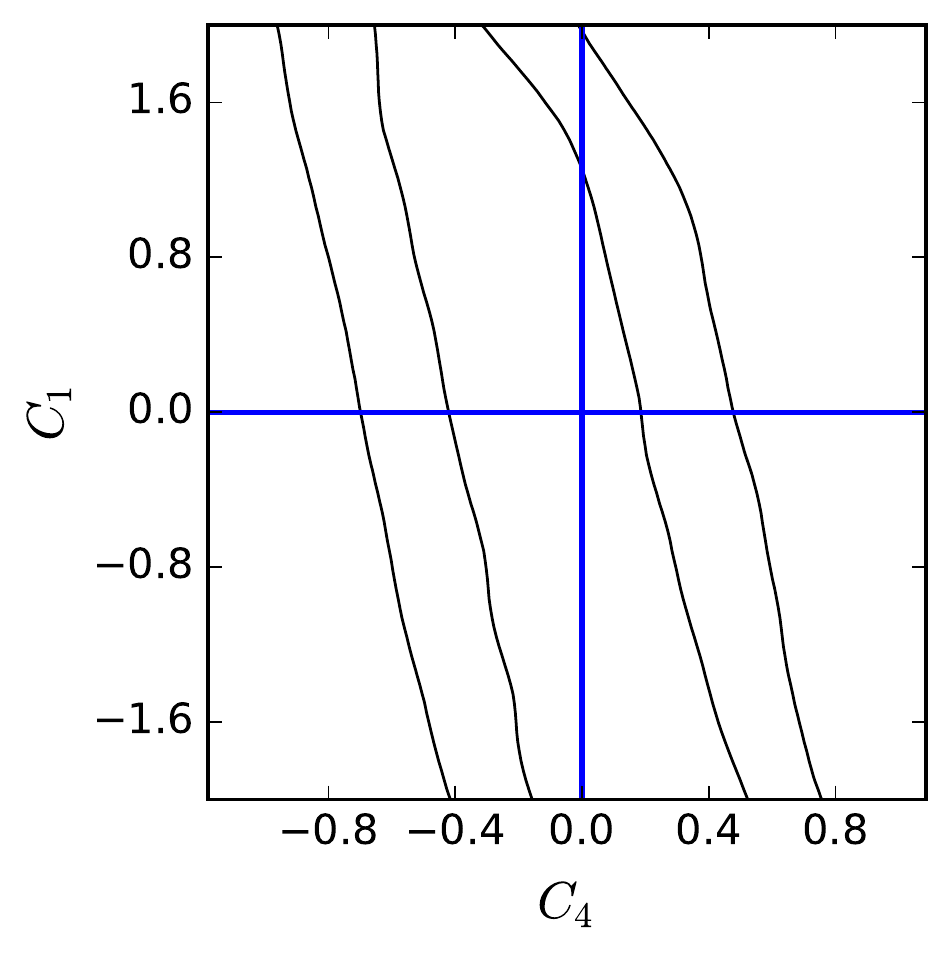}} 
\resizebox{120pt}{120pt}{\includegraphics{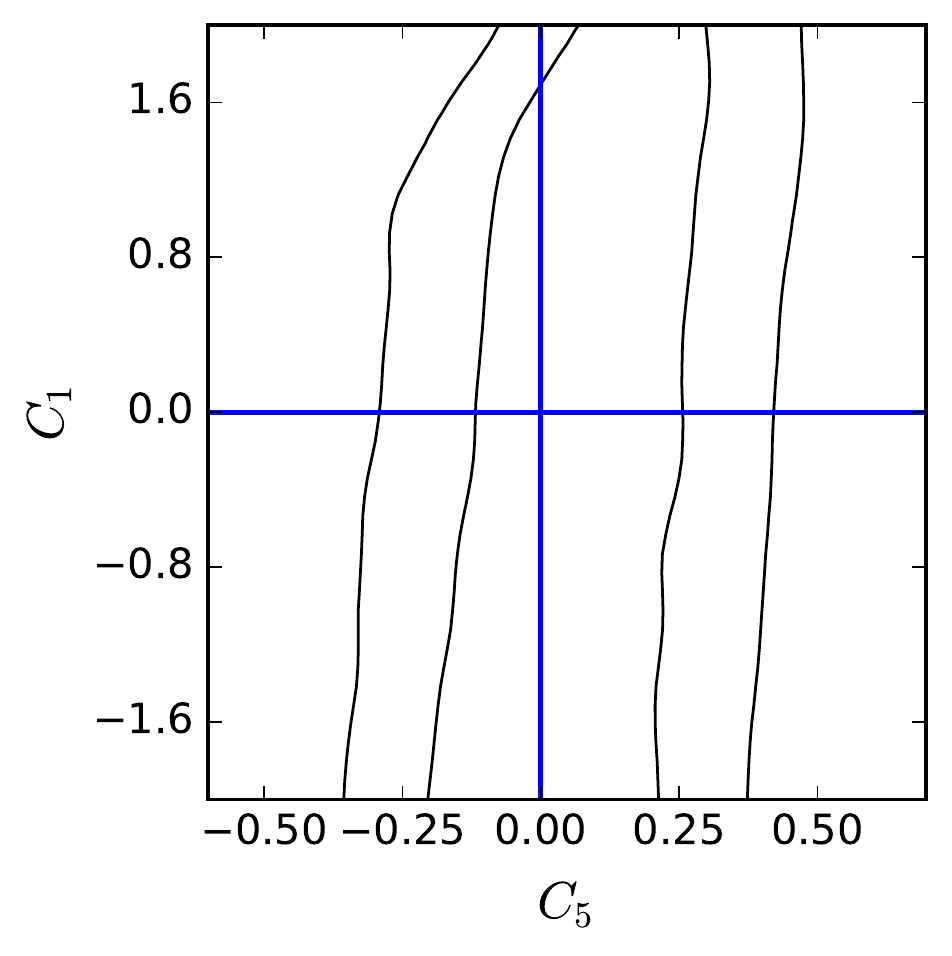}} 
\resizebox{120pt}{120pt}{\includegraphics{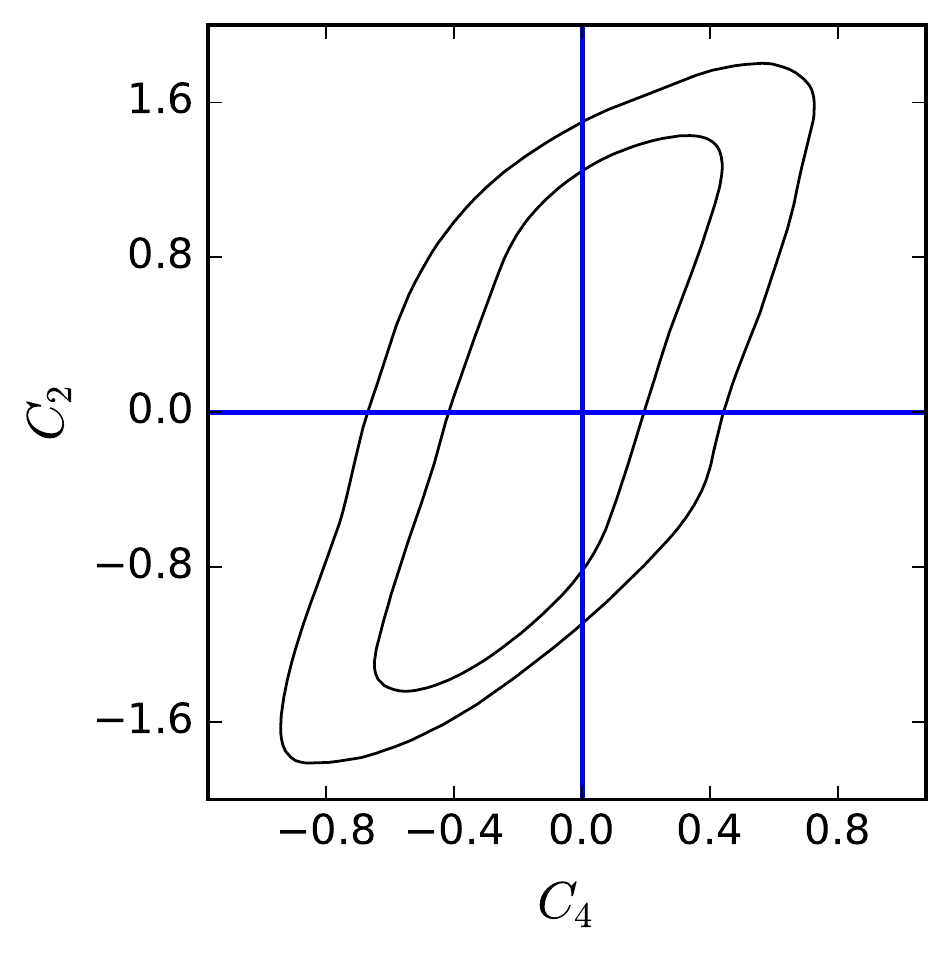}} 

\resizebox{120pt}{120pt}{\includegraphics{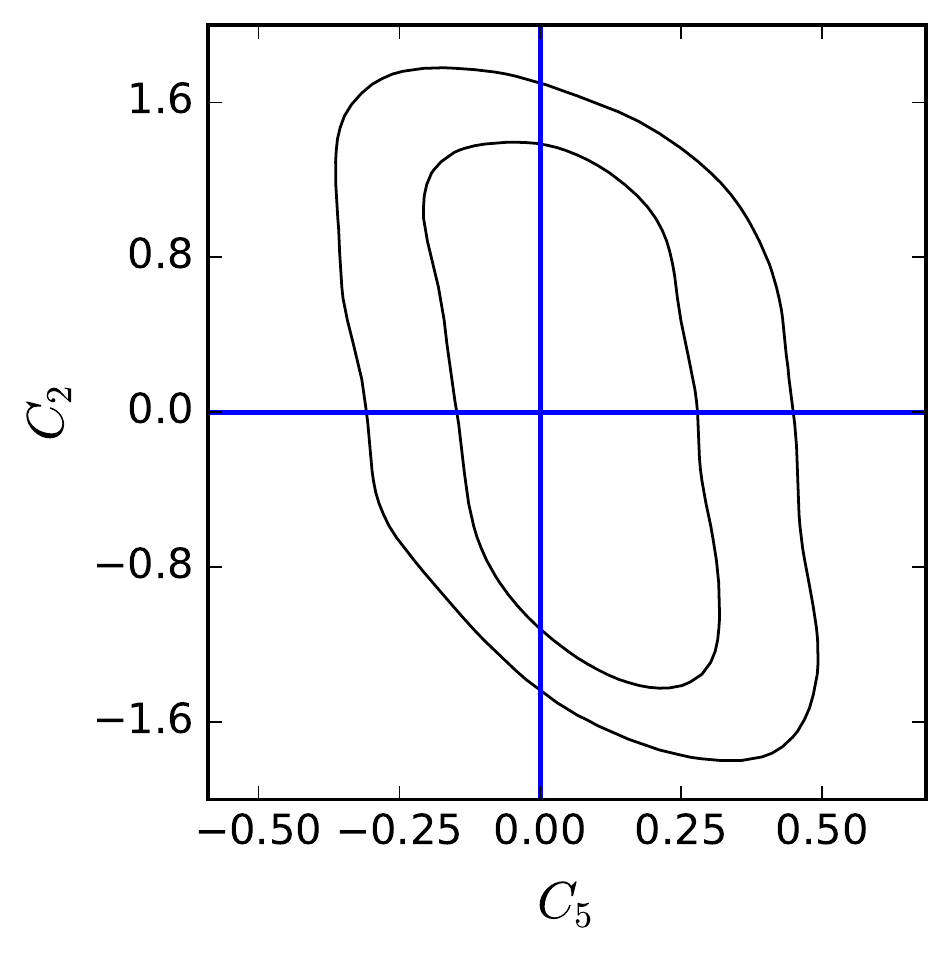}} 
\resizebox{120pt}{120pt}{\includegraphics{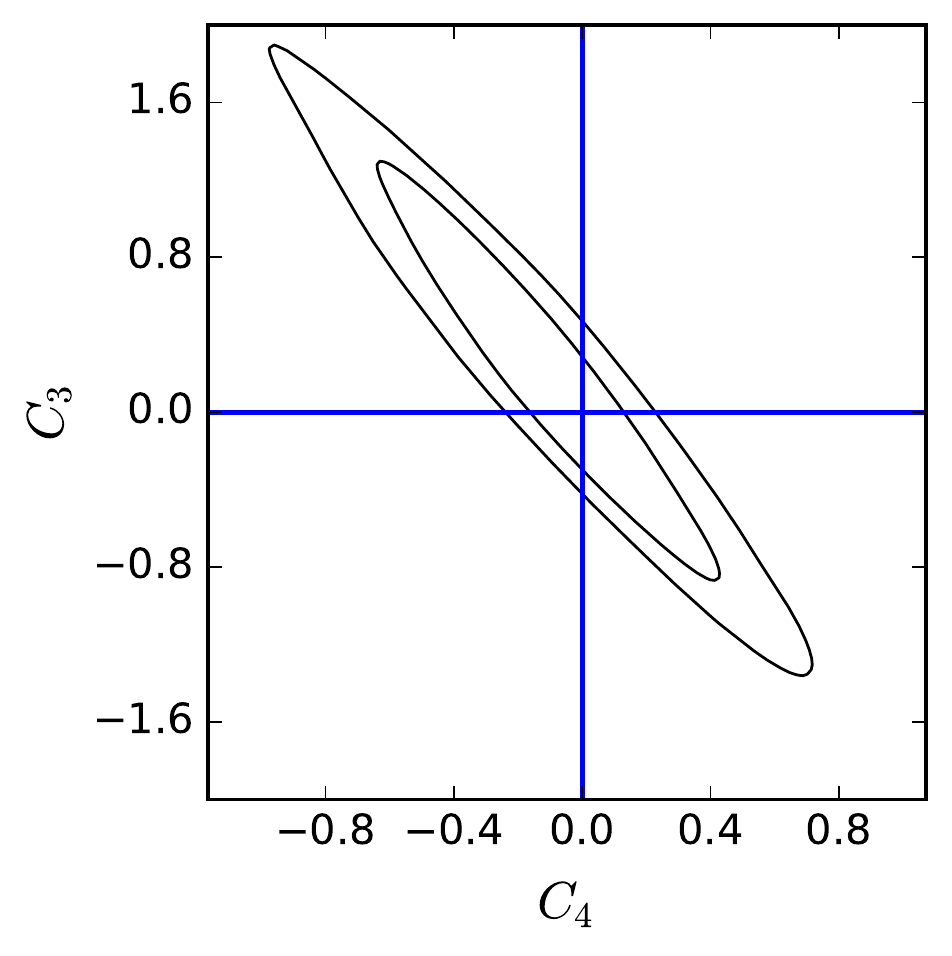}} 
\resizebox{120pt}{120pt}{\includegraphics{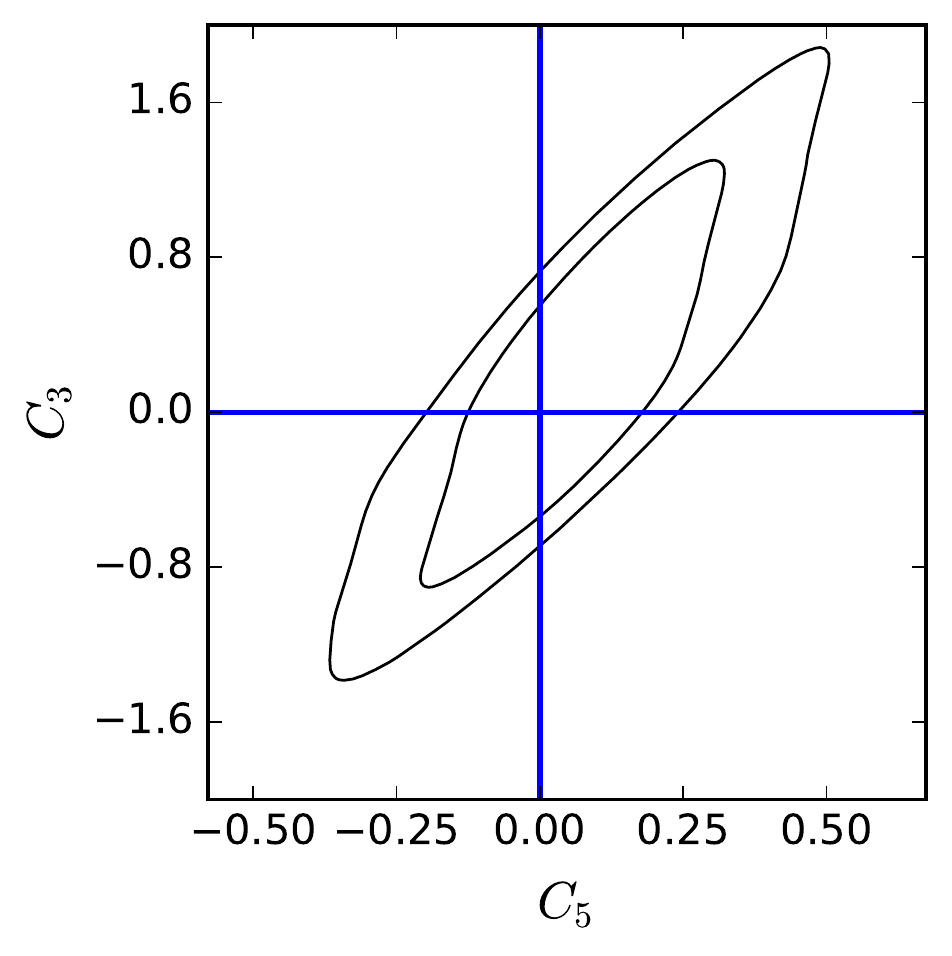}} 

\resizebox{120pt}{120pt}{\includegraphics{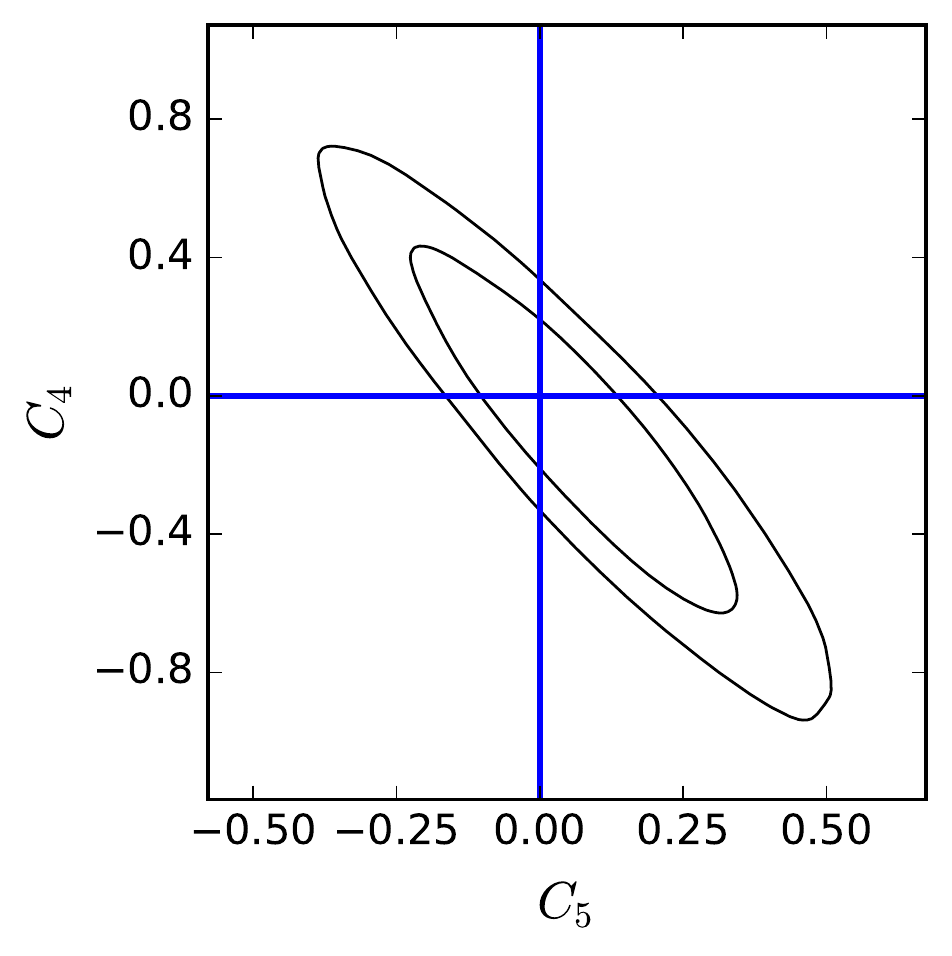}} 

\end{center}
\caption{\footnotesize\label{fig:TTtoEEC5} Confronting TT + lowT best fit $\Lambda$CDM model as a mean function to EE + lowTEB data considering fifth order Crossing function. The marginalized contours of the crossing hyperparameters are plotted. Similar to the third order results, here too we find proper consistency as the mean function is located within 1$\sigma$ confidence ball of the hyperparameters.}
\end{figure*}

However, as mentioned earlier, since the uncertainties in the polarization data is significantly larger compared to the
TT data, only with this result we cannot conclude on the issue of internal consistency between Planck datasets. We need to compare the 
best fit power spectrum from a more precise dataset with a less precise data to understand whether the uncertainties in the 
less precise data can allow the best fit theory from the high precision data make a reasonable fit to the whole data samples or not. We perform the reciprocative analysis where we use the best fit parameter values from TT + lowT data to generate the EE angular power spectrum and compare with EE + lowTEB data considering modifications by the Crossing functions. With the third order functions, the hyperparameter contours are shown in Fig.~\ref{fig:TTtoEEC3}. Note that unlike Fig.~\ref{fig:EEtoTTC3} and Fig.~\ref{fig:EEtoTTC5}, here the mean function is within the 1$\sigma$ region of the confidence ball of the hyperparameters. This result implies that the TT + lowT best fit model is completely in agreement with the EE + lowTEB datasets and it does not require any modification. To search for possible higher order deviation, we reanalyze the above with upto fifth order Crossing function and plot the marginalized hyperparameter contours in Fig.~\ref{fig:TTtoEEC5}. Similar to the third order results, we find that the mean function is inside 1$\sigma$ interval of all the contours. Considering both TT to EE and EE to TT consistency results, we can summarize that the two data have a proper consistency with each other. We should restate the obvious again here that the best fit $\Lambda$CDM model from the EE data alone should not be given high importance while uncertainties in EE data is very large. 

%completely in agreement with the TT data. The inconsistency of the best fit , the opposite is not true. The inference that 
%can be drawn is since, TT angular power spectra are significantly better signal-to-noise detection compared to EE, the statistical
%fluctuations allowed by EE data is not compatible with TT data, and hence the results may not hint any inconsistencies within these 
%two datasets. 

\begin{figure*}[!htb]
\begin{center} 
\resizebox{120pt}{100pt}{\includegraphics{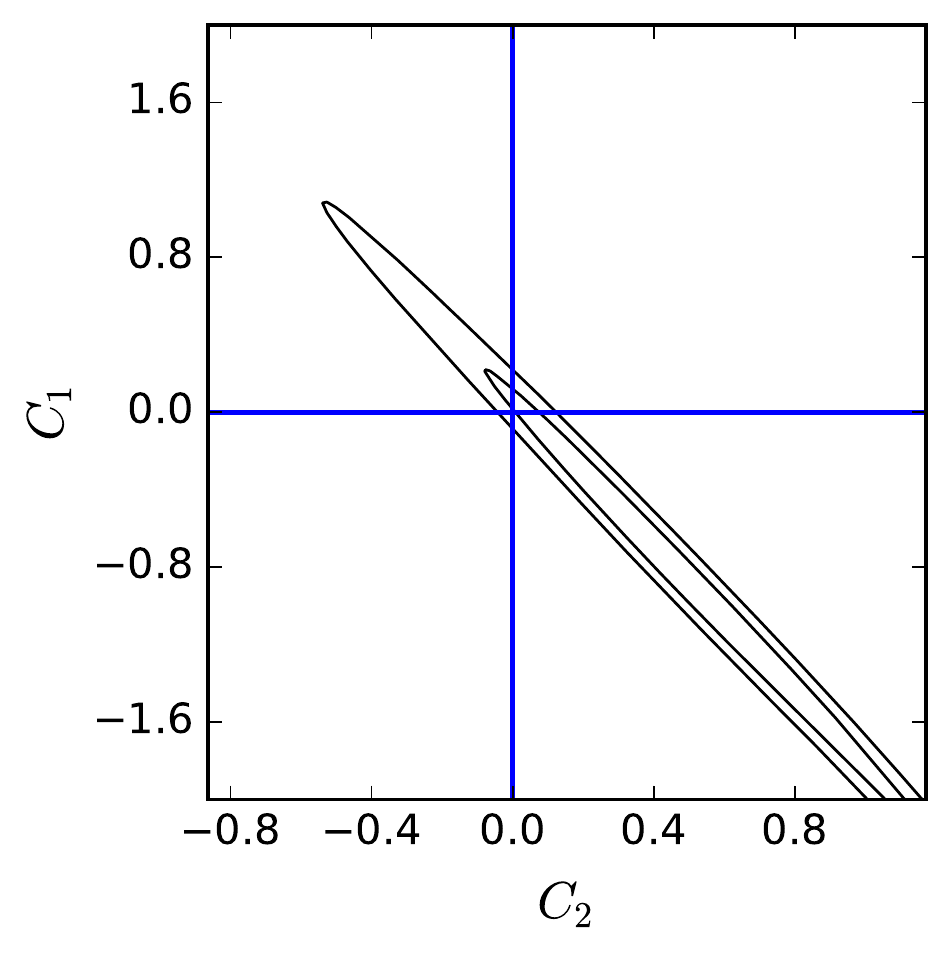}} 
\hskip 15pt\resizebox{120pt}{100pt}{\includegraphics{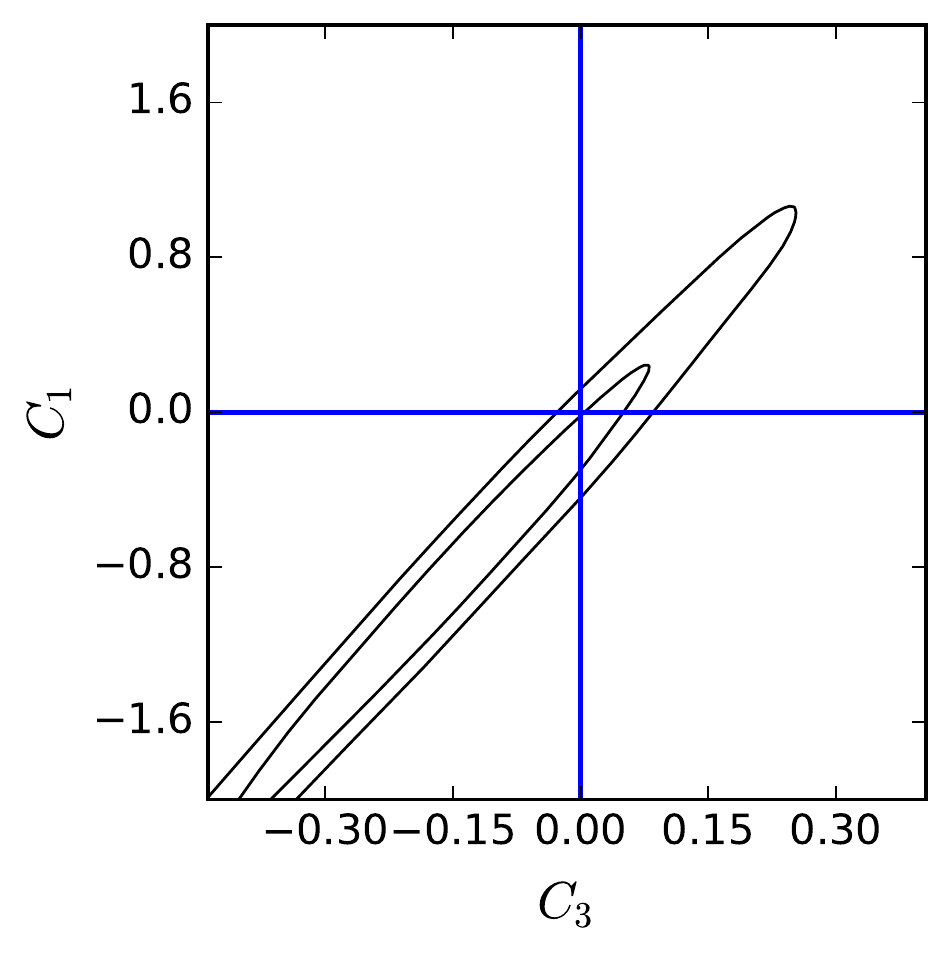}} 
\hskip 15pt\resizebox{120pt}{100pt}{\includegraphics{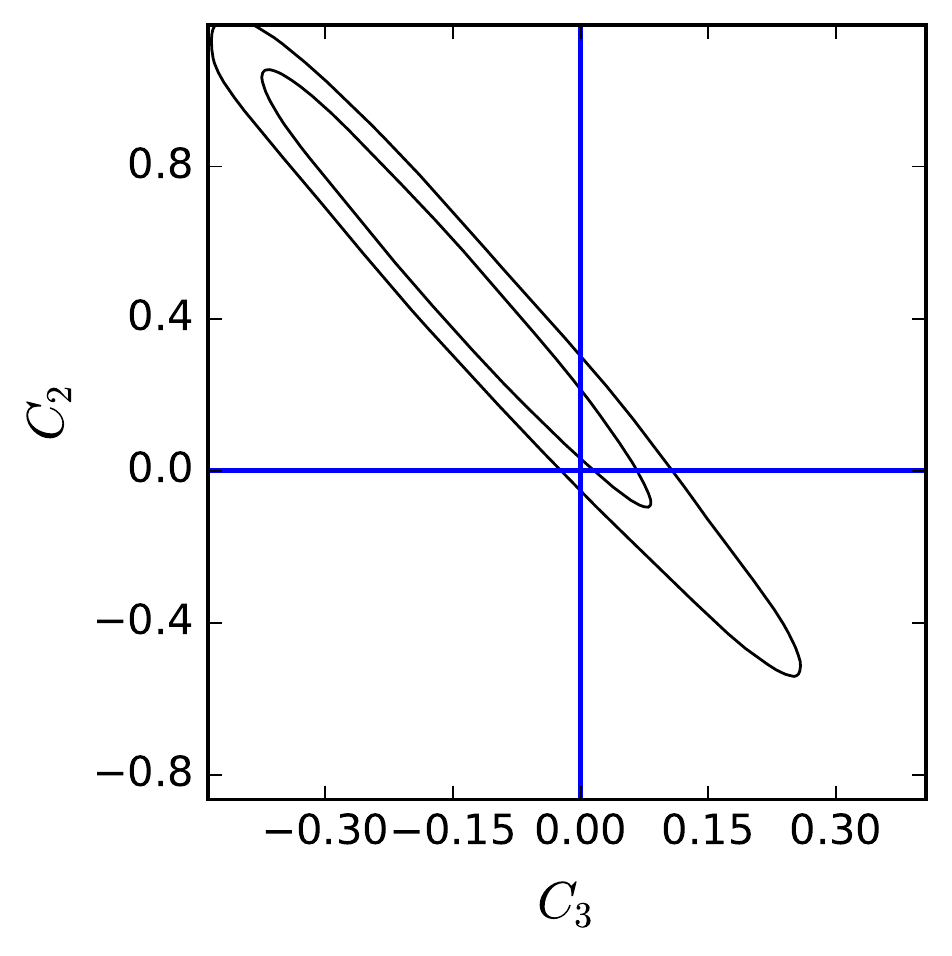}} 

\resizebox{120pt}{100pt}{\includegraphics{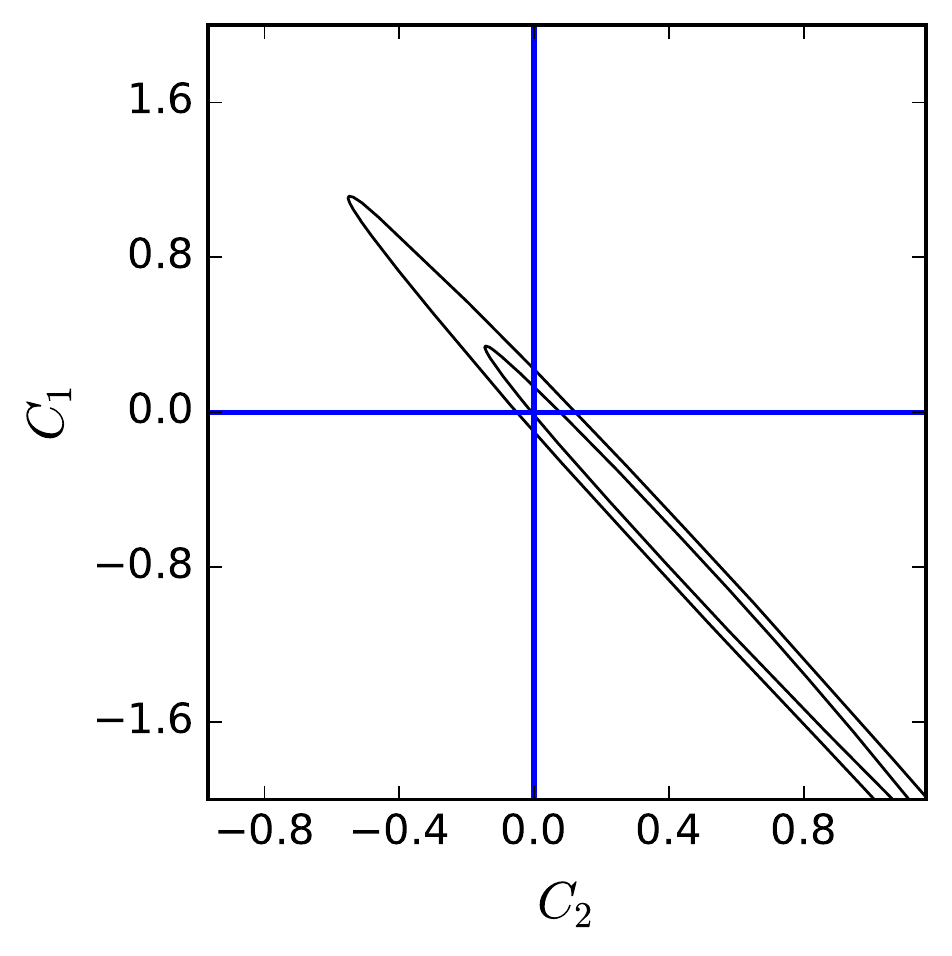}} 
\hskip 15pt\resizebox{120pt}{100pt}{\includegraphics{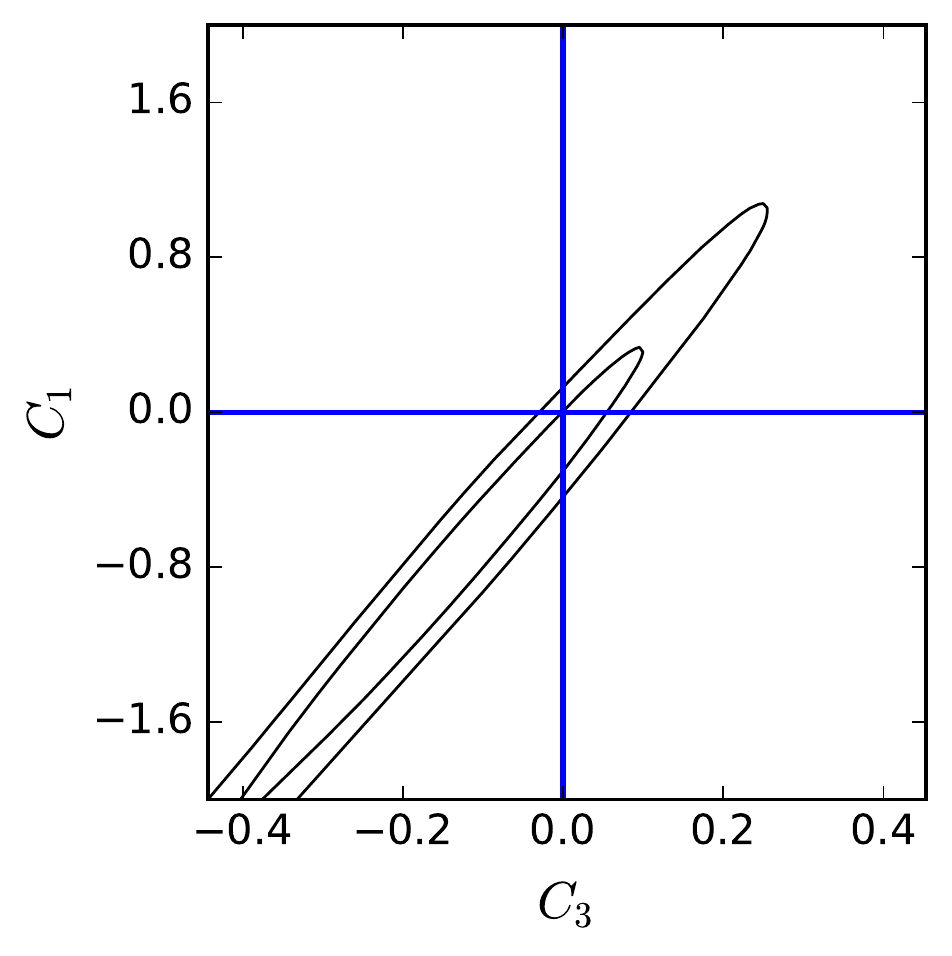}} 
\hskip 15pt\resizebox{120pt}{100pt}{\includegraphics{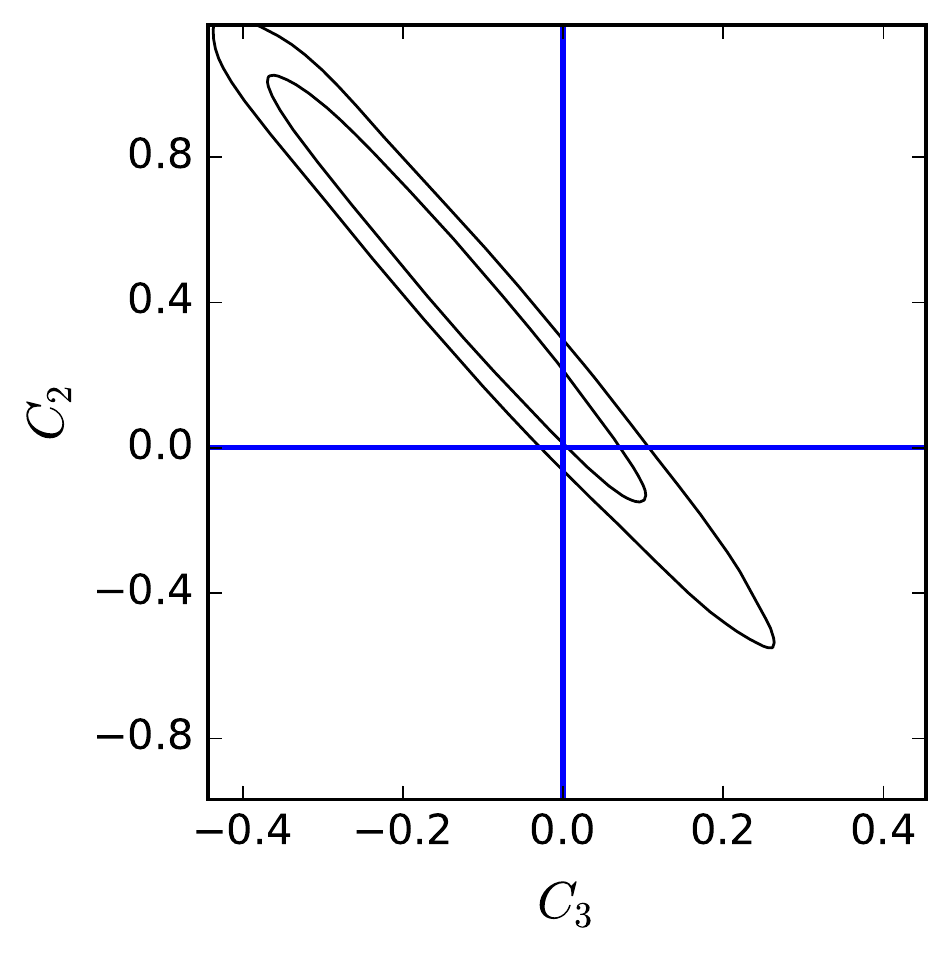}} 

\end{center}
\caption {\footnotesize\label{fig:TTtoTE}Confronting TT + lowT best fit $\Lambda$CDM model as a mean function to TE + lowTEB data (top panels) and TE data (bottom panels) 
considering third order Crossing function. The marginalized contours of the crossing hyperparameters are plotted. Note that while TE + lowTEB data indicate about 2$\sigma$ deviation
from the best fit $\Lambda$CDM model temperature power spectrum, removing constraints from lowTEB, we find slightly better agreement.}
\end{figure*}

In Fig.~\ref{fig:TTtoTE} we compare the best fit mean function from TT + lowT with TE + lowTEB (in the top panels) and TE data (bottom panels) using upto third order crossing function. 
The marginalized crossing hyperparameters show small deviations beyond 1$\sigma$. This explains the differences in the cosmological parameter constraints, particularly the reionization 
optical depth, $\tau$ ($0.112^{+0.036}_{-0.032}$ from TT + lowT and $0.061{\pm0.021}$ from TE + lowTEB) and the normalization, $\ln\l[10^{10}A_{\rm s}\r]$ ($3.154^{+0.068}_{-0.060}$ from
TT + lowT and $3.048\pm0.045$ from TE + lowTEB). The differences arise from the better constraints on the reionization from large scale TE angular power spectrum data that breaks the 
degeneracies between the CMB normalization and the optical depth. As can be seen from the crossing hyperparameters, TT + lowT best fit model however agree with the TE + lowTEB data at $2\sigma$
level. To confirm the above, we performed the analysis again without using lowTEB constraints (lower panel). Note that without lowTEB data, we find slightly better agreement between the two datasets.

\begin{figure*}[!htb]
\begin{center} 
\resizebox{213pt}{160pt}{\includegraphics{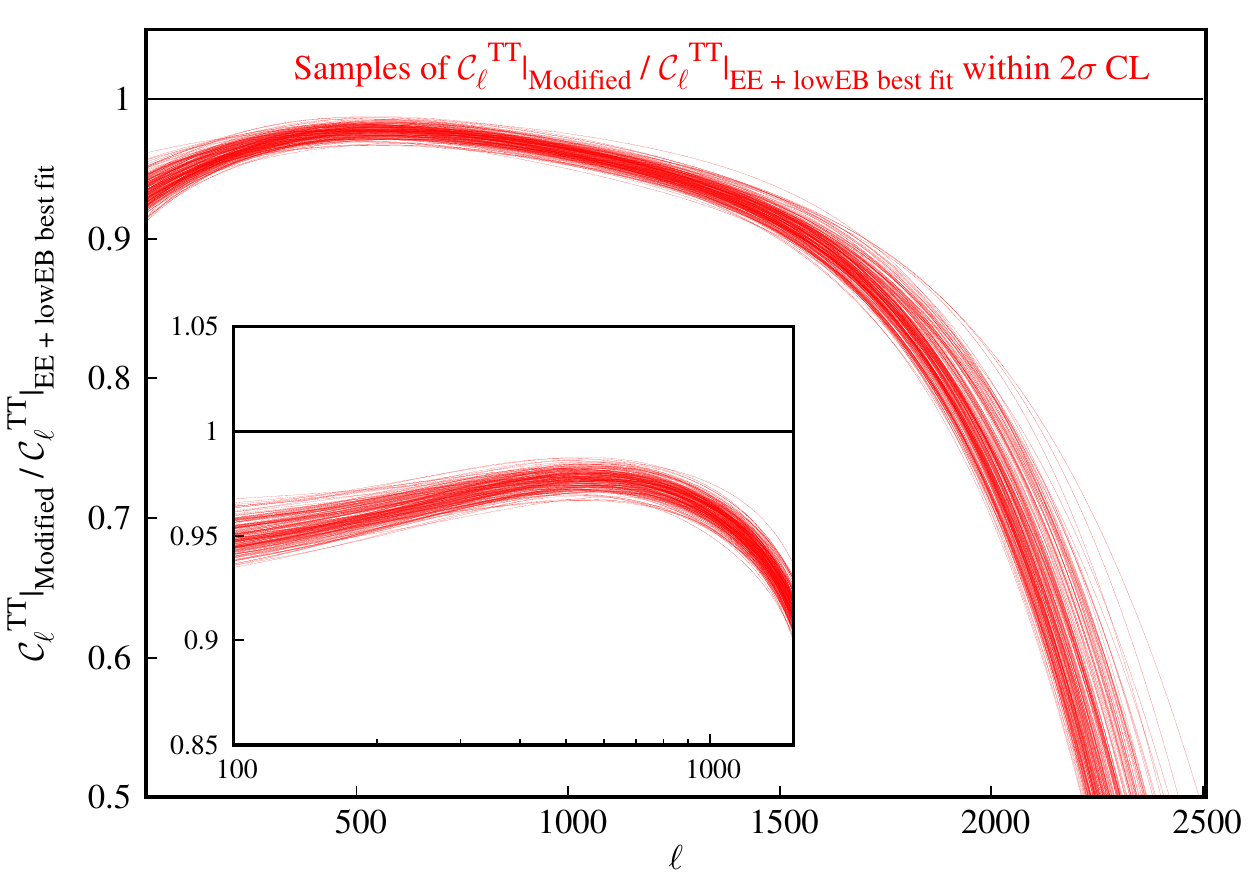}} 
\resizebox{213pt}{160pt}{\includegraphics{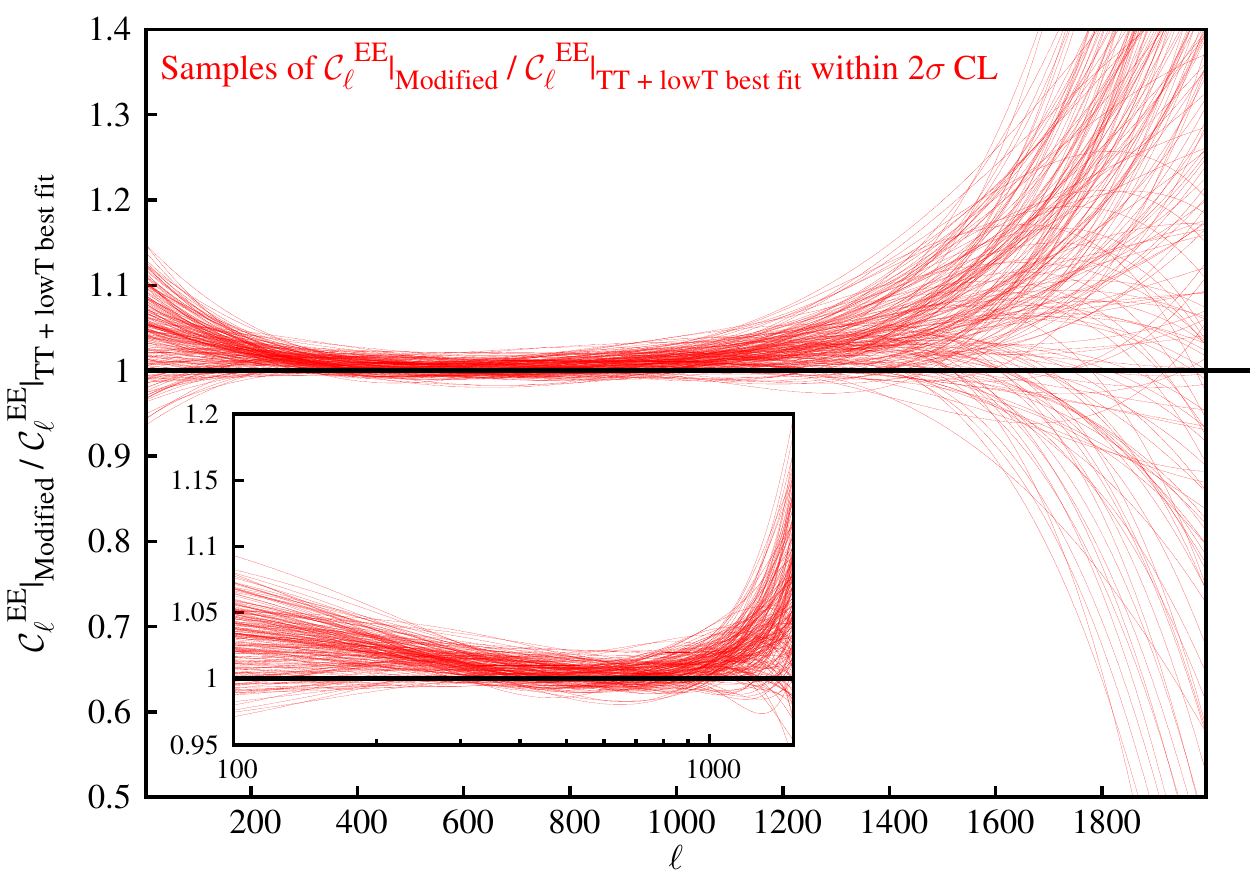}} 
\end{center}
\caption{\footnotesize\label{fig:samples}Crossing functions within 2$\sigma$ allowed range when best fit power spectrum of temperature 
anisotropy from EE + lowEB data is confronted with TT + lowT data (left); and best fit power spectrum of EE from TT + lowT data is 
confronted with EE + lowTEB data (right). Note that TT data is not in agreement with the best fit model from EE data, while EE data is 
completely consistent with best fit model from TT data.}
\end{figure*}

%%%%%%%%%%%%%%%%%%%%%%%%%%%%%%%%%%%%%%%%%%%%%%%%%%%%%%%%%%%%%%%%%%%%%%%%%%%%%%%
At this point it is important to understand the modifications in the multipole space that are indicated by the crossing functions. 
In Fig.~\ref{fig:samples} we plot the required Crossing modifications when we attempt to use EE + lowEB best fit model as a mean function 
to fit TT + lowT data (left panel) and when we use TT + lowT best fit model as a mean function to fit EE + lowTEB data(right panel). We have 
plotted the Crossing hyperfunctions up to fifth order, within 2$\sigma$ confidence intervals from our MCMC chains. The black vertical lines in 
both plots indicate no modification. The left plot shows significant differences between the best fit model to the EE data and the TT data. An 
amplitude difference here is very much evident. This plot suggests that, the best fit model from EE + lowEB data provides a TT angular power spectrum 
which is at least 2.5\% (around $\ell\sim750$) higher in amplitude as required by TT data. All the modifications also indicate that the power difference 
increases at smaller and larger scales. However, the plot at the right suggests that when we use the best fit model to the temperature data as a mean 
function to fit EE data, we do not need any particular modification to this mean function. A mild amplitude difference can be seen on the right panel 
plot (that most Crossing functions are above one consistently at all multiples) but this does not look very significant and we can conclude here that 
there is no considerable tension between the data sets. 

%Comparing both the plots, we can conclude that due to the noise fluctuations in the EE data, the best fit cosmological parameters from EE + lowEB can not be trusted as it is incompatible with a more precise TT data. 

%However, it does not give rise to any significant tension since the best fit from TT + lowT is compatible with EE and hence, results from a joint analysis can be trusted. 

In a recent analysis~\cite{Hazra:2016fkm} we have found that some modification in the primordial power spectrum can improve the fit to the joint
TTTEEE + lowTEB datasets to more than 13 improvement in the $\chi^2$ values, though we found the modifications not to be significant enough to 
rule out the concordance model. The modified primordial power spectra were able to decrease the differences between the estimated values of the 
cosmological parameters in  the joint analysis. This is in accordance with our results in this paper where we also find by making some modifications
in the angular power spectra, as can be done considering Crossing functions, we can improve the fit to both data, but this looks not to be very
significant statistically.  We should also note that the extent of improvement in the likelihood value in the Crossing analysis can not be compared 
to the results of the modification of the primordial spectrum in~\cite{Hazra:2016fkm} since here we use only upto fifth order of Crossing functions 
that is not able to capture fine features in the data.

\begin{figure*}[t]
\begin{center} 
\resizebox{120pt}{120pt}{\includegraphics{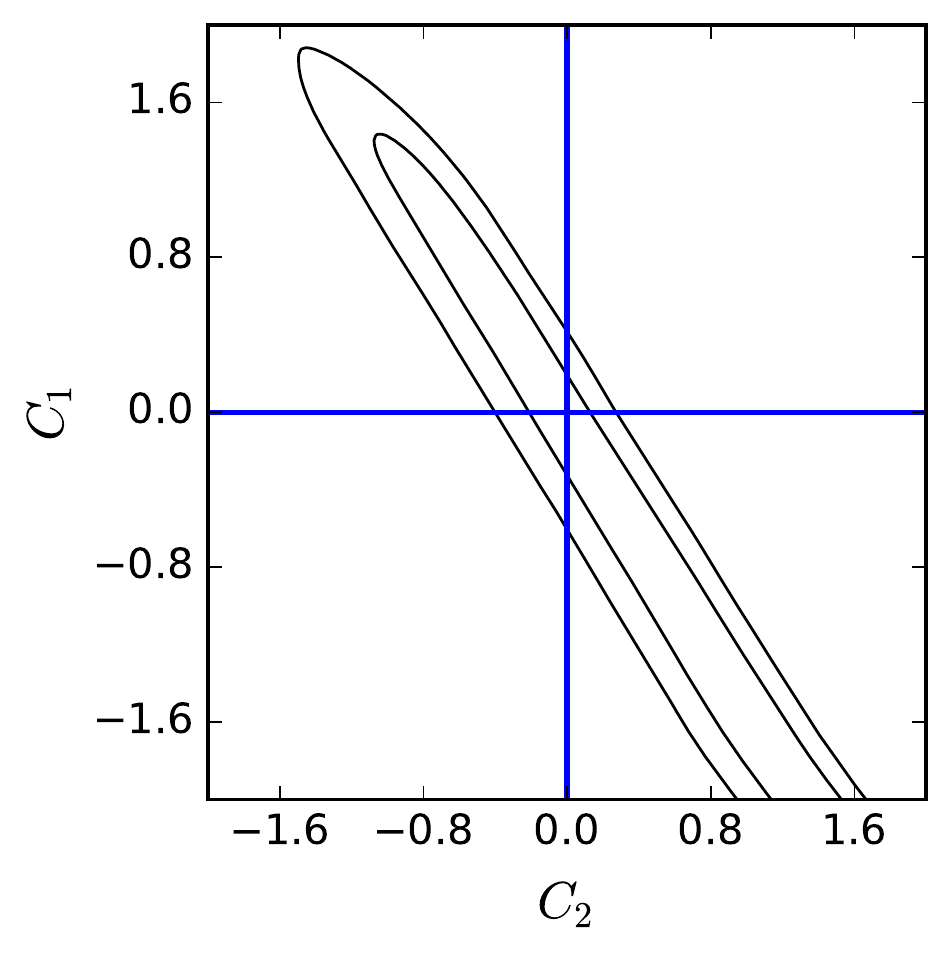}} 
\resizebox{120pt}{120pt}{\includegraphics{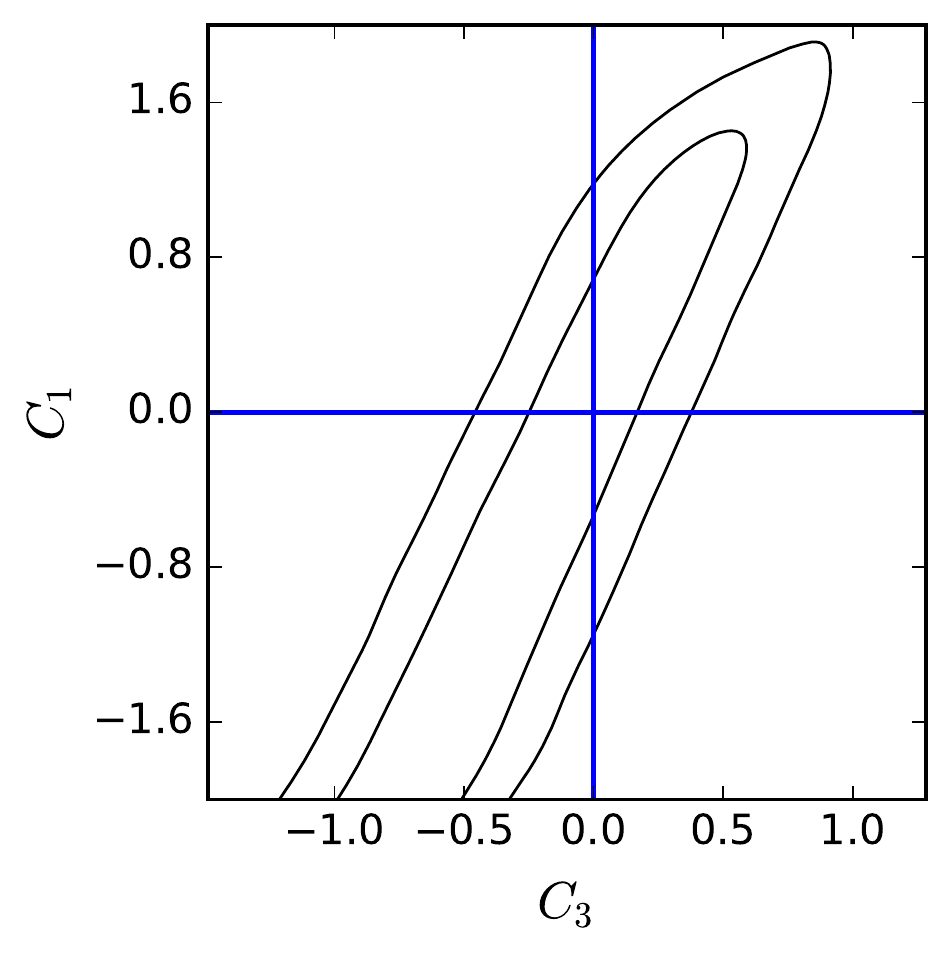}} 
\resizebox{120pt}{120pt}{\includegraphics{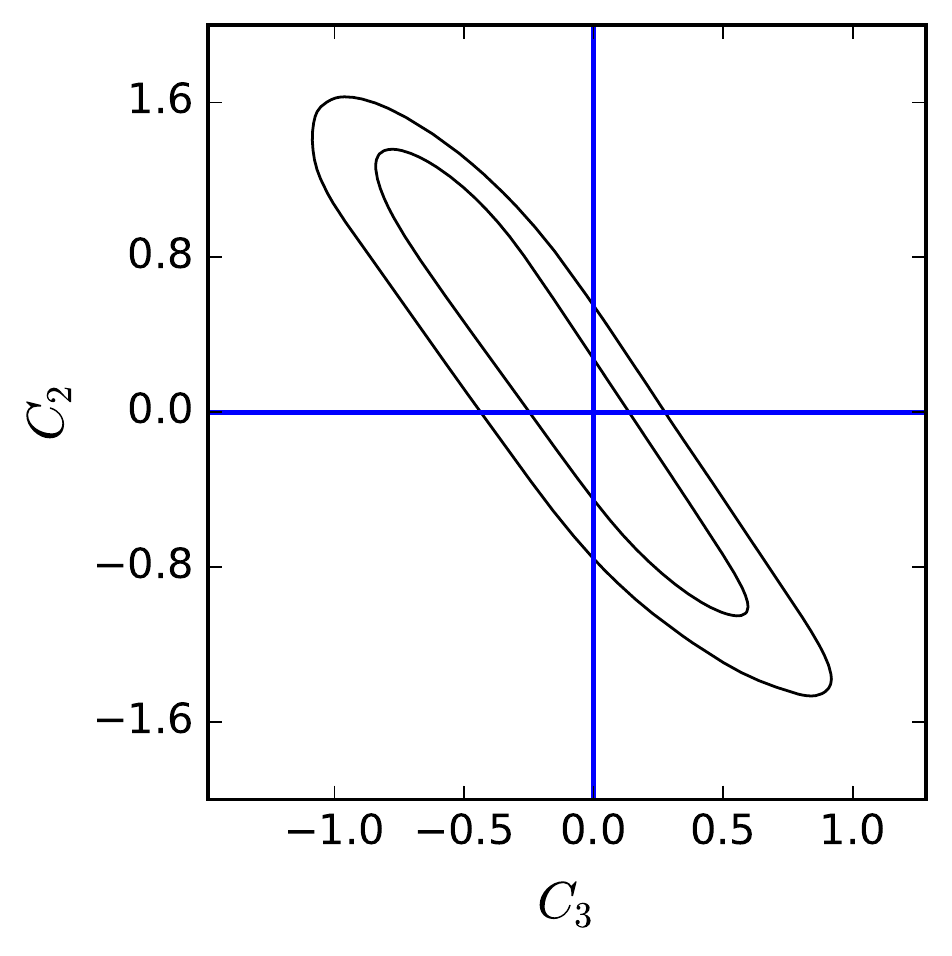}}

\resizebox{120pt}{120pt}{\includegraphics{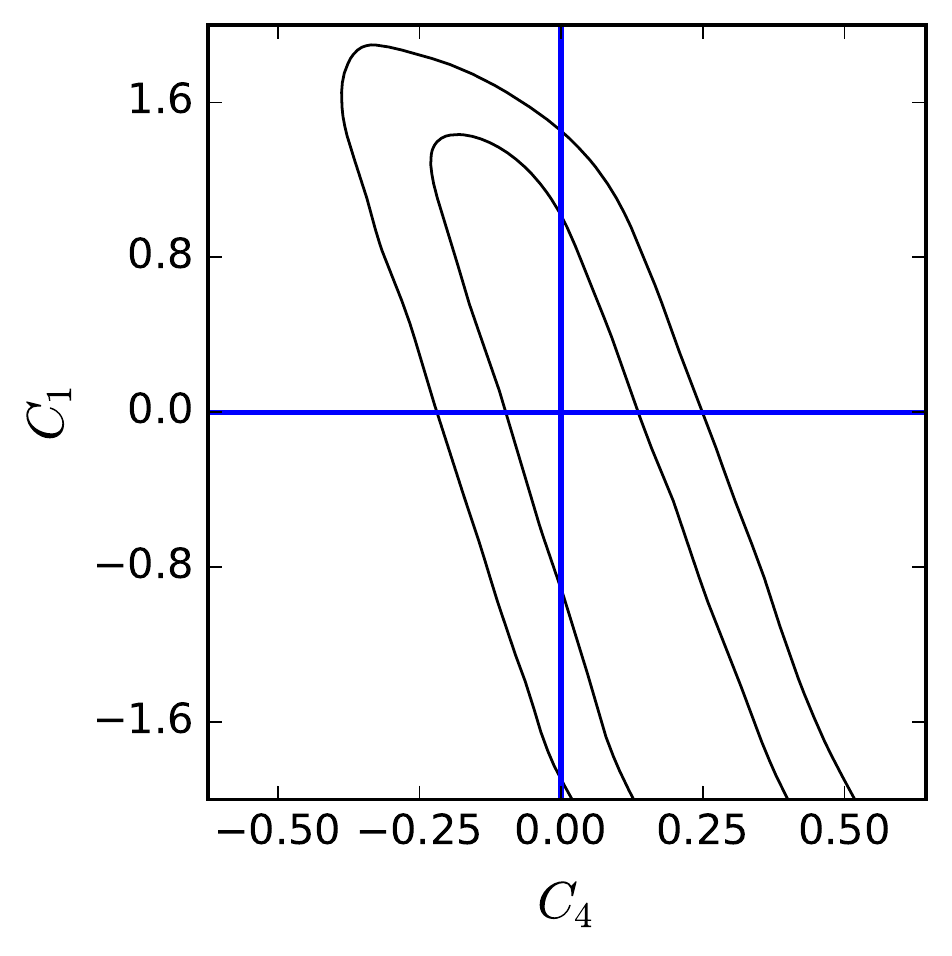}} 
\resizebox{120pt}{120pt}{\includegraphics{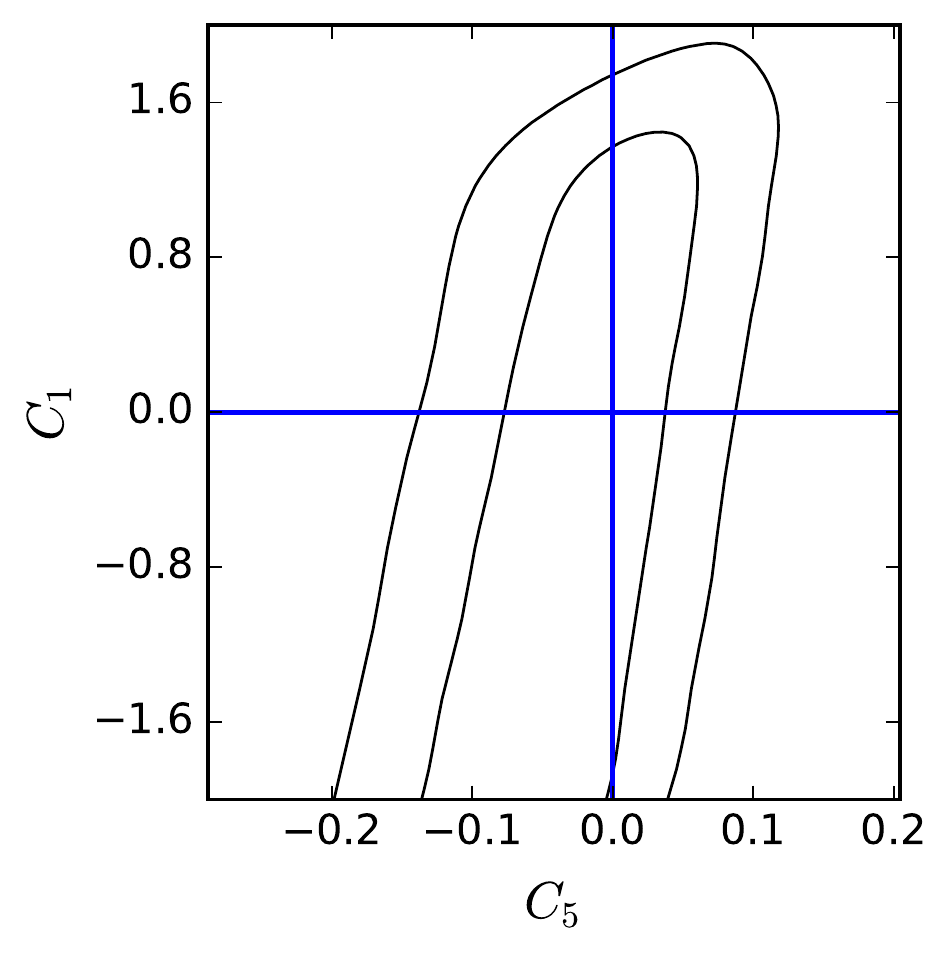}} 
\resizebox{120pt}{120pt}{\includegraphics{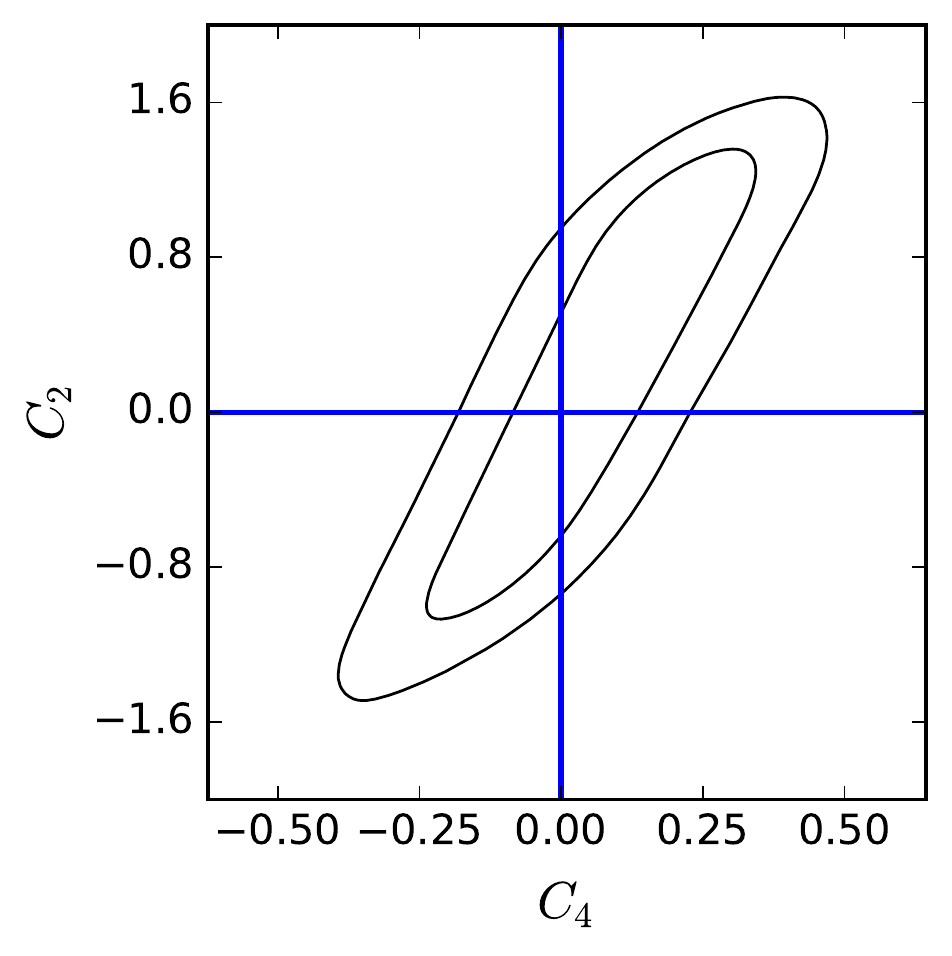}} 

\resizebox{120pt}{120pt}{\includegraphics{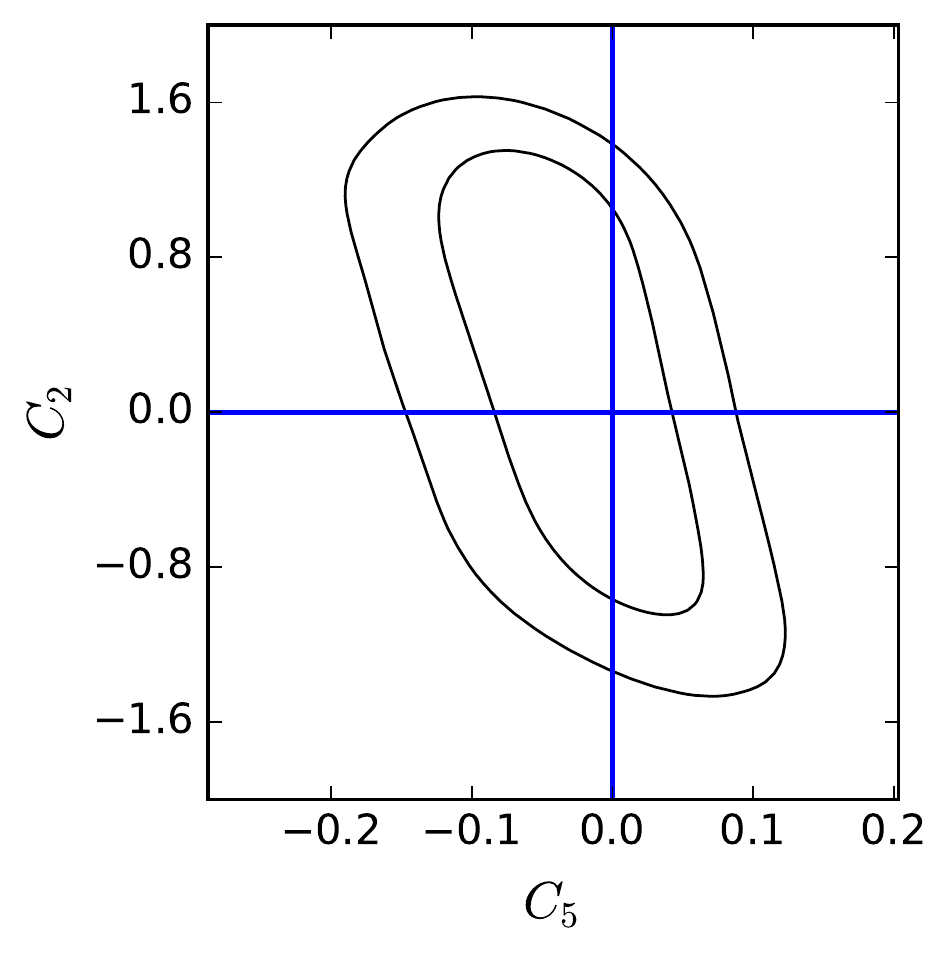}} 
\resizebox{120pt}{120pt}{\includegraphics{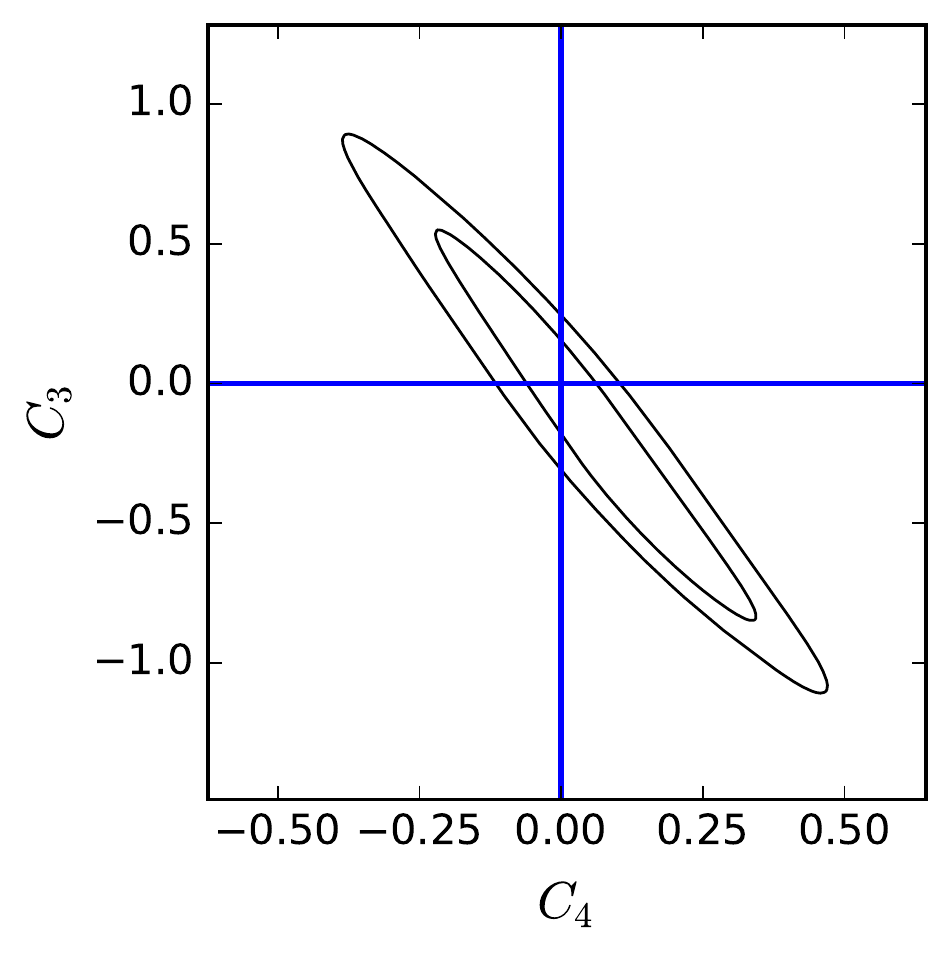}} 
\resizebox{120pt}{120pt}{\includegraphics{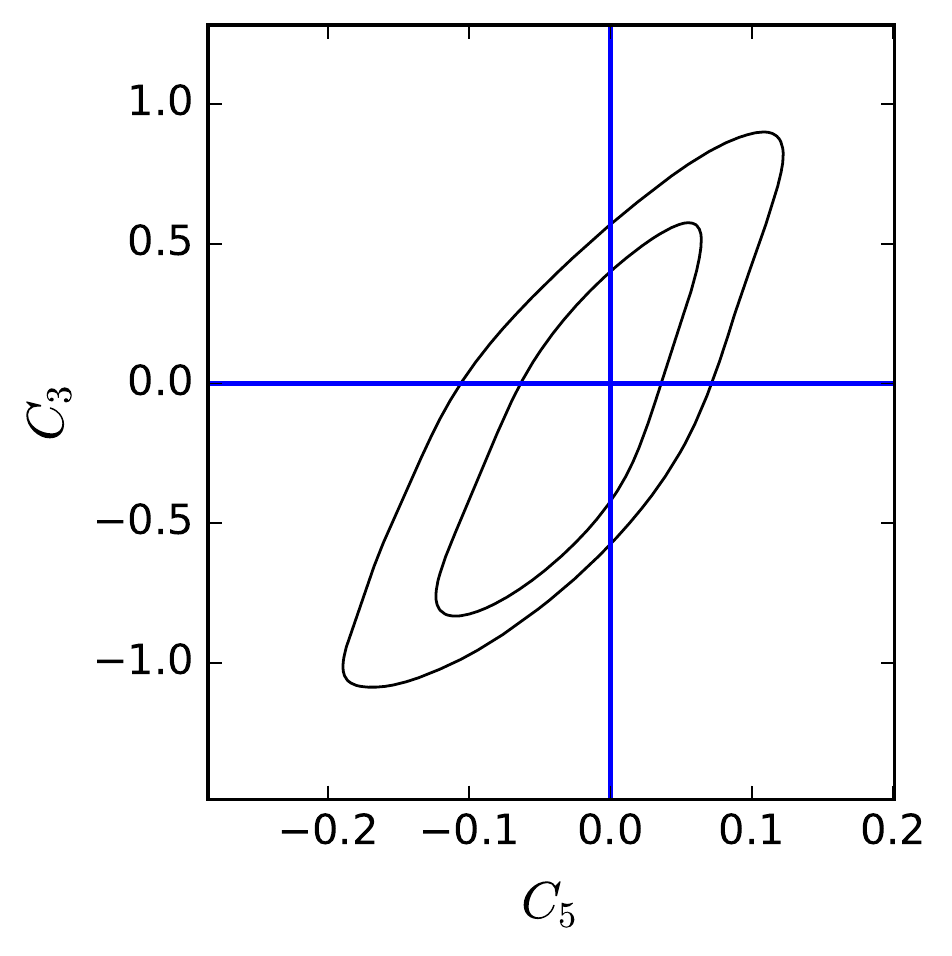}} 

\resizebox{120pt}{120pt}{\includegraphics{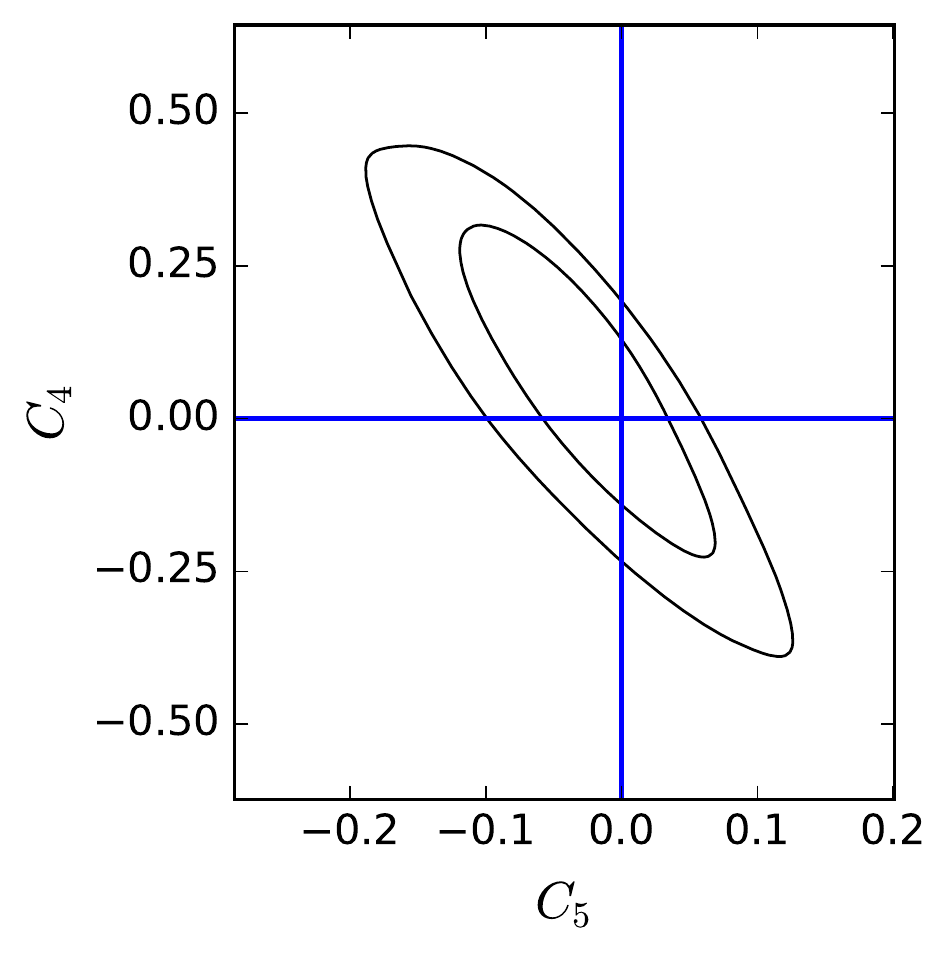}} 

\end{center}
\caption{\footnotesize\label{fig:TTtoTTC5} Confronting concordance model to TT + lowT data considering fifth order Crossing function. 
Note that, here, along with the Crossing hyperparameters we also allow the 
background cosmological parameters and the primordial power spectrum parameters to vary. The contours do not show any
deviation from the standard model represented by $C_{1-5}=0$.}
\end{figure*}

\begin{figure*}[t]
\begin{center} 
\resizebox{120pt}{120pt}{\includegraphics{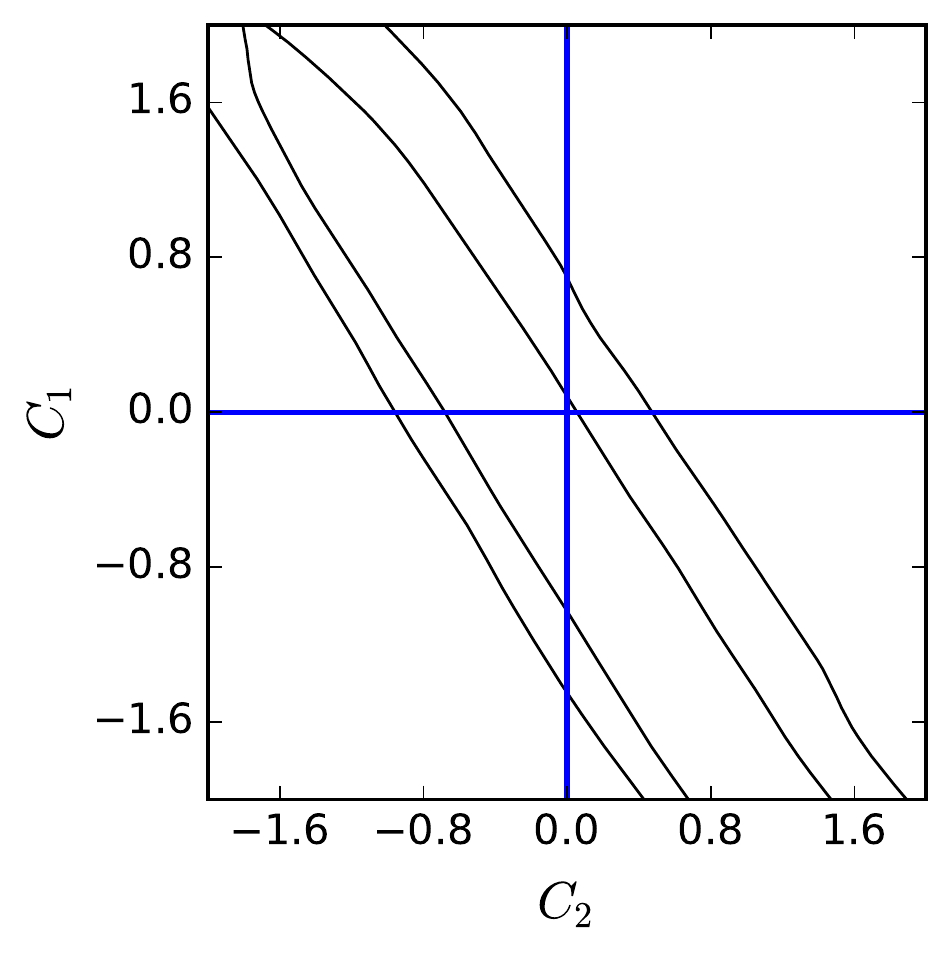}} 
\resizebox{120pt}{120pt}{\includegraphics{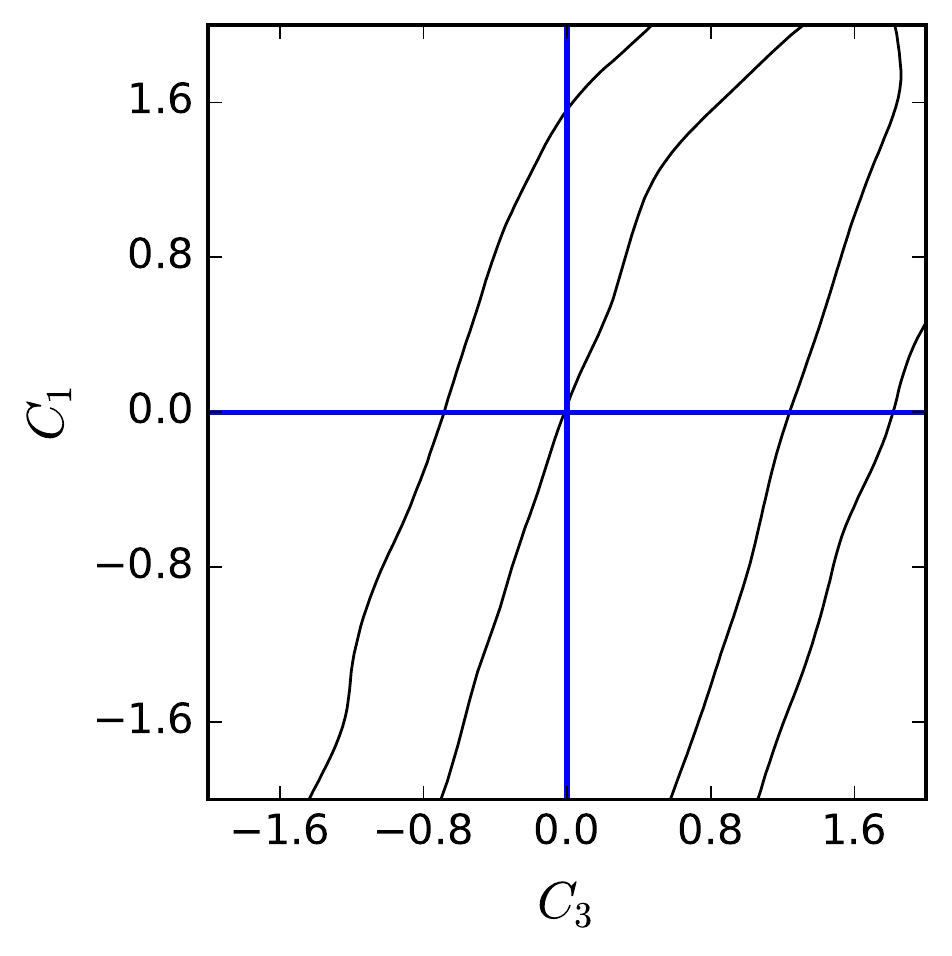}} 
\resizebox{120pt}{120pt}{\includegraphics{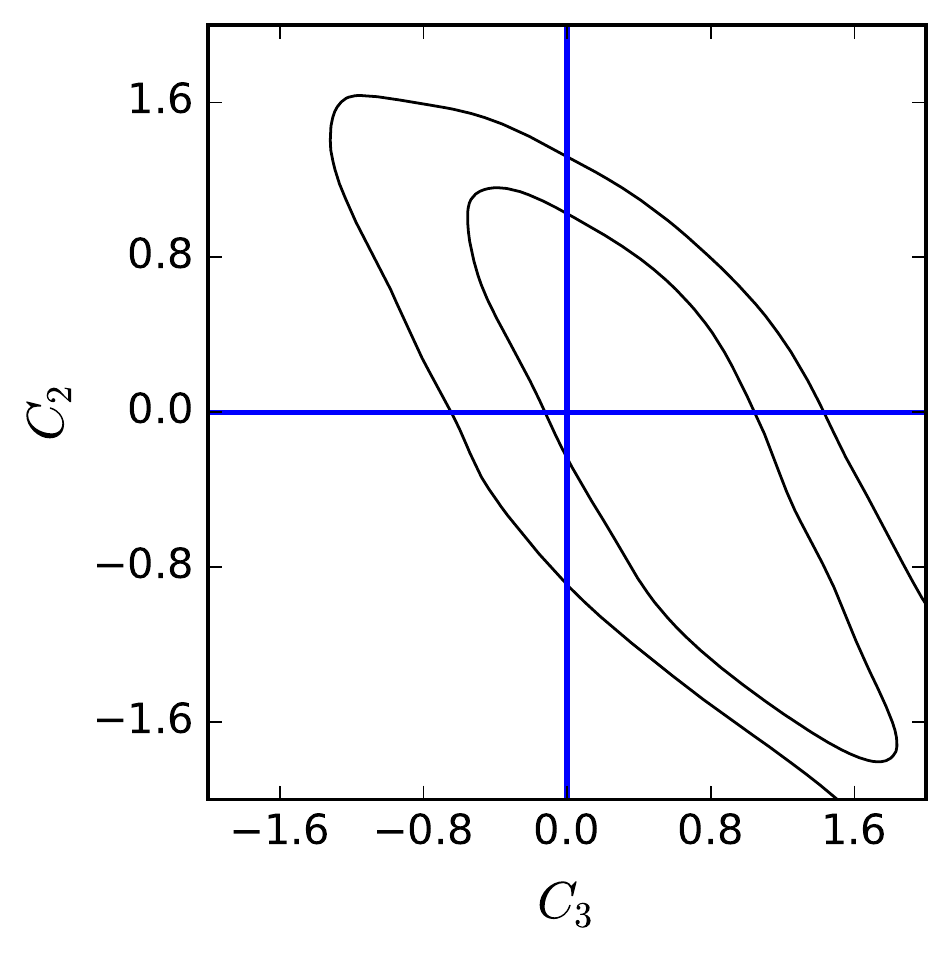}}

\resizebox{120pt}{120pt}{\includegraphics{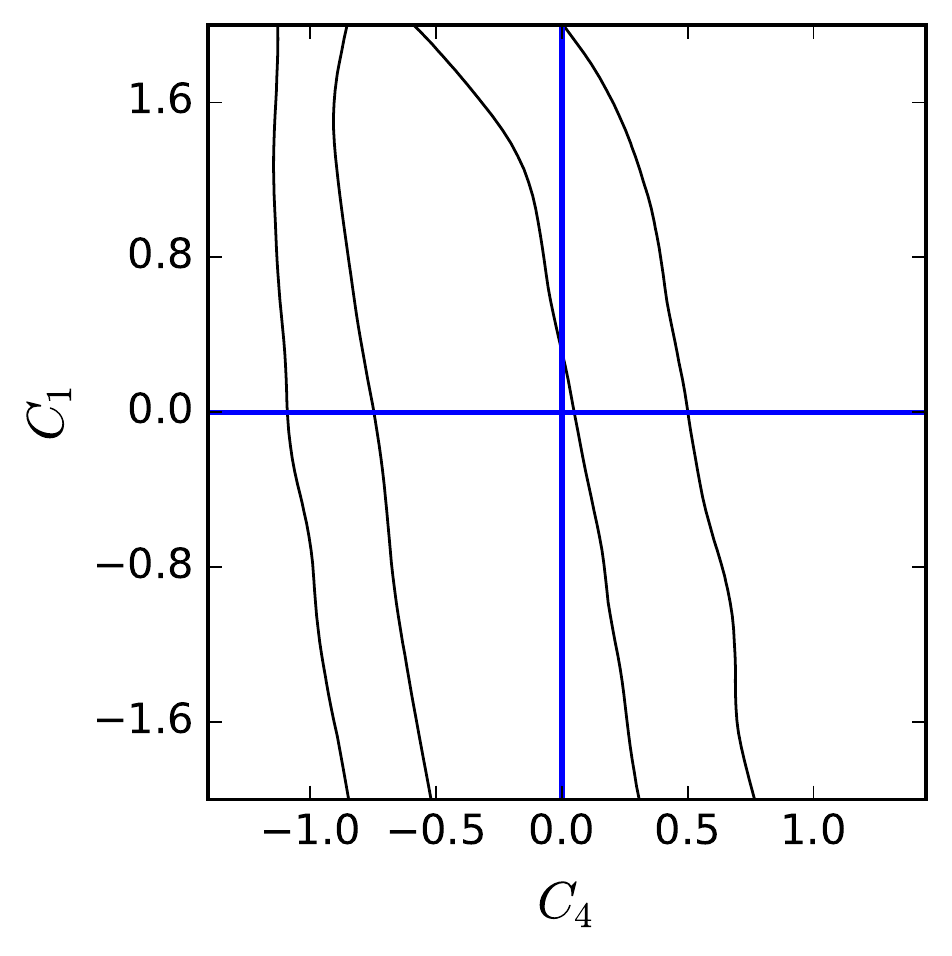}} 
\resizebox{120pt}{120pt}{\includegraphics{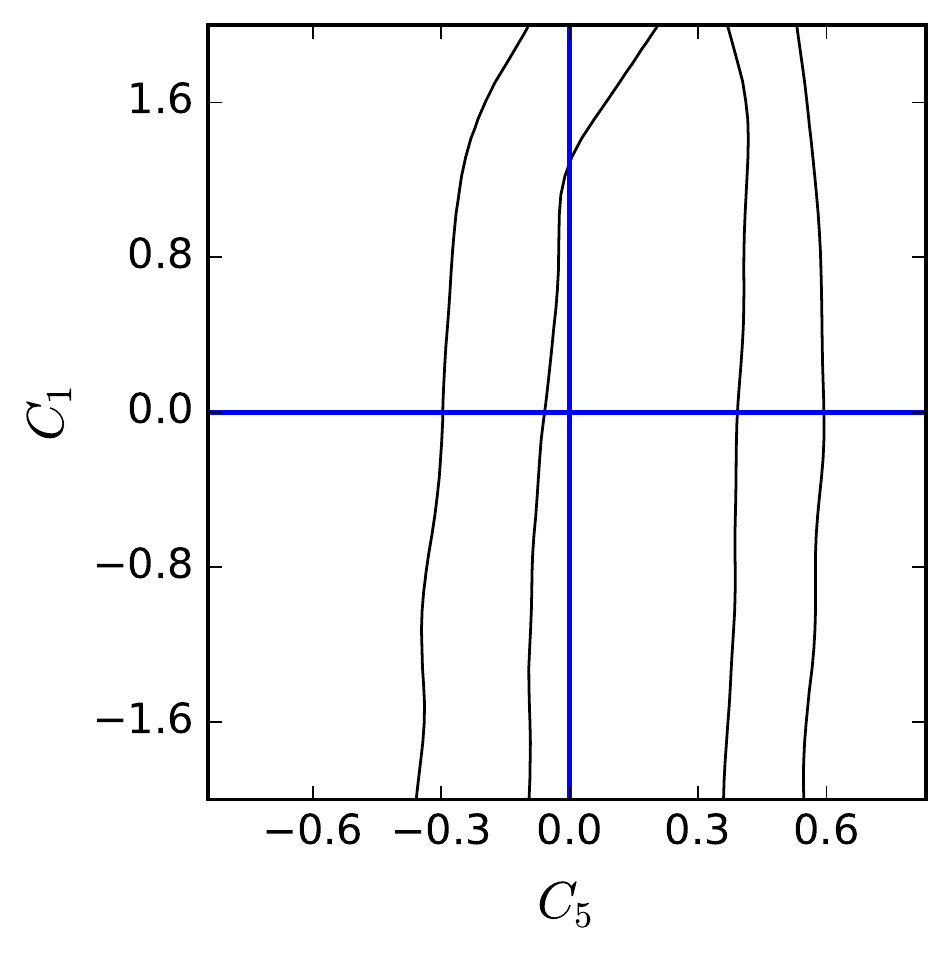}} 
\resizebox{120pt}{120pt}{\includegraphics{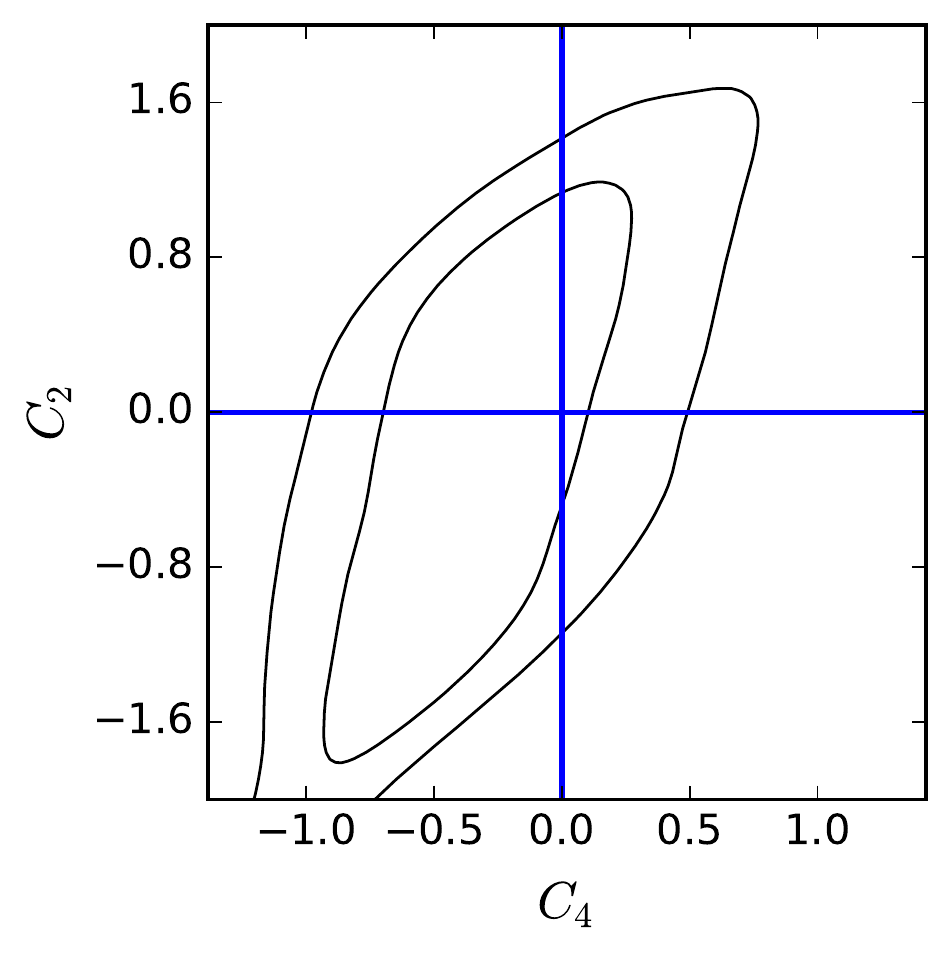}} 

\resizebox{120pt}{120pt}{\includegraphics{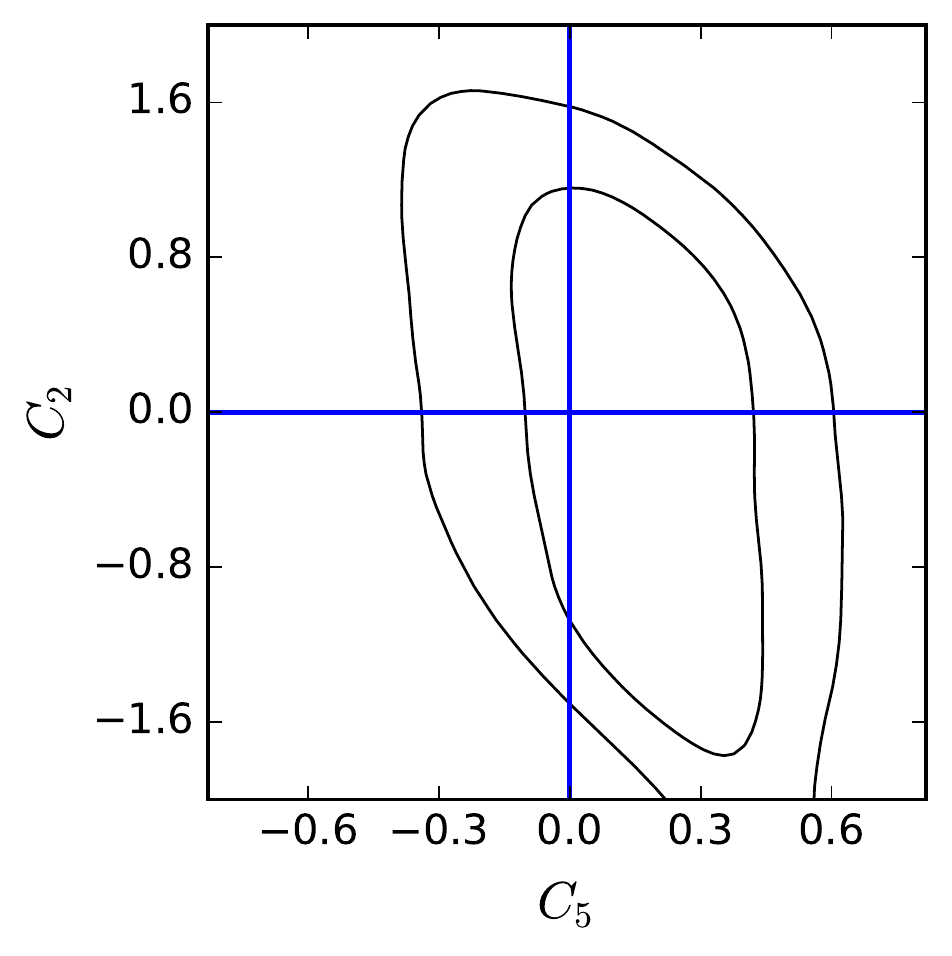}} 
\resizebox{120pt}{120pt}{\includegraphics{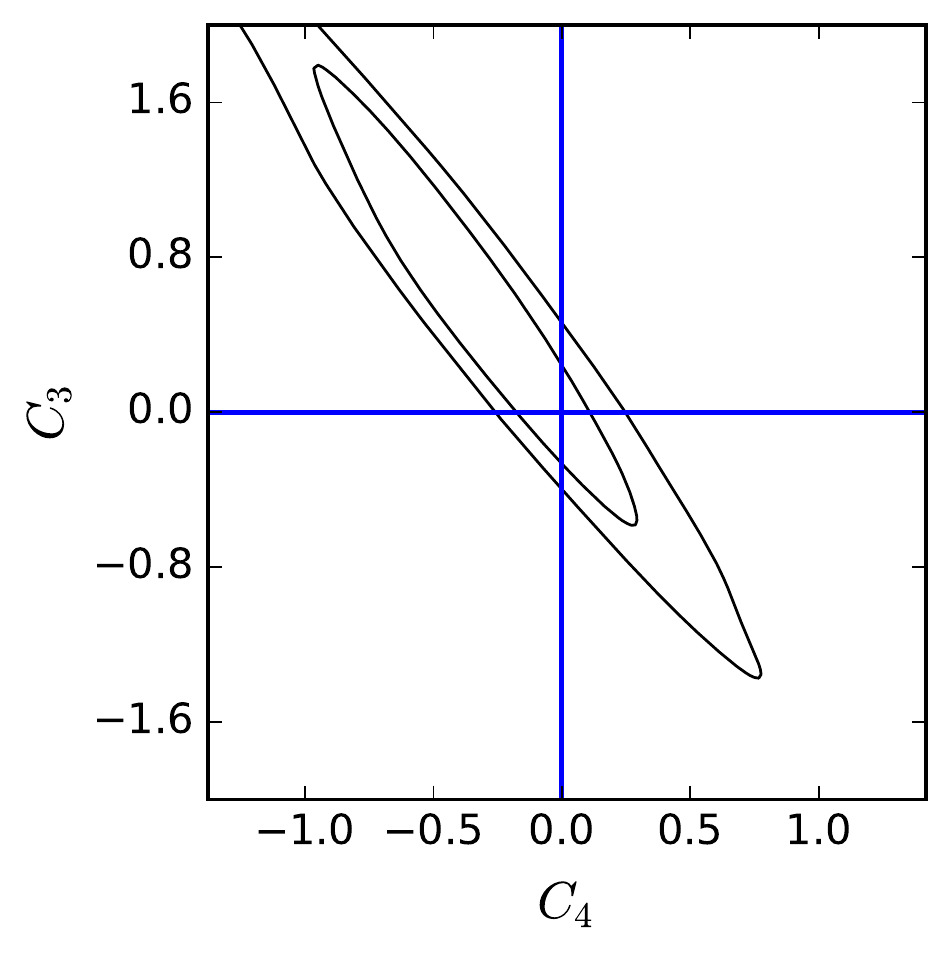}} 
\resizebox{120pt}{120pt}{\includegraphics{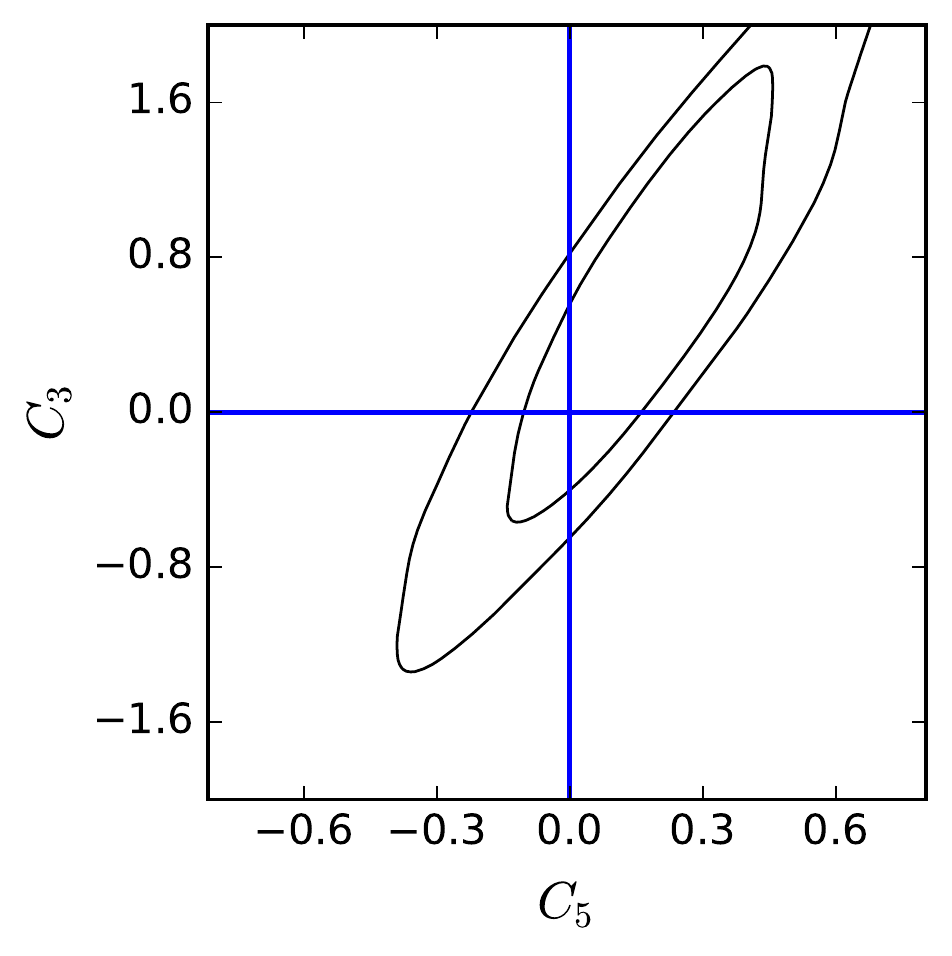}} 

\resizebox{120pt}{120pt}{\includegraphics{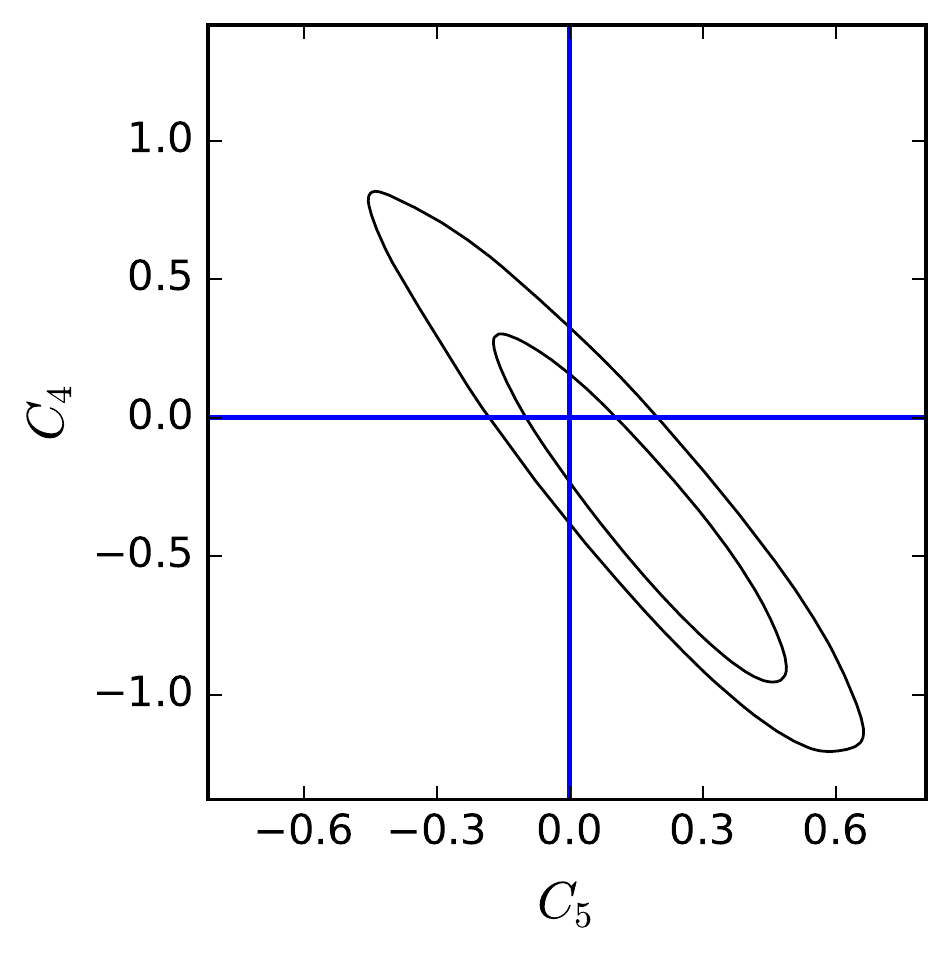}} 

\end{center}
\caption{\footnotesize\label{fig:EEtoEEC5} Confronting concordance model to EE + lowTEB data considering fifth order Crossing function. 
Similar to Fig.~\ref{fig:TTtoTTC5}, here too we do not find any indication for departure from the concordance model.}
\end{figure*}

\begin{figure*}[t]
\begin{center} 
\resizebox{120pt}{120pt}{\includegraphics{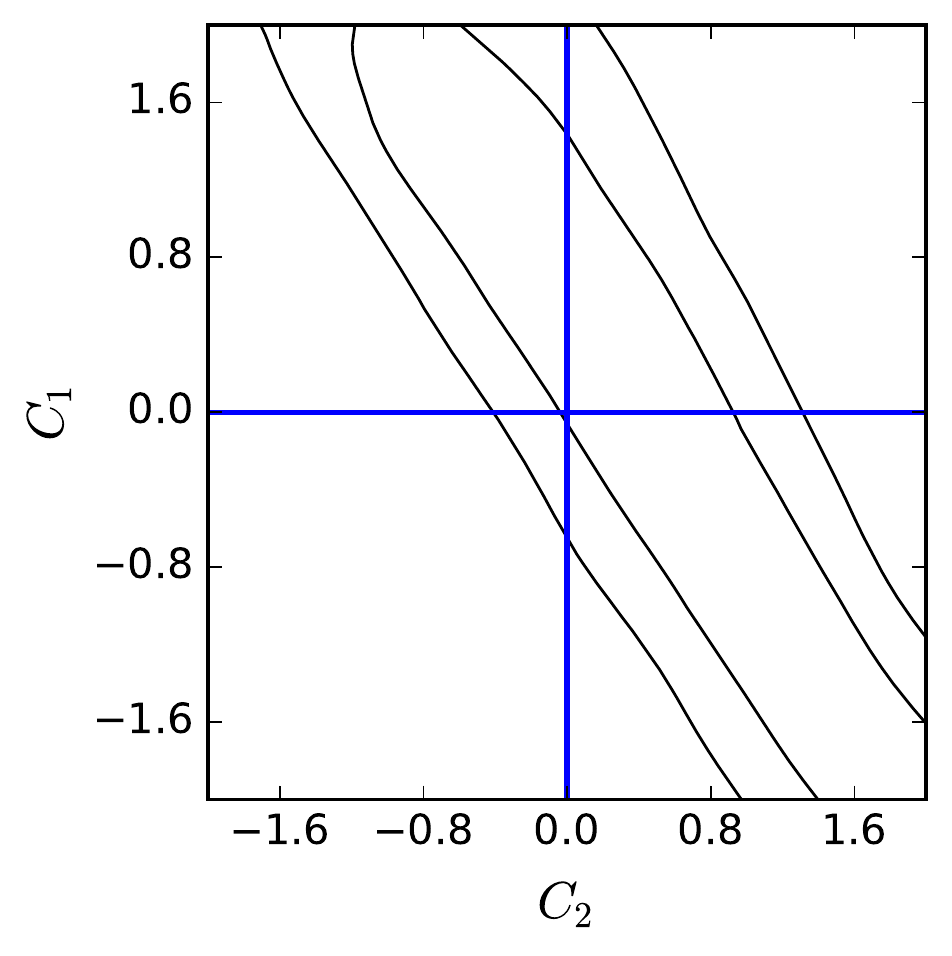}} 
\resizebox{120pt}{120pt}{\includegraphics{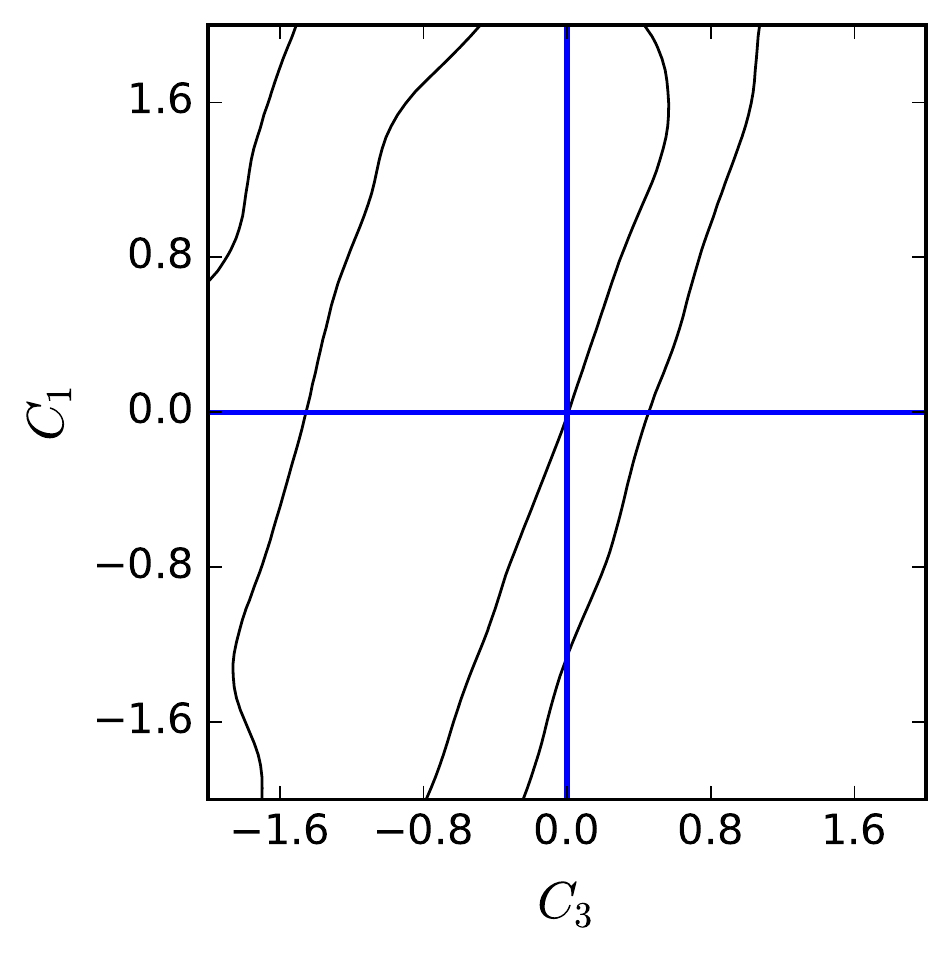}} 
\resizebox{120pt}{120pt}{\includegraphics{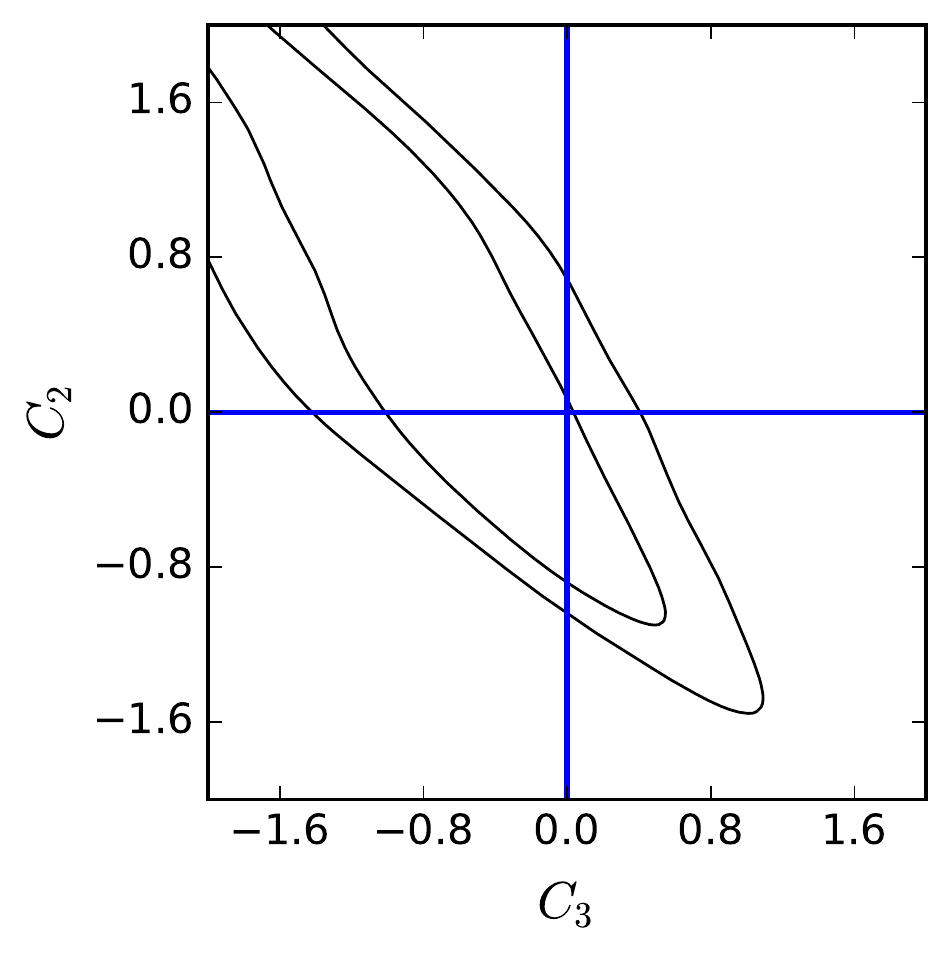}}

\resizebox{120pt}{120pt}{\includegraphics{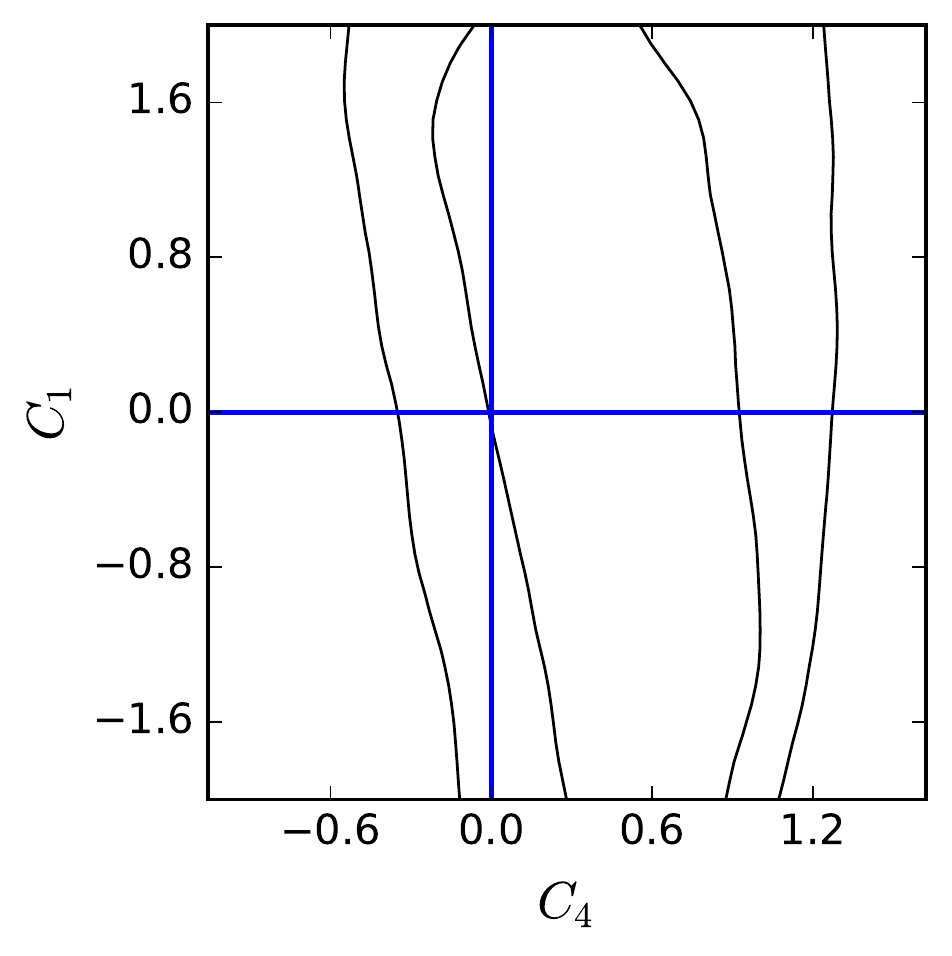}} 
\resizebox{120pt}{120pt}{\includegraphics{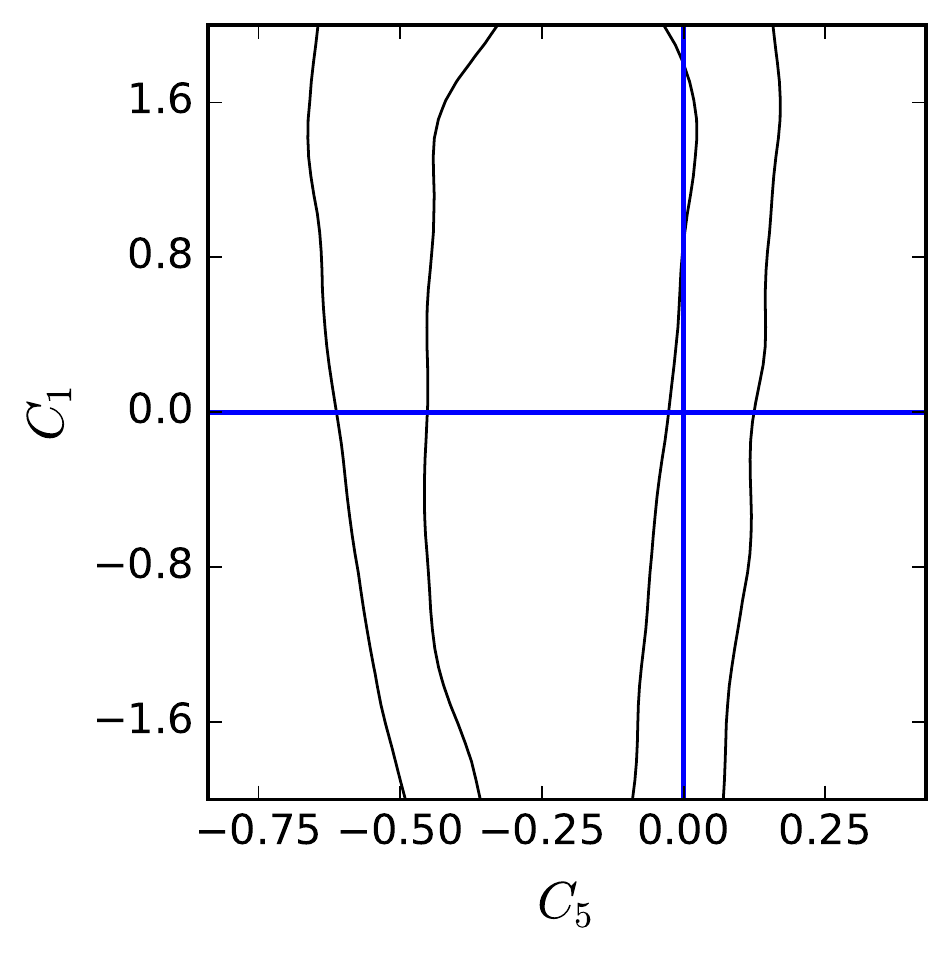}} 
\resizebox{120pt}{120pt}{\includegraphics{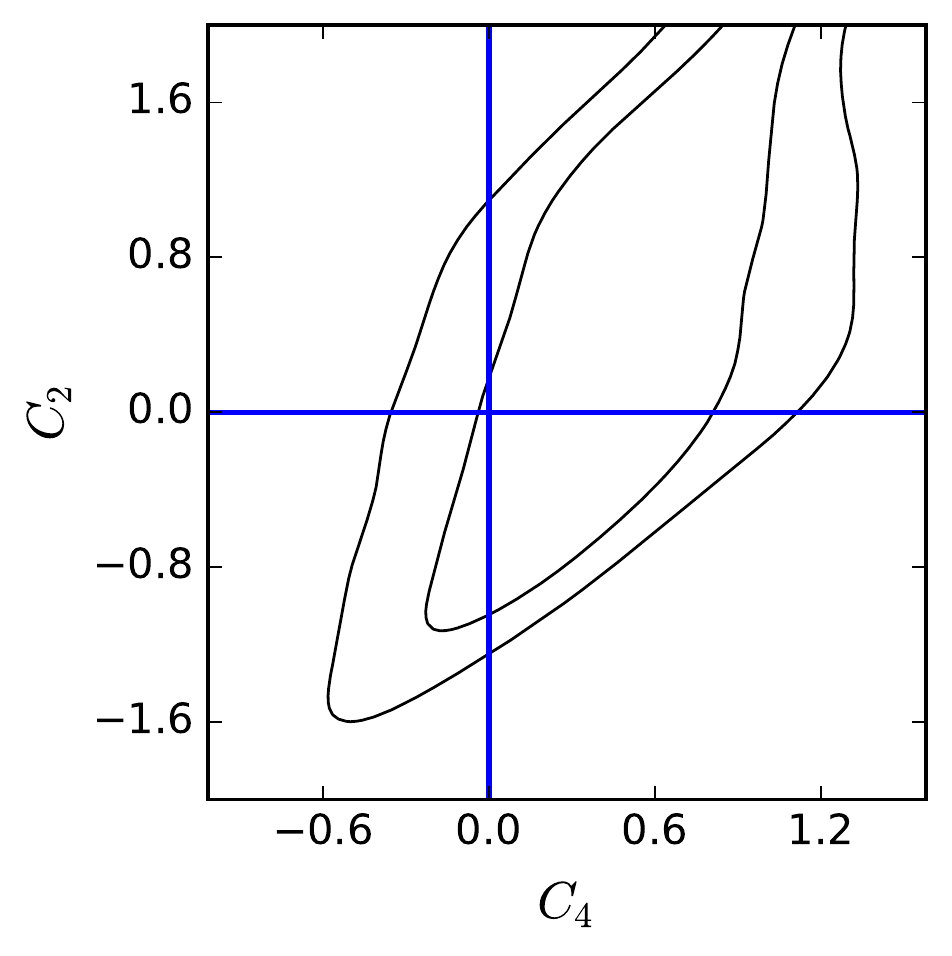}} 

\resizebox{120pt}{120pt}{\includegraphics{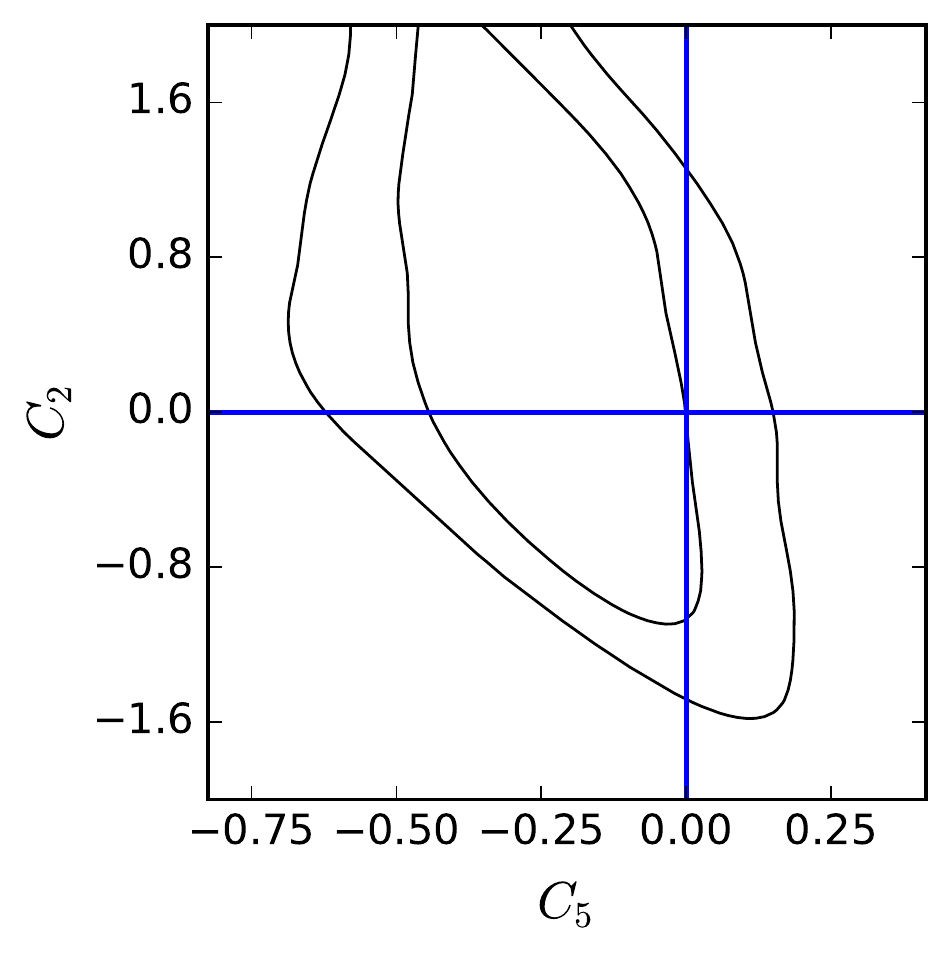}} 
\resizebox{120pt}{120pt}{\includegraphics{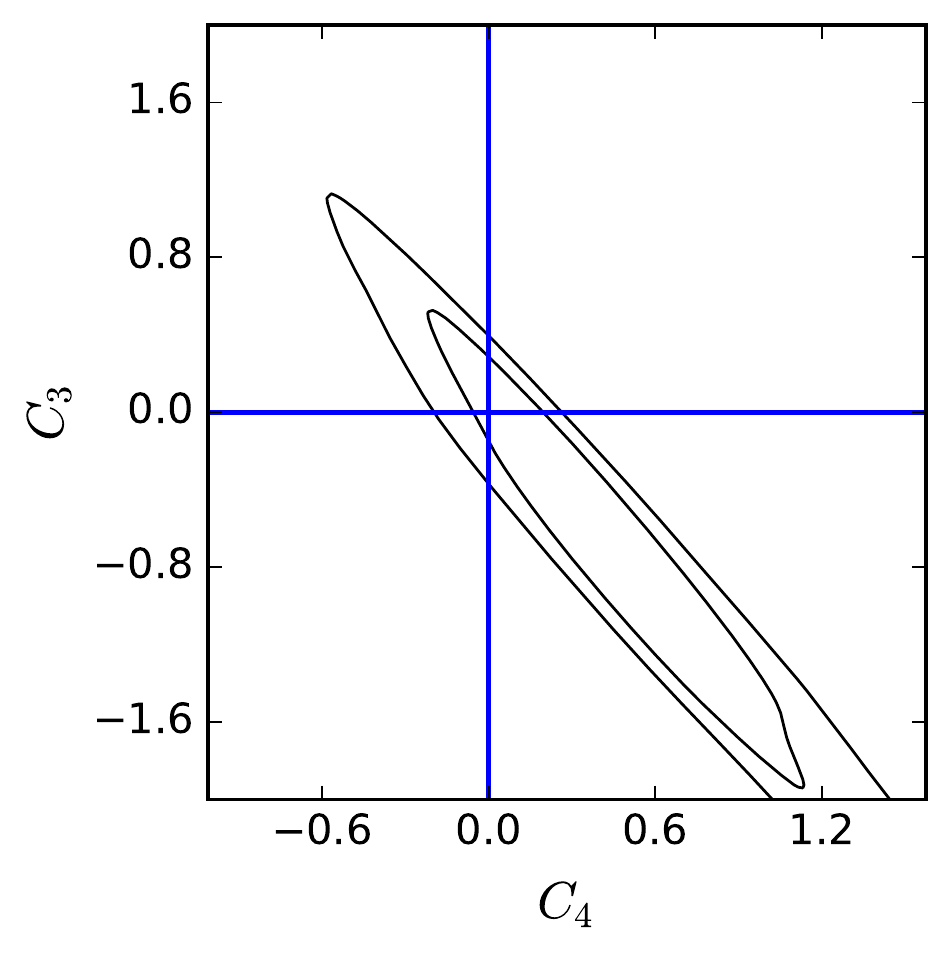}} 
\resizebox{120pt}{120pt}{\includegraphics{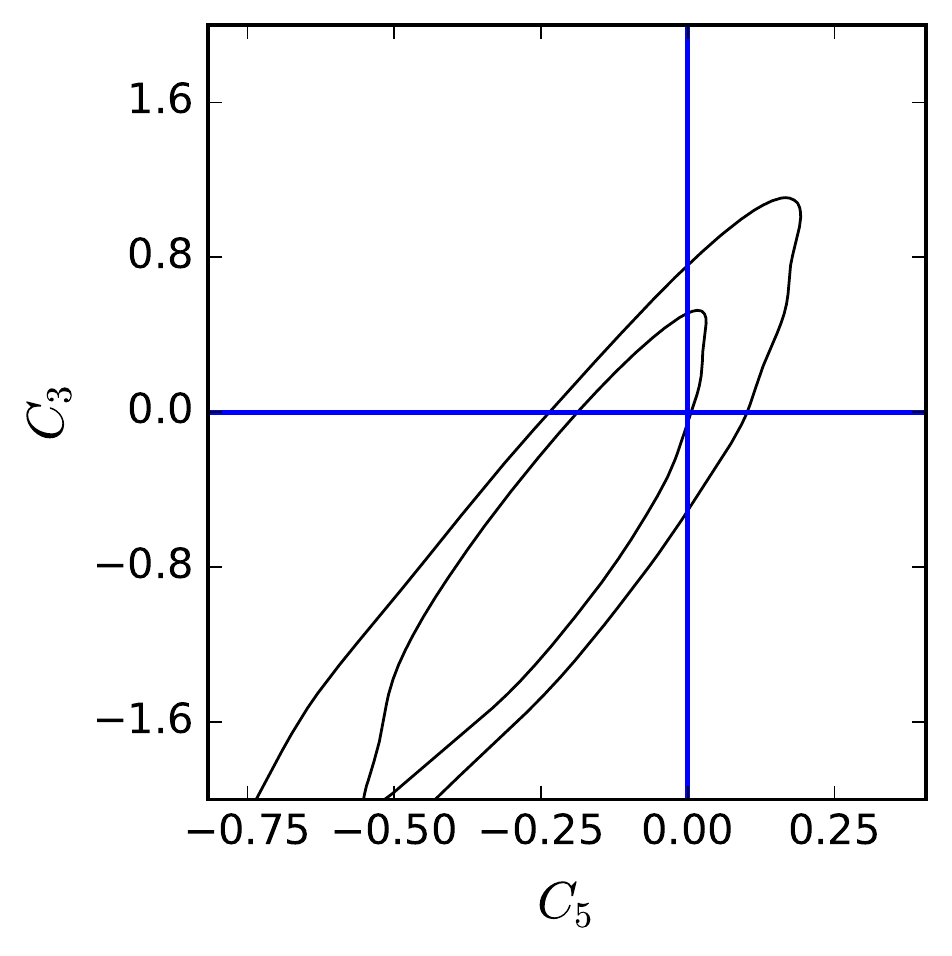}} 

\resizebox{120pt}{120pt}{\includegraphics{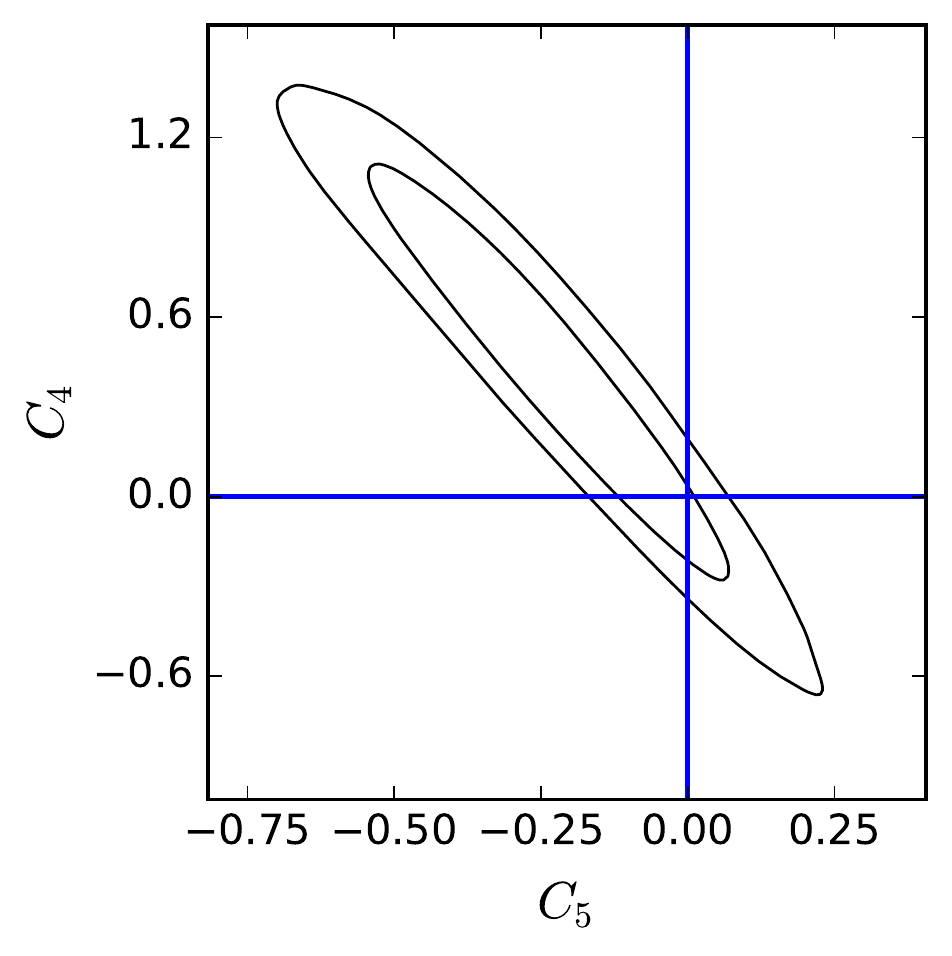}} 

\end{center}
\caption{\footnotesize\label{fig:TEtoTEC5} Confronting concordance model to TE + lowTEB data considering fifth order Crossing function. 
Unlike Fig.~\ref{fig:TTtoTTC5} and  Fig.~\ref{fig:EEtoEEC5} we find 1$\sigma$ discordance from $\Lambda$CDM.}
\end{figure*}

%%%%%%%%%%%%%%%%%%%%%%%%%%%%%%%%%%%%%%%%%%%%%%%%%%%%%%%%%%%%%%%%%%%%%%%%%%%%%%%

\begin{figure*}[!htb]
\begin{center} 
\resizebox{120pt}{100pt}{\includegraphics{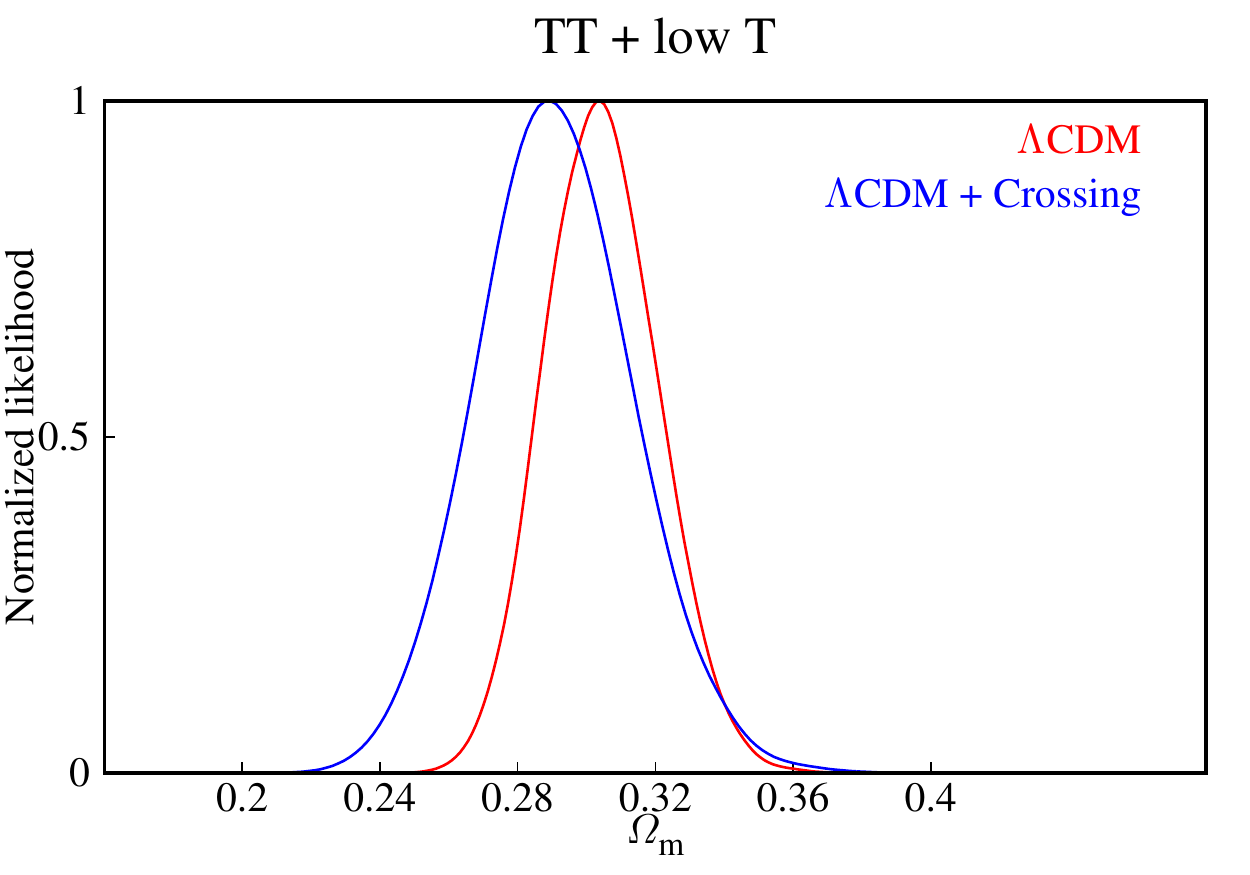}} 
\hskip 15pt\resizebox{120pt}{100pt}{\includegraphics{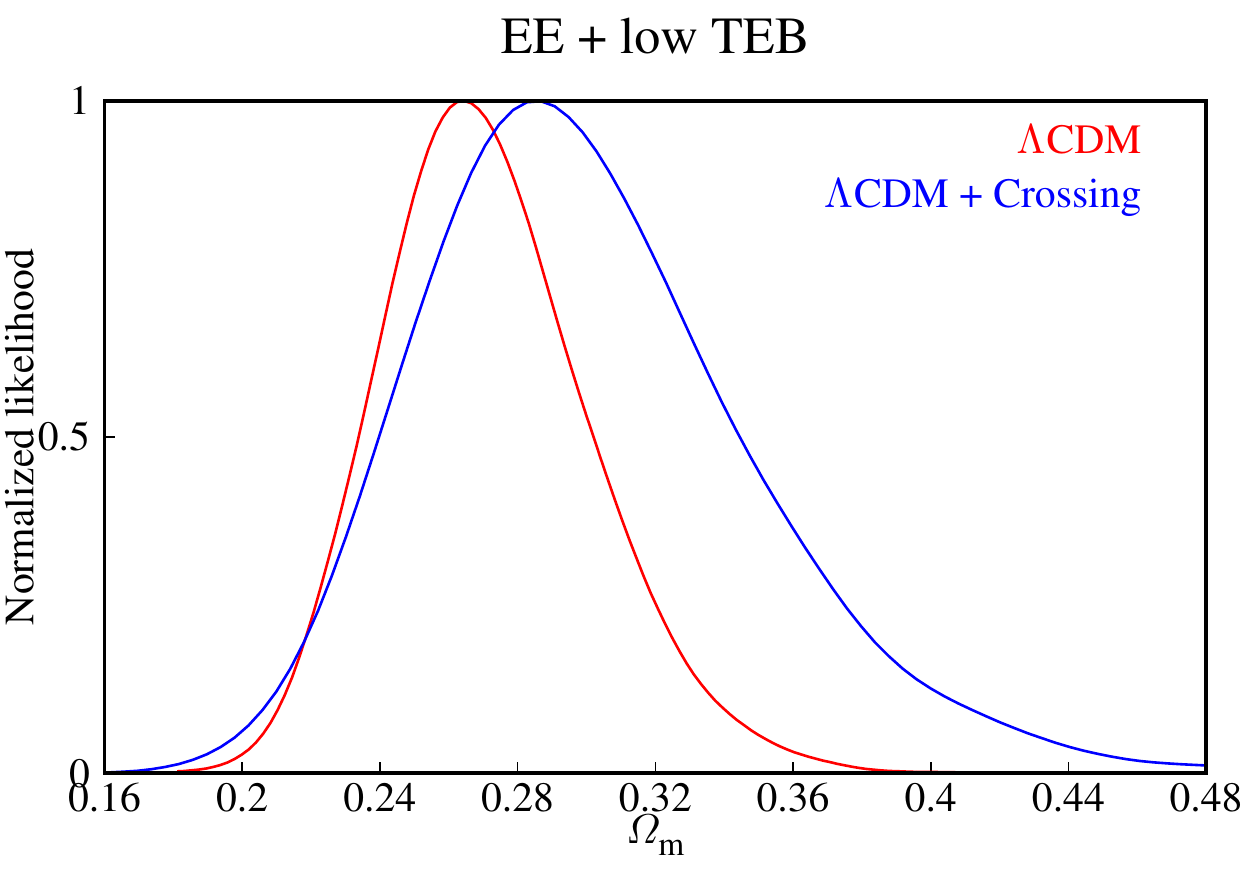}} 
\hskip 15pt\resizebox{120pt}{100pt}{\includegraphics{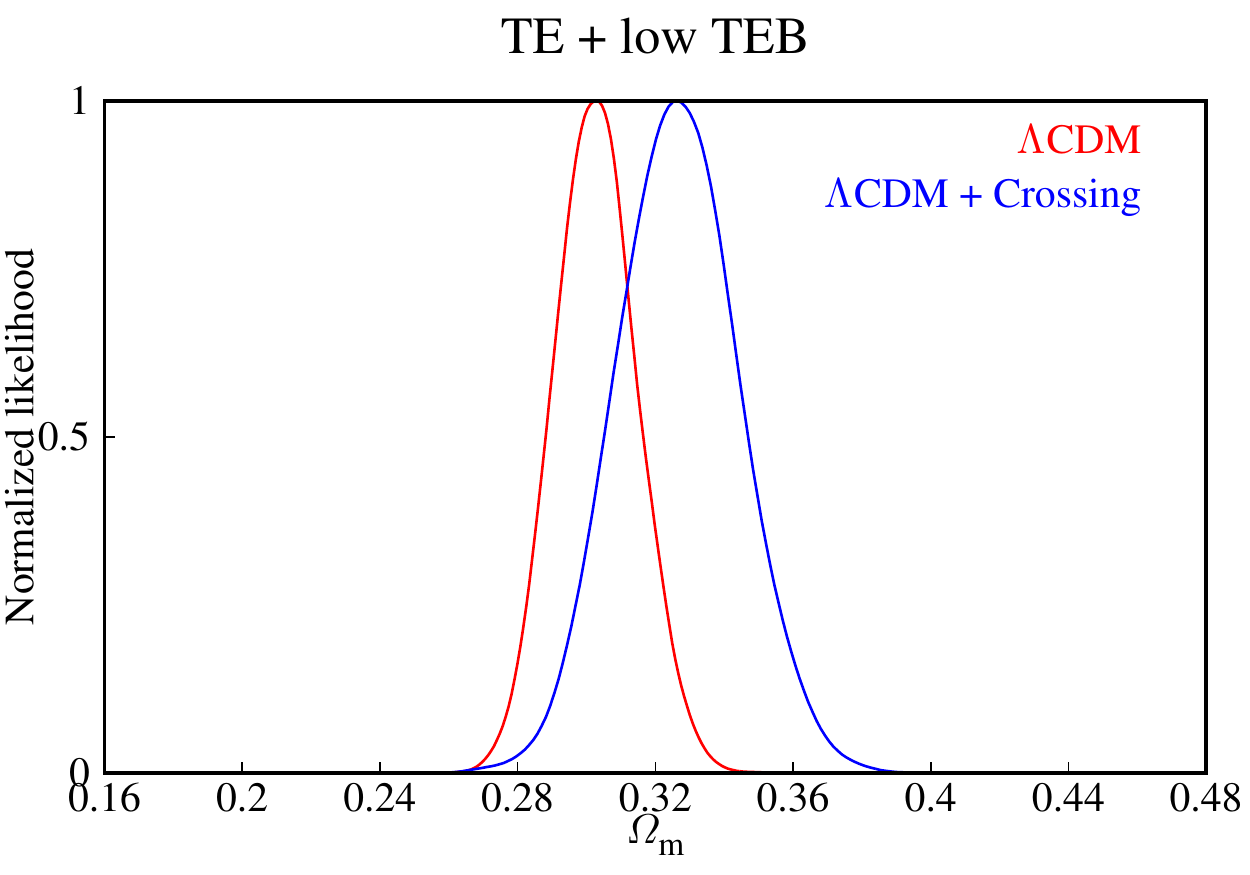}} 

\resizebox{120pt}{100pt}{\includegraphics{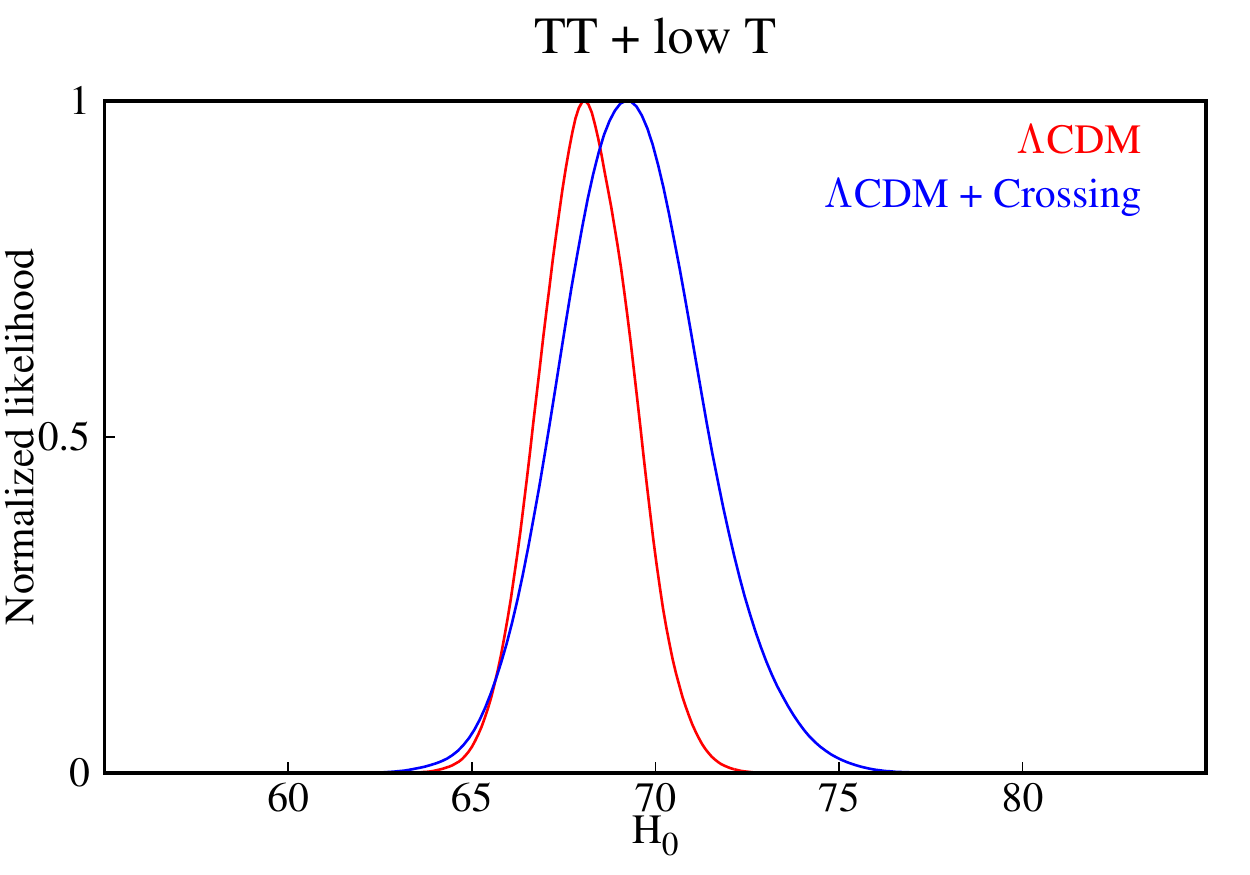}} 
\hskip 15pt\resizebox{120pt}{100pt}{\includegraphics{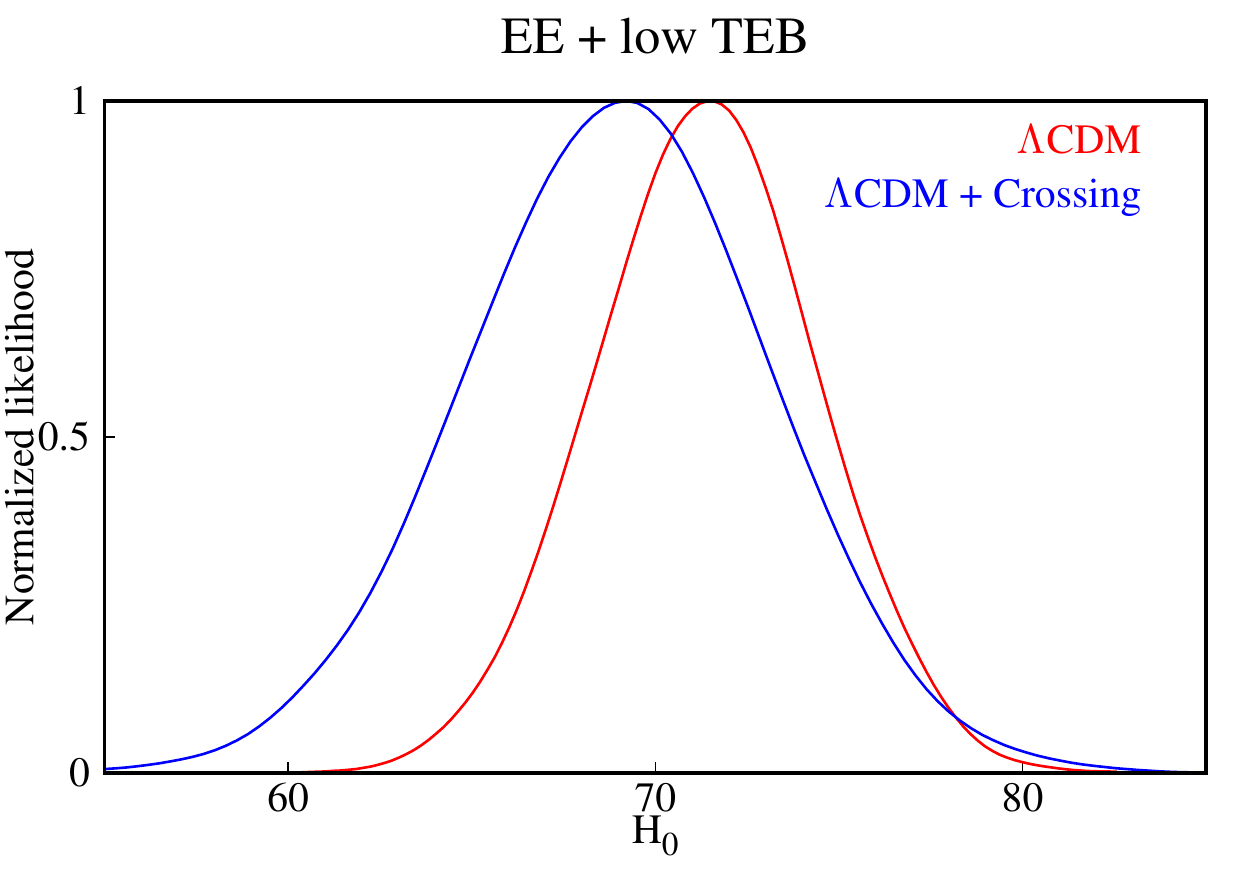}} 
\hskip 15pt\resizebox{120pt}{100pt}{\includegraphics{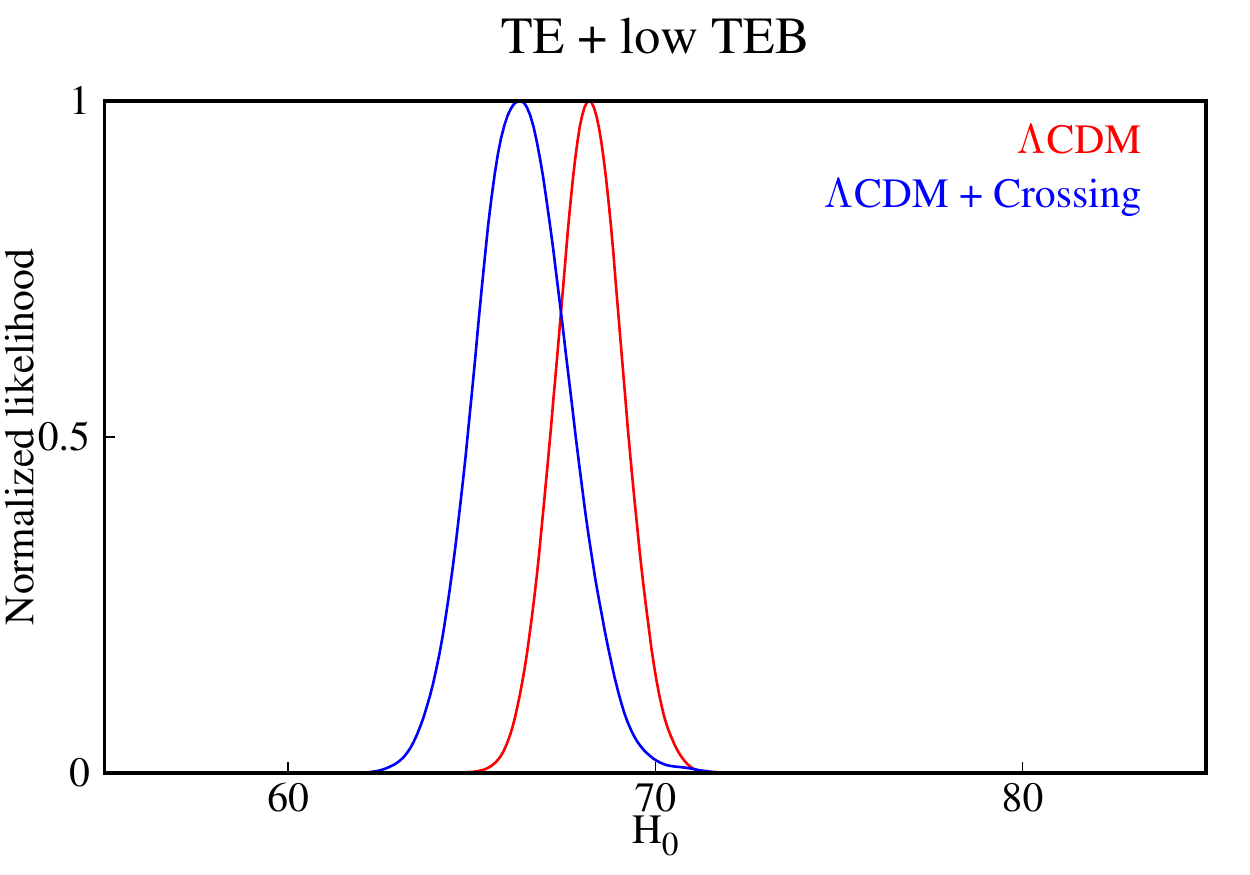}} 

\resizebox{135pt}{135pt}{\includegraphics{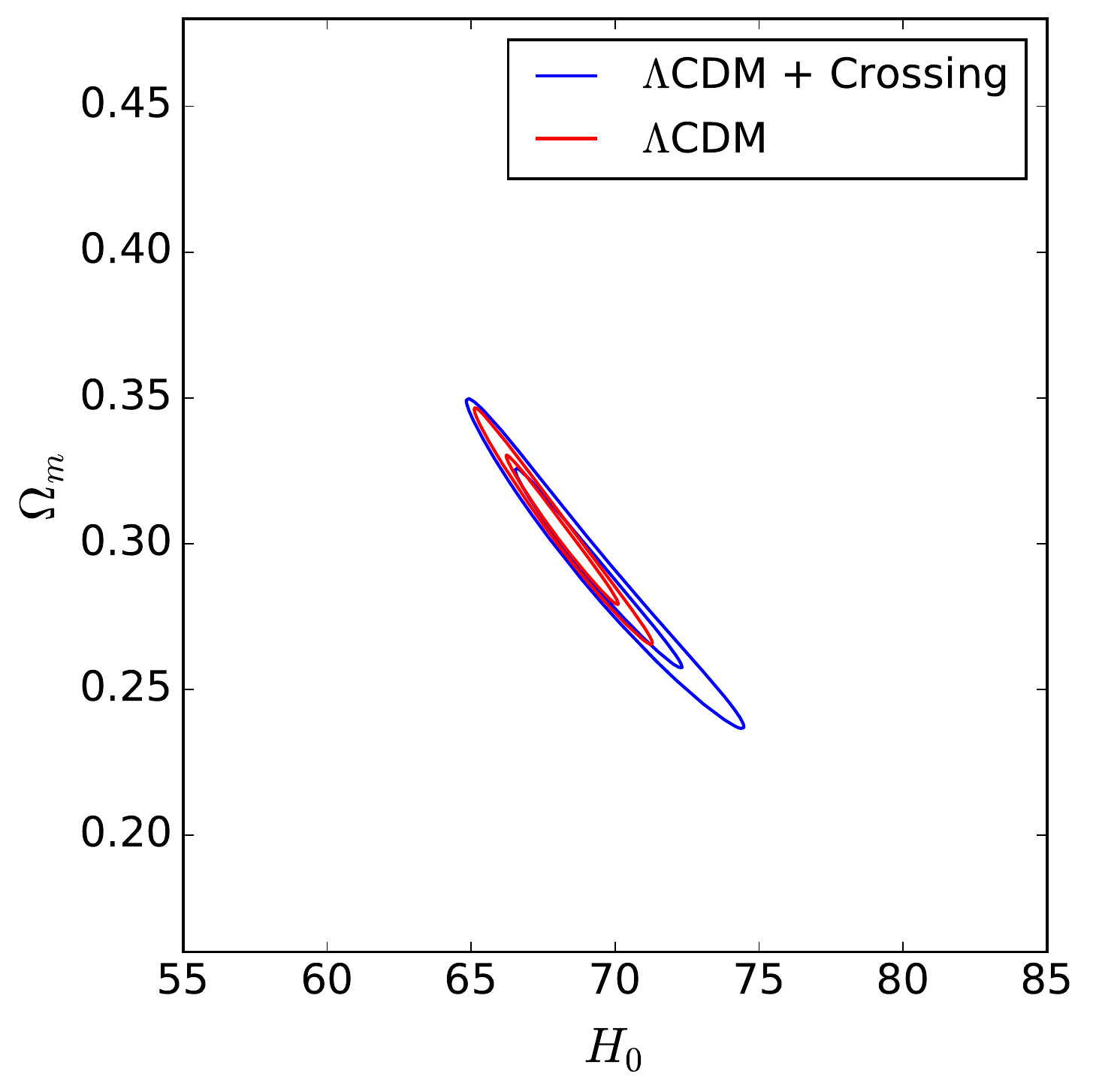}} 
\resizebox{135pt}{135pt}{\includegraphics{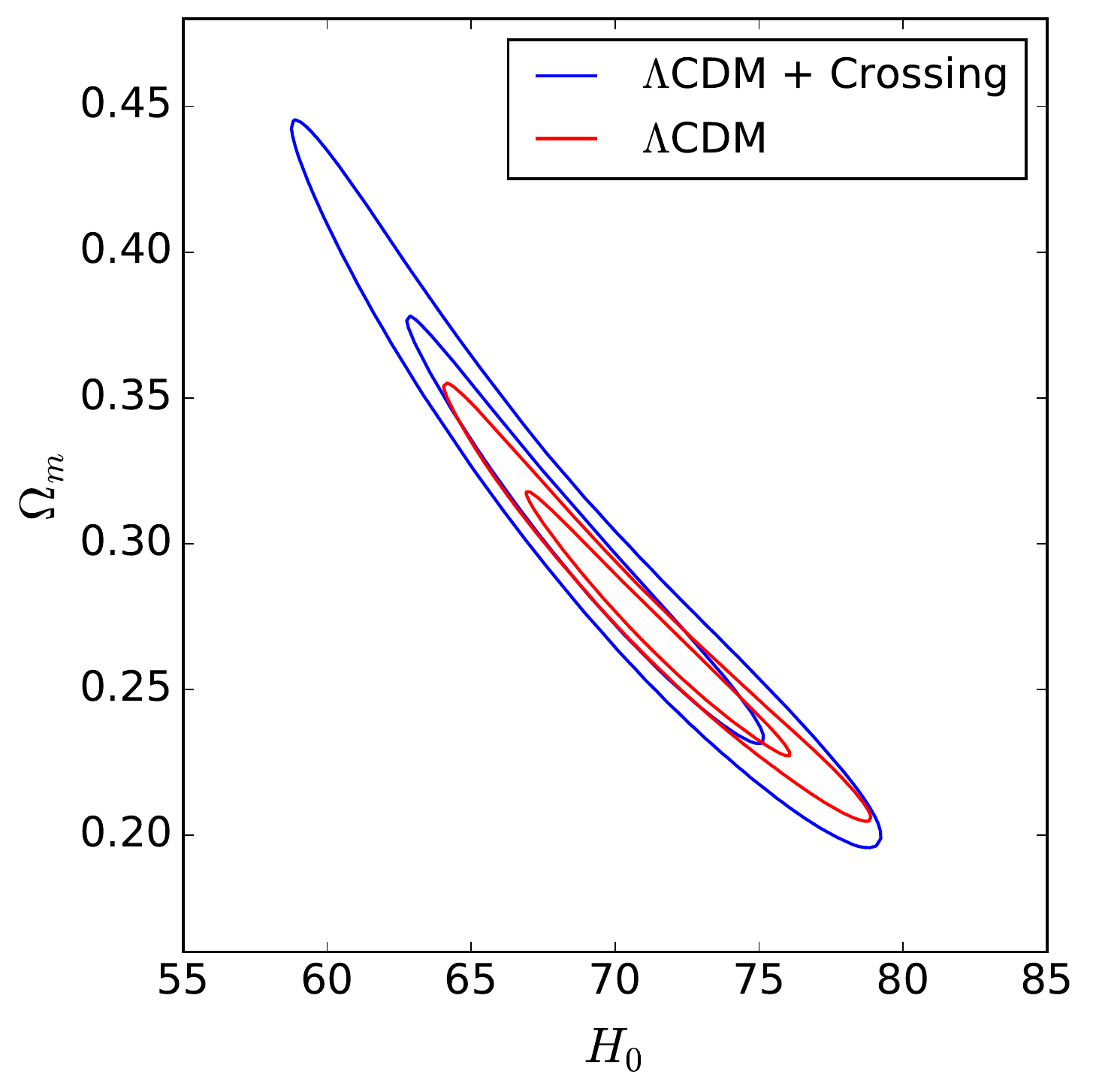}} 
\resizebox{135pt}{135pt}{\includegraphics{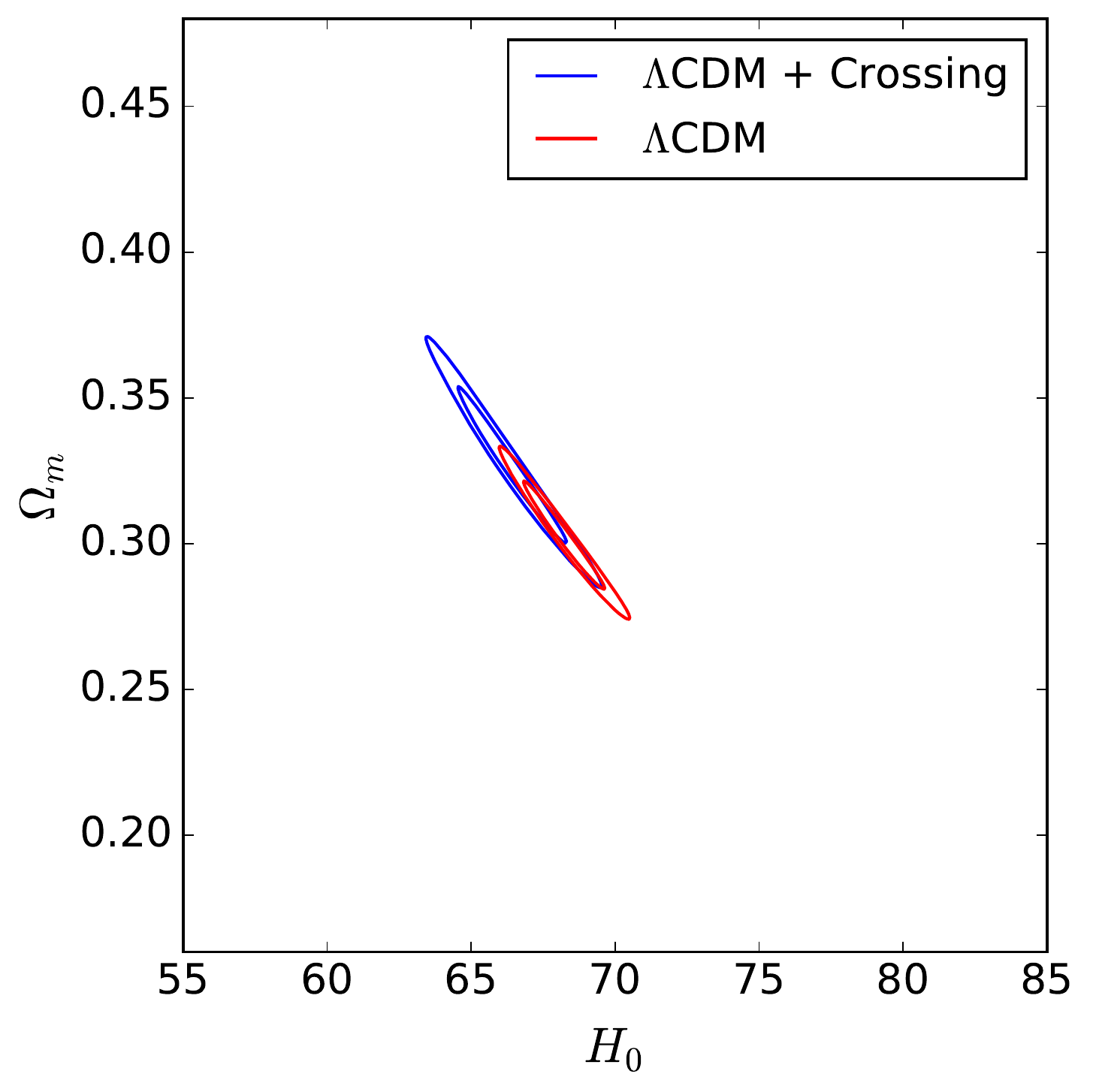}} 
\end{center}
\caption{\footnotesize\label{fig:OmmH0}Marginalized contours of matter density ($\Omega_{\rm m}$) and Hubble parameter 
($H_0$) when background parameters are allowed to vary along with Crossing hyperparameters. Here we provide the results 
considering fifth order Crossing function. Left panel displays the results from Planck TT and lowT datasets, middle panel provides the contours from EE + lowTEB  
and the right panel demonstrates TE and lowTEB constraints. The red and the blue contours represent the constraints on $\Lambda$CDM model and the 
$\Lambda$CDM model modified by Crossing functions, respectively.}
\end{figure*}

\subsection{Consistency of the $\Lambda$CDM}\label{subsec:conlcdm}

Testing consistency of the $\Lambda$CDM model in the same framework of Crossing statistics had indicated towards some required modifications at small scales in our previous
analysis~\cite{Hazra:2014hma}. With the Planck 2015 data we can test this consistency against both TT and EE data. Here, we allow all the cosmological parameters to vary. 
Hence here the mean function is no longer the best fit model. Instead $\Lambda$CDM model with different combination of cosmological parameters act as a group of mean functions. 
The angular power spectra from CAMB after adding lensing effects are further modified with the Crossing functions. 
The modified TT/EE spectra is then compared with TT + lowT/EE + lowTEB datasets. The marginalized contours of the 
crossing hyperparameters determines whether we need modifications in the concordance model of cosmology. 

In Fig.~\ref{fig:TTtoTTC5} we provide the results of the test against TT dataset using upto fifth order of crossing functions. Note that we find no indication of any deviations from 
the standard model as the $C_0=1$ and $C_{1-5}=0$ stays inside the 1$\sigma$ region of all the confidence contours. 

%The concordance model shows proper consistency with the Planck TT data. 

Similar analyses were carried out for EE data and we provide the results in Fig.~\ref{fig:EEtoEEC5}. Since the EE data has larger uncertainties, the contour
sizes are bigger compared to the TT case. Here too, we do not find any deviation from the standard model. Hence, both TT and EE datasets indicate that the concordance model 
of cosmology is well suited to explain the temperature and polarization data from the Planck CMB measurements. 

 We have also tested the consistency of the concordance model with TE + lowTEB data and the marginalized crossing hyperparameters' contours are provided in Fig.~\ref{fig:TEtoTEC5}.
TE data interestingly supports some deviation from the concordance model, though not at a significant level. In~\cite{Hazra:2016fkm}
our results points out some models of inflation that provides modifications to the power-law primordial power spectra such that it provides improvements in all (TT, TE, EE) datasets compared to 
the concordance model. In particular, we found improvements in fit to the TE data in almost all the cases of models with features. We believe the deviations found in the contour of the 
crossing hyperparameters might indicate toward such phenomena.

\begin{figure*}[!htb]
\begin{center} 
\resizebox{213pt}{160pt}{\includegraphics{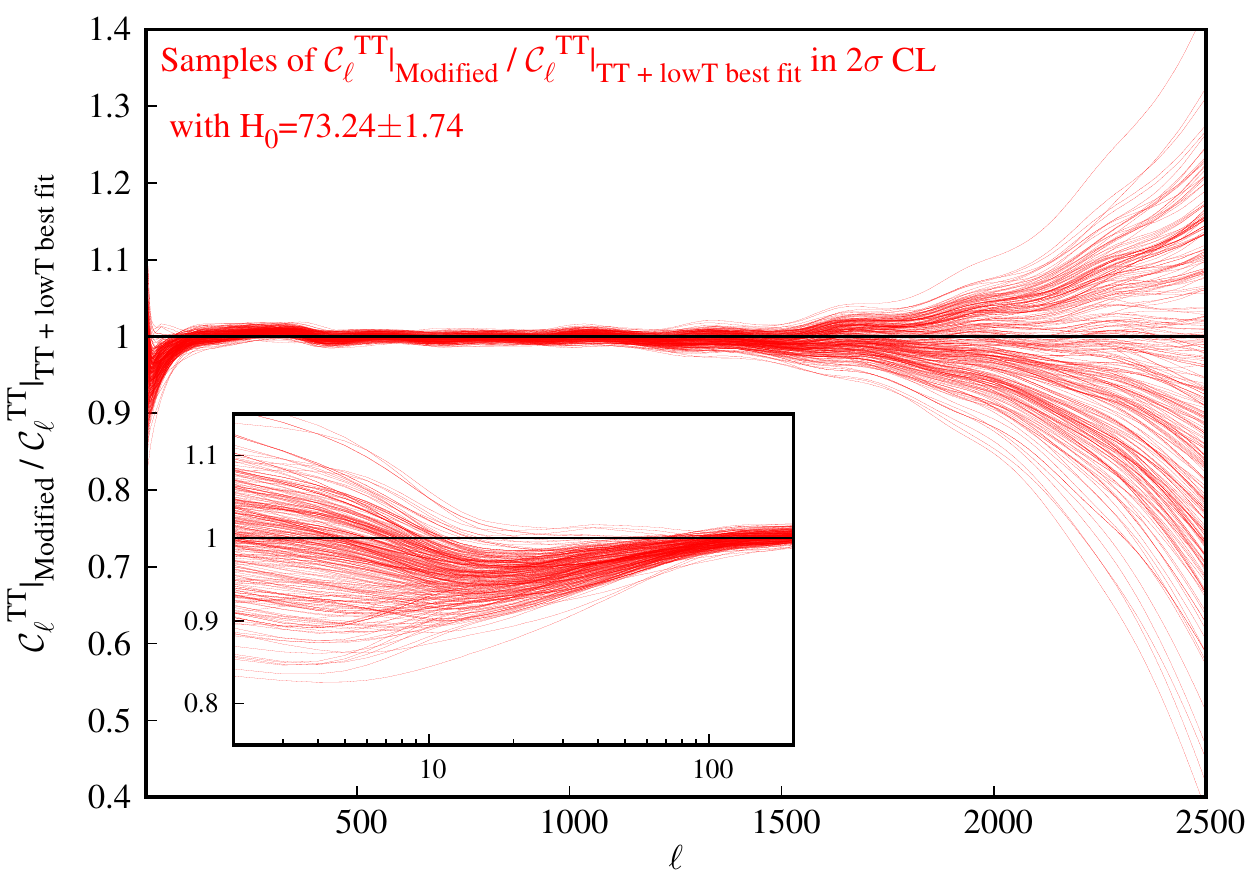}} 
\resizebox{213pt}{160pt}{\includegraphics{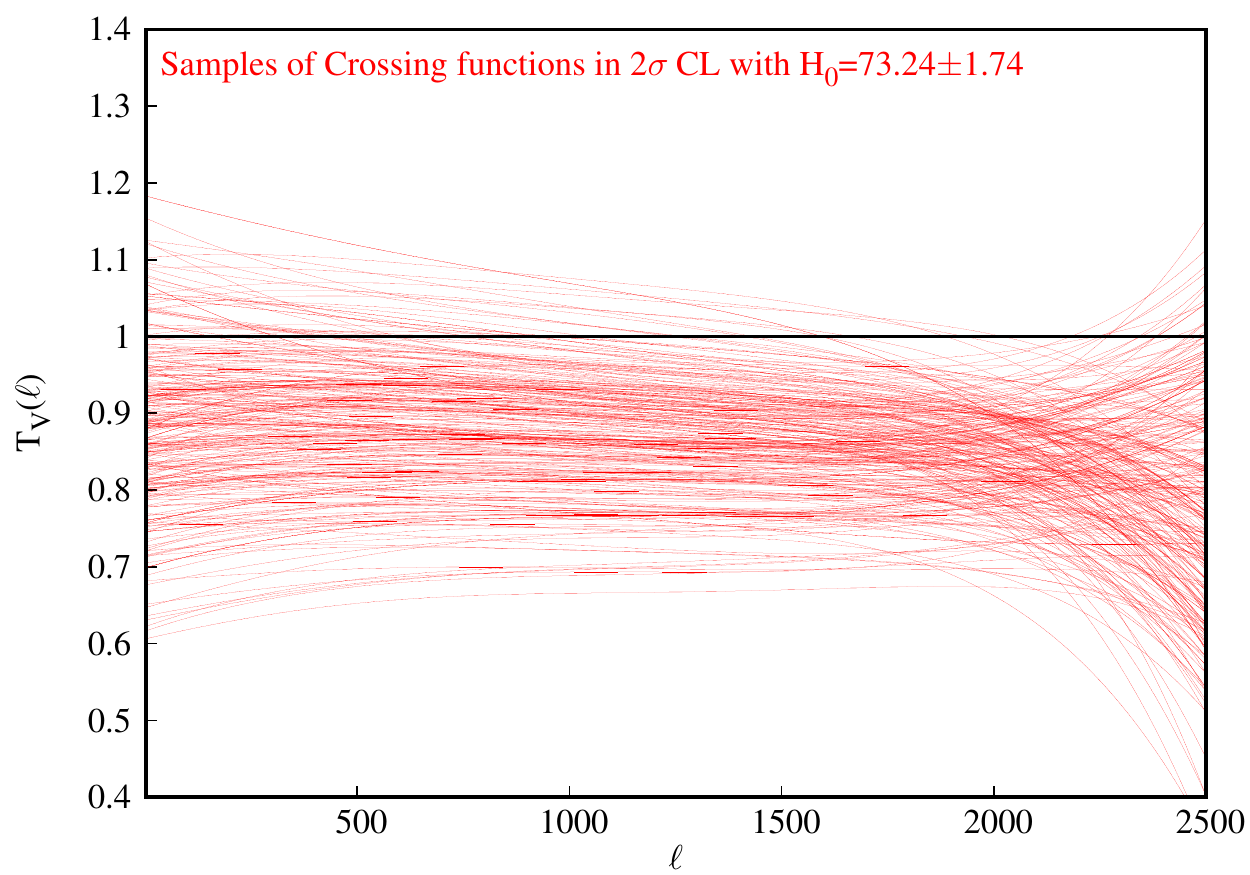}} 

\resizebox{213pt}{160pt}{\includegraphics{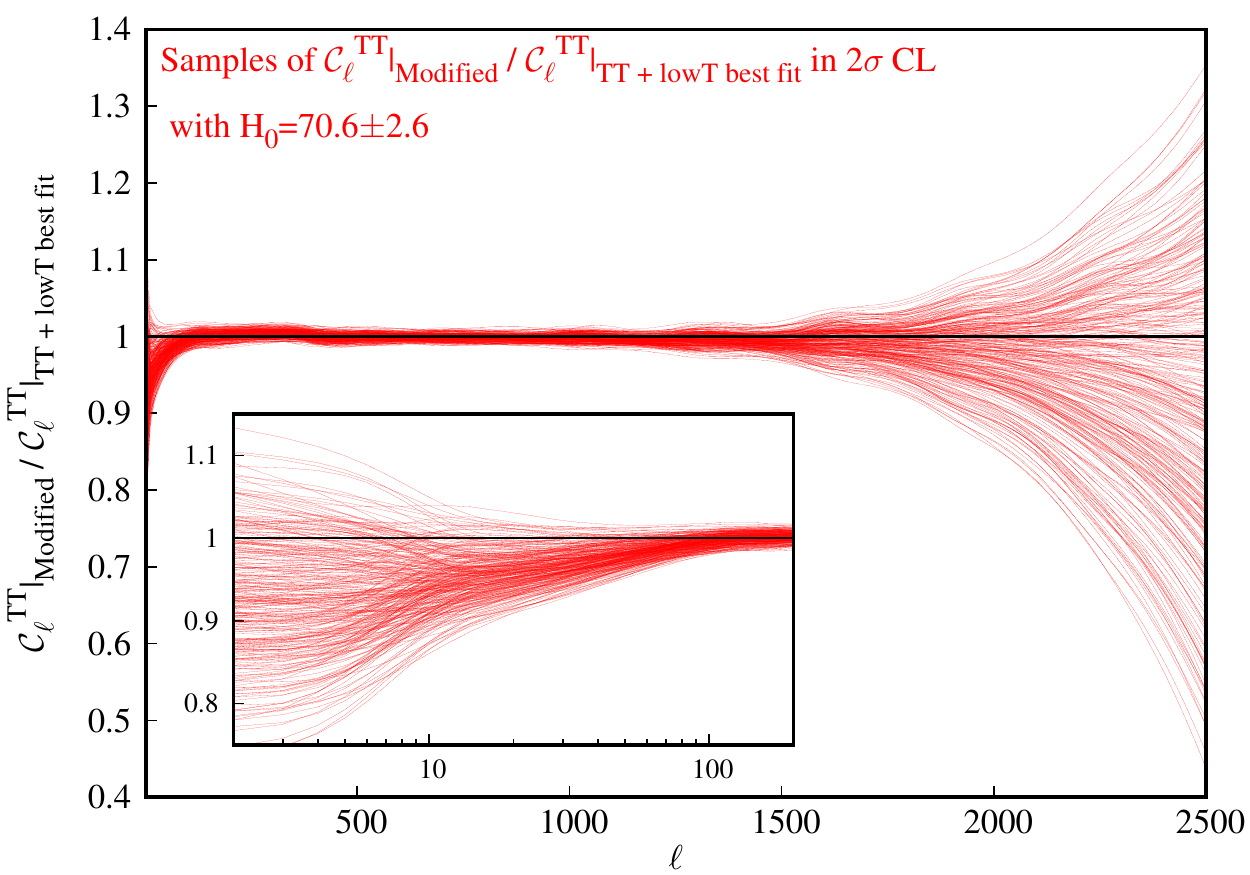}} 
\resizebox{213pt}{160pt}{\includegraphics{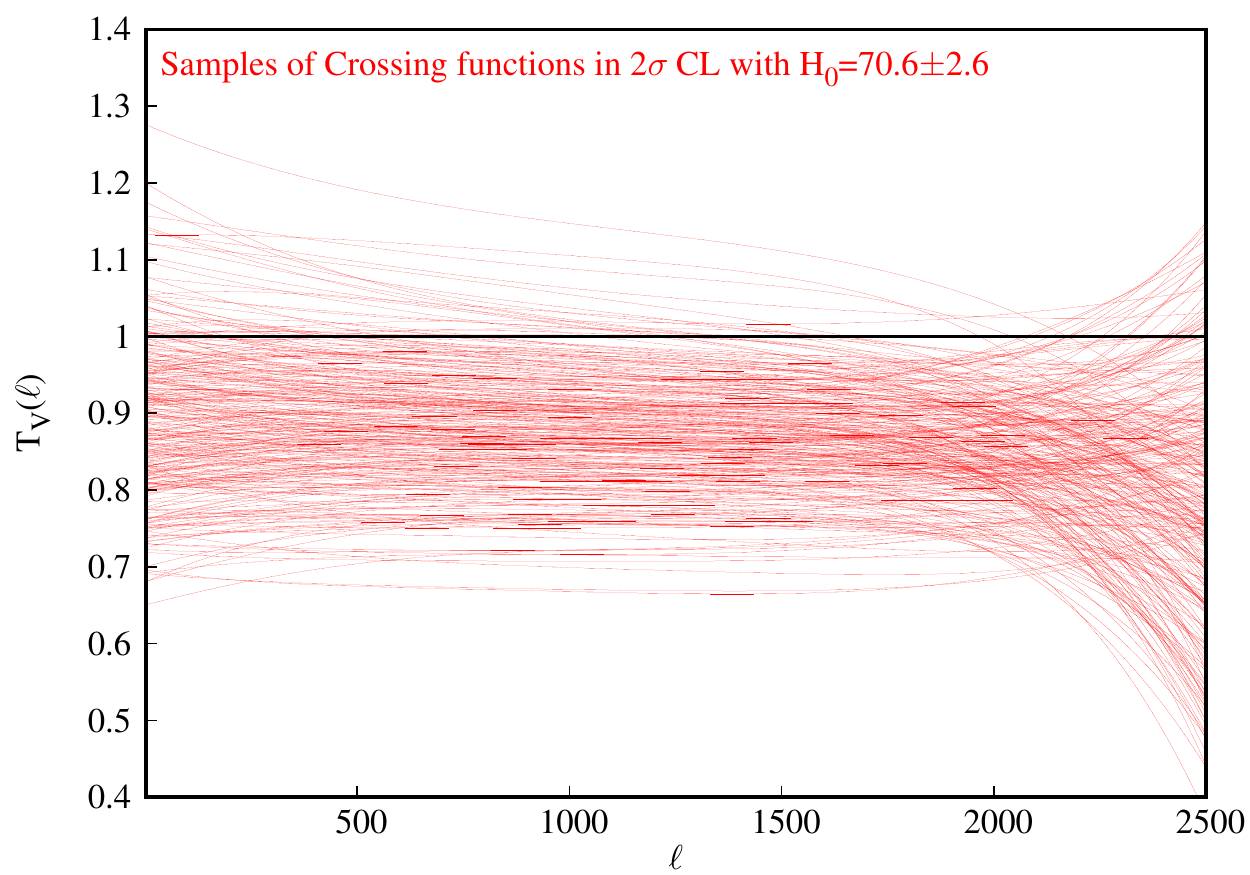}} 
\end{center}
\caption{\footnotesize\label{fig:samplesHST}[Top Left] Samples of modifications {\it w.r.t} the Planck TT + lowT best fit $\Lambda$CDM model
within 2$\sigma$ CL obtained when concordance model is compared to TT + lowT data considering fifth order Crossing function. Here we have only 
plotted the samples with $H_0=73.24\pm1.74$ in order to demonstrate the modifications required compared to the best fit model. Note that within
the 2$\sigma$ CL, most of the modifications to the concordance model require a suppression at small scales. [Top Right] For the same samples 
plotted in the left, we have plotted only the samples of Crossing functions. [Bottom Left] Similar to top left panel but only samples with 
$H_0=70.06\pm2.6$ ~\cite{Rigault:2014kaa} are depicted. [Bottom Right] Similar to top right panel but only the samples with $H_0=70.06\pm2.6$
are plotted. Note that we can see better consistency between the standard model and Planck data in the lower plots.} 
\end{figure*}

Considering Crossing functions introduce degeneracies in the parameter space of the background model as well. Here we show the results 
for the marginalized likelihoods of matter density ($\Omega_{\rm m}$) and the Hubble constant ($H_0$). We plot the 1D likelihood and 2D 
correlation contours of $\Omega_{\rm m}$ and $H_0$ in Fig.~\ref{fig:OmmH0}. On the left we plot the results from TT + lowT, at the centre we provide
the results from EE + lowTEB and on the right TE + lowTEB results are displayed. We also plot the results for $\Lambda$CDM model for comparison. 
The increase in the degeneracy is evident in the Crossing analysis.
However, it is interesting to note that, for TT + lowT, the contours extend in the direction of higher $H_0$ and lower $\Omega_{\rm m}$. As the tension
in the measurement of the Hubble parameter $H_0$, between Planck CMB for the concordance model and the local Hubble measurements~\cite{Riess:2016jrr} is significant, 
results from the Crossing modification may provide possible hints to solve the problem. Note that since the Crossing functions do not directly relate to any particular
physical parameter, it is not possible to determine the role of any beyond the standard model physics from this analysis. However, the modifications in the angular power
spectrum at different scales can indicate towards some plausible physical effects. To understand these modifications, in Fig.~\ref{fig:samplesHST} top panels we show 
the ratio of the ${\cal C}_{\ell}^{TT}$ and the best fit ${\cal C}_{\ell}^{TT}$ analyzing TT + lowT data when we only consider cases with the value of the 
Hubble parameter within $73.24\pm1.74$.

This can show us that if we set the value of Hubble parameter to what has been derived by local observations from ~\cite{Riess:2016jrr}, what 
Crossing modifications will be required to adjust the concordance model with Planck data. As we can see in Fig.~\ref{fig:samplesHST} top-left panel, 
the modified forms of the angular power spectrum at low multipoles show some mild deviation from the best fit $\Lambda$CDM model. To have a more clear
picture, on the right panel we plot the Crossing functions for the samples that are shown on the left panel. The Crossing functions represent the modifications
required to adjust the background model, here concordance $\Lambda$CDM model, with the given data. Note that majority of the Crossing functions indicate a 
suppression at almost all scales, though they are strongly scale variant. This can hint us towards possible effects than can be originated by some well 
defined physical processes. This can include an extra relativistic species, a higher neutrino mass compared to the baseline model assumption of $0.06$eV, 
running in the primordial power spectrum and lower amplitude of the primordial fluctuations. We should emphasize here that this is only for a case that we
set a strong prior on the $H_0$ measurements from ~\cite{Riess:2016jrr}.

Considering a prior on $H_0 = 70.6 \pm 2.6 ~{\rm Km/s/MPc}$ from~\cite{Rigault:2014kaa} can result to a better consistency between the concordance
model and Planck data as we can see in the bottom plots of Fig.~\ref{fig:samplesHST}. 

 In Fig.~\ref{fig:OmmH0}, for EE and TE data, we do not find the similar trend in the Hubble parameter. In fact both EE and TE supports higher $\Omega_{\rm m}$ and lower 
$H_0$ when crossing hyperparameters are used as the degeneracies increase towards the opposite direction of TT. However we find that in both the cases $\Lambda$CDM has a 
substantial overlap with the $\Lambda$CDM + Crossing case at the 1$\sigma$ level and thereby do not provide any strong support for beyond the concordance model.

\section{Conclusions}\label{sec:conclusions}
In this work we repeat our previous analysis of Planck 2013 data~\cite{Hazra:2013oqa,Hazra:2014hma} on complete Planck 2015 temperature and polarization angular power 
spectrum data. We have implemented the Crossing statistic to test the internal consistency between Planck temperature and polarization data as well as testing the 
consistency of the concordance model with the full Planck data. 

As we did in our previous work, here also we model the deviation from the concordance model using different orders of the Crossing function. Deviation from the concordance model 
is estimated by deriving the confidence contours of the Crossing hyperparameters as we marginalized over parameters of the cosmological model (mean function). In this approach if
we derive Crossing function to be consistent with `one' at all multiples we can conclude that the given mean function (here the concordance model) has proper consistency to the data. 
Crossing function of `one' basically represent that the data does not suggest any deviation from the given mean function. In the first part of this work we compared the temperature
and polarization (EE and TE) data and we found proper concordance between the two data. We then confronted the concordance model to TT, TE and EE data and again we found 
the model is in proper agreement with both data. 

While temperature and polarization data from Planck show clear agreement with each other, we noticed a systematic amplitude difference between the two spectra where the temperature
data seems to have lower power than what we expect in comparison with the EE polarization data (at all $\ell$). While our finding is not about a significant inconsistency, it might 
hint towards a possible calibration issue that can be corrected further. 

In our analysis of the Planck 2013 temperature data we found 2-3$\sigma$ deviation from predictions of the concordance $\Lambda$CDM model suggested by confidence contours of the
Crossing hyperparameters~\cite{Hazra:2014hma}. Excluding the 217x217 GHz channel data, suggested by some colleagues from Planck collaboration, the discrepancy was removed. Planck 
2015 seems to have corrected for all the systematics in 217x217 GHz and our analysis in this paper show that the concordance model has now a proper consistency with the data.

At last, our analysis reveals something interesting that seems to be important interpreting the Planck CMB angular power spectrum data. Considering Crossing functions along with the
concordance model (as the mean function) and using Planck 2015 temperature data the probability distribution function of the Hubble parameter $H_0$ broadens towards larger $H_0$ values resulting to $H_0=69.4^{+1.76}_{-2}$ which shows proper agreement with all local observations ~\cite{Riess:2016jrr,Rigault:2014kaa}. This clearly reflects the fact that Planck CMB temperature data has no fundamental tension with the local $H_0$ measurements and there are in fact forms of the angular power spectrum that are fully consistent with Planck data as well as with local $H_0$ measurements. Setting a strong prior on $H_0$ from~\cite{Riess:2016jrr} while our mean functions are from the  background concordance $\Lambda$CDM model, data suggest a systematic suppression at low and high
multiples. This can hints towards some possible physical effects such as having running in the spectrum of the primordial fluctuations, having an additional relativistic species 
or a massive neutrino.  However, analysis of the TE and EE data suggest something different as the marginalized confidence contours of $H_0$ broadens towards lower values (and higher values of matter density). In other words the tension between the Planck 2015 polarization data (EE and TE) and local measurement of $H_0$ from~\cite{Riess:2016jrr} seems to be serious and beyond model assumption or choice of parametric fitting.

We should note at the end that one can also consider that there might be some systematics in the measurement of $H_0$ from ~\cite{Riess:2016jrr} since we could see less 
deviations from expectations of the concordance model considering $H_0$ priors from ~\cite{Rigault:2014kaa}. This highlights the importance of independent and accurate measurements 
of $H_0$ for precision cosmology~\cite{L'Huillier:2016lzh,Farooq:2016zwm,Bernal:2016gxb}.

%%%%%%%%%%%%%%%%%%%%%%%%%%%%%%%%%%%%%%%%%%%%%%%%%%%%%%%%%%%%%%%%%%%%%%%%%%%%%%%

\section*{Acknowledgments}
A.S. would like to acknowledge the support of the National Research Foundation of Korea (NRF-2016R1C1B2016478). We also acknowledge the use of publicly available CAMB and 
CosmoMC in our analysis. DKH acknowledges Laboratoire APC-PCCP, Universit\'e Paris Diderot and Sorbonne Paris Cit\'e (DXCACHEXGS) and also the financial support of the 
UnivEarthS Labex program at Sorbonne Paris Cit\'e (ANR-10-LABX-0023 and ANR-11-IDEX-0005-02).

%%%%%%%%%%%%%%%%%%%%%%%%%%%%%%%%%%%%%%%%%%%%%%%%%%%%%%%%%%%%%%%%%%%%%%%%%%%%%%%

%%%%%%%%%%%%%%%%%%%%%%%%%%%%%%%%%%%%%%%%%%%%%%%%%%%%%%%%%%%%%%%%%%%%%%%%%%%%%%%
\end{document}